\documentclass[prd,amssymb,amsmath,amsfonts,nofootinbib,reprint,showpacs,longbibliography]{revtex4-1}

\usepackage{graphicx}
\usepackage{lmodern}
\usepackage{amsmath,amssymb}
\usepackage{mathrsfs}
\usepackage{amsfonts}
\usepackage[utf8]{inputenc}
\usepackage{url}
\usepackage[colorlinks]{hyperref}
\usepackage[dvipsnames]{xcolor}
\usepackage{multirow}
\usepackage[normalem]{ulem}
\usepackage{lipsum}

\usepackage{marvosym}
\usepackage{enumerate}
\usepackage{color,soul}
\usepackage{acronym}

\usepackage{bm}
\usepackage{color}
\usepackage{commath}
\allowdisplaybreaks
\usepackage{multirow}

\newcommand{\mbf}[1]{\mathbf{#1}}

\newcommand{\Lhat}{\bm{\ell}}

\newcommand{\Sa}{\mbf{S}_1}
\newcommand{\Sb}{\mbf{S}_2}
\newcommand{\Lhatdot}{\dot{\bm{\ell}}}

\newcommand{\Sadot}{\dot{\mbf{S}}_1}
\newcommand{\Sbdot}{\dot{\mbf{S}}_2}

\newcommand{\ud}{\mathrm{d}}

\newcommand{\Jz}{\hat{L}_0}
\newcommand{\Ez}{\hat{E}_0}

\def\TEOBResumS{\texttt{TEOBResumS}}
\def\RIT{\texttt{RIT}}
\def\MAYA{\texttt{MAYA}}
\def\TEOBResumSGIOTTO{\texttt{TEOBResumS-GIOTTO}}
\def\TEOBResumSDali{\texttt{TEOBResumS-Dalì}}
\def\SEOBNRE{\texttt{SEOBNRE}~}
\def\SEOBNRvE{\texttt{SEOBNRv4EHM}~}

\newacro{adm}[ADM]{Arnowitt-Deser-Misner}
\newacro{bbh}[BBH]{binary black hole}
\newacro{bh}[BH]{black hole}
\newacroplural{bh}[BHs]{black holes}
\newacro{bhns}[BHNS]{black hole-neutron star}
\newacro{bns}[BNS]{binary neutron star}
\newacro{bf}[BF]{Bayes' factor}
\newacro{cbc}[CBC]{compact binary coalescence}
\newacro{ce}[CE]{Cosmic Explorer}
\newacro{da}[DA]{data analysis}
\newacro{et}[ET]{Einstein Telescope}
\newacro{eob}[EOB]{Effective-One-Body}
\newacro{eom}[EOM]{equations of motion}
\newacro{fd}[FD]{frequency domain}
\newacro{fft}[FFT]{Fast Fourier transform}
\newacro{gw}[GW]{gravitational-wave}
\newacroplural{gw}[GWs]{Gravitational-waves}
\newacro{gr}[GR]{general relativity}
\newacro{grb}[GRB]{gamma-ray burst}
\newacro{grhd}[GRHD]{general-relativistic hydrodynamics}
\newacro{gwosc}[GWOSC]{Gravitational Wave Open Science Center}
\newacro{gwtc1}[GWTC-1]{the first gravitational-wave transients catalog}
\newacro{gsf}[GSF]{Gravitational Self Force}
\newacro{hm}[HM]{Higher mode}
\newacroplural{hm}[HMs]{Higher modes}
\newacro{ifo}[IFO]{interferometer}
\newacro{imr}[IMR]{inspiral-merger-ringdown}
\newacro{im}[IMR]{inspiral-to-merger}
\newacro{kagra}[KAGRA]{Kamioka Gravitational Wave Detector}
\newacro{ligo}[LIGO]{Laser Interferometer Gravitational-Wave Observatory}
\newacro{lso}[LSO]{Last Stable Orbit}
\newacro{lvc}[LVC]{LIGO-Virgo Collaboration}
\newacro{lvk}[LVK]{LIGO-Virgo-Kagra Collaboration}
\newacro{lo}[LO]{leading order}
\newacro{ns}[NS]{neutron star}
\newacroplural{ns}[NSs]{neutron stars}
\newacro{nr}[NR]{numerical relativity}
\newacro{nqc}[NQCs]{Next-to-quasicircular corrections}
\newacro{nlo}[NLO]{next-to-leading order}
\newacro{nnlo}[NNLO]{next-to-next-to-leading order}
\newacro{n3lo}[N3LO]{next-to-next-to-next-to-leading order}
\newacro{n4lo}[N3LO]{next-to-next-to-next-to-next-to-leading order}
\newacro{ode}[ODE]{Ordinary Differential Equation}
\newacroplural{ode}[ODEs]{Ordinary Differential Equations}
\newacro{pn}[PN]{post-Newtonian}
\newacro{pm}[PM]{post-Minkowskian}
\newacro{pe}[PE]{parameter estimation}
\newacro{psd}[PSD]{power spectral density}
\newacro{pa}[PA]{post-adiabatic}
\newacro{qnm}[QNM]{quasi-normal mode}
\newacro{qc}[QC]{quasi-circular}
\newacro{rwz}[RWZ]{Regge-Wheeler-Zerilli}
\newacro{snr}[SNR]{signal-to-noise ratio}
\newacro{spa}[SPA]{stationary-phase approximation}
\newacro{sxs}[SXS]{Simulating eXtreme Spacetimes}
\newacro{td}[TD]{time domain}
\newacro{ng}[NG]{Nect Generation}

\definecolor{cyan}{rgb}{0,0.9,0.9}
\definecolor{orange}{rgb}{0.9,0.5,0}
\definecolor{magenta}{rgb}{1,0,1}
\definecolor{purple}{rgb}{0.8,0.4,0.8}
\definecolor{gray}{rgb}{0.8242,0.8242,0.8242}
\definecolor{dodgerblue}{rgb}{0.12, 0.56, 1.0}

\begin{document}

\title{Towards efficient Effective One Body models for generic, non-planar orbits}
\author{Rossella \surname{Gamba}${}^{1,2,3}$}
\author{Danilo \surname{Chiaramello}${}^{4}$}
\author{Sayan \surname{Neogi}${}^{3,5}$}

\affiliation{${}^{1}$ Institute for Gravitation \& the Cosmos, The Pennsylvania State University, University Park PA 16802, USA}
\affiliation{${}^{2}$ Department of Physics, University of California, Berkeley, CA 94720, USA}
\affiliation{${}^{3}$ Theoretisch-Physikalisches Institut, Friedrich-Schiller-Universit{\"a}t Jena, 07743, Jena, Germany}
\affiliation{${}^{4}$ INFN sezione di Torino, Torino, 10125, Italy}
\affiliation{${}^{5}$ Indian Institute of Science Education and Research, Homi Bhabha Road, Pashan, Pune 411008, India.}

\begin{abstract}
Complete waveform models able to account for arbitrary non-planar orbits
represent a holy grail in current gravitational-wave astronomy.
Here, we take a step towards this direction and present a simple yet efficient 
prescription to obtain the evolution of the spin vectors and of the orbital angular 
momentum along non-circularized orbits, that can be applied to any eccentric aligned-spins waveform model.
The scheme employed is motivated by insights gained from the \ac{pn} regime.
We investigate the phenomenology of the Euler angles characterizing the time-dependent rotation
that connects the co-precessing frame to the inertial one, gauging the importance of non-circular 
terms in the evolution of the spins of a precessing binary. We demonstrate that such terms are 
largely negligible, irrespectively of the details of the orbit.
Such insights are confirmed by studying the radiation-frame of a few eccentric, precessing \ac{nr} simulations. 
Our investigations confirm that the usual ``twisting'' technique
employed for quasi-spherical systems can be safely applied to non-circularized binaries.
By then augmenting a state-of-the-art \ac{eob} model for non-circular planar orbits 
with the prescription discussed, we obtain an \ac{imr} model
for eccentric, precessing \acp{bbh}.
We validate the model in the quasi-spherical limit via mismatches and present one phasing comparison against
a precessing, eccentric simulation from the \RIT\  catalog.

\end{abstract}

\date{\today}
\maketitle

\acresetall

\section{Introduction}

Ever since the first historic \ac{bbh} detection, the field of \ac{gw} astronomy has attracted
much interest from the broader physics community. The wealth of information that can be garnered from the 
detection and analysis of \acp{cbc} has been proven time and time again~\cite{Abbott:2017xzu,
TheLIGOScientific:2017qsa, Abbott:2018exr, Annala:2017llu, Radice:2018pdn, De:2018uhw, Fattoyev:2017jql, Most:2018hfd, Raithel:2018ncd, Tews:2018chv,
Meidam:2014jpa, TheLIGOScientific:2016src, LIGOScientific:2018dkp}, with the 
growth in the number of observed events being accompanied by detection of exceptional systems, each 
characterized by peculiar features such as very unequal masses~\cite{LIGOScientific:2020ibl, LIGOScientific:2020stg}, hints of spins 
precession~\cite{KAGRA:2021vkt, LIGOScientific:2020ibl, Varma:2022pld}, large total mass~\cite{Abbott:2020tfl,Abbott:2020mjq}, the presence of one 
or more \acp{ns}~\cite{TheLIGOScientific:2017qsa, LIGOScientific:2017ync, LIGOScientific:2020aai, LIGOScientific:2021qlt} and more.
One such exceptional event, GW190521~\cite{Abbott:2020tfl,Abbott:2020mjq}, has been the center of attention for many 
groups, with a large number of possible astrophysical interpretations having been 
suggested~\cite{Gayathri:2020coq, Romero-Shaw:2020thy, CalderonBustillo:2020odh, Shibata:2021sau, Nitz:2020mga, Gamba:2021gap}. 
Ref.~\cite{Gayathri:2020coq}, in particular, suggested that GW190521 could be interpreted 
as the merger of two precessing \acp{bh} coalescing along highly eccentric orbits.
While successive studies have shown that the effects of precession and non-circularity on the detected signal are mostly degenerate for 
binaries with large total mass~\cite{CalderonBustillo:2020odh, Romero-Shaw:2022fbf} and that prior choices strongly 
affect the outcome of the analysis itself~\cite{Chandra:2023nge}, it is nonetheless of paramount importance for 
the~\ac{gw} modelling community to be able to deliver models that can quickly and 
reliably generate waveforms for these kinds of binaries, thus covering a portion of 
the parameter space that has been up to now largely ignored. 
In fact, only by relying on complete models one can hope to fully understand the interplay between the two effects, 
and break the degeneracies discussed.

The history of the development of models containing precession is rather rich~\cite{Apostolatos:1994mx, Buonanno:2002fy, Schmidt:2010it, Schmidt:2012rh, Boyle:2011gg, OShaughnessy:2011pmr}. 
Most models now employ a ``twist'' technique~\cite{Schmidt:2010it, Schmidt:2012rh, Boyle:2011gg, OShaughnessy:2011pmr}, 
coupled with a way of obtaining the evolution of the so-called ``co-precessing frame'', in which waveforms appear as if they were emitted 
by a quasi-aligned system~\cite{Buonanno:2002fy,Pan:2013rra}. While many non-trivial effects have just started to be properly 
understood and modelled, such as mode asymmetries or the merger-ringdown emission~\cite{Thompson:2023ase, Ghosh:2023mhc, Hamilton:2023znn}, 
it is safe to say that \ac{gw} models of precessing binaries 
have reached a mature state~\cite{London:2017bcn,Garcia-Quiros:2020qpx,Khan:2019kot,Hannam:2013oca,Schmidt:2014iyl,Khan:2018fmp,Pratten:2020ceb, Ossokine:2020kjp, Akcay:2020qrj,Ramos-Buades:2023ehm}, 
and they are now routinely employed in \ac{pe} of \ac{gw} data.

The inclusion of eccentricity, instead, is much more recent -- and largely limited to the
\ac{eob}~\cite{Buonanno:1998gg,Buonanno:2000ef,Damour:2000we,Damour:2001tu,Damour:2008qf,Nagar:2011fx,Damour:2015isa}
family of models\footnote{Though note also the \ac{pn} models of ~\cite{Hinder:2017sxy,Islam:2024tcs} and the non-spinning, eccentric
surrogate of~\cite{Islam:2021mha}.} that can rely on a Hamiltonian formulation of the dynamics not restricted to
quasi-circular or quasi-spherical systems~\cite{Damour:2014afa,Hinderer:2017jcs}. 
Three main families of models exist: the \SEOBNRE~\cite{Cao:2017ndf, Liu:2021pkr, Liu:2023dgl}, 
\TEOBResumSDali~\cite{Chiaramello:2020ehz, Nagar:2021gss, Nagar:2021xnh, Nagar:2023zxh} 
and \SEOBNRvE~\cite{Ramos-Buades:2021adz} models.
All differ in the way eccentricity is incorporated into the equations of motion
-- and in particular, in the radiation reaction driving the dissipative dynamics.
While the \SEOBNRE models include non-circular effects in the radiation reaction up to $2$\ac{pn} as an additive correction 
to the circular terms in the energy-balance equation, the \TEOBResumSDali~models employ a different strategy, and include the non-circular terms in the radiation reaction as a multiplicative
Newtonian correction to the factorized EOB waveform. The \SEOBNRvE model does not include any explicit
non-circular terms in the radiation reaction, which is driven by the circular terms only, but a-posteriori
includes 2\ac{pn} order corrections in the computation of the waveform~\cite{Khalil:2021txt}.
Comparisons of the validity of these approaches have been carried out in the test-mass limit~\cite{Placidi:2021rkh},
showing that for particles moving in Schwarzschild spacetimes the waveforms obtained with the \TEOBResumSDali~prescription
are in excellent agreement with the exact numerical waveforms obtained solving the \ac{rwz} and Teukolsky equations~\cite{Albanesi:2022ywx}, 
and on average closer to the numerical results than the ones obtained by including explicit $2$\ac{pn} expressions.

The inclusion of both eccentricity and precession is a regime that is still mostly unexplored.
A few studies have recently investigated the interplay between the two effects~\cite{Klein:2018ybm,Phukon:2019gfh,Klein:2021jtd,Ireland:2019tao,Fumagalli:2023hde,Arredondo:2024nsl}, but the
considerations are typically (i) limited to purely \ac{pn} (and at times Newtonian) arguments, (ii) based on
orbit-averaged \ac{pn} expressions, (iii) not immediately applicable
to the full \ac{imr} regimes or (iv) not validated against full-fledged \ac{nr} simulations.
The lack of such simulations spanning a large number of
orbital cycles and covering the parameter space of interest has so far limited the development of models for these systems. 
The only exception is the recent work of~\cite{Liu:2023ldr}, who have extended the 
\SEOBNRE model to include the evolution of the spins along non-circular orbits.
In this work, the authors solve the full \ac{eob} equations, with a spherical (rather than planar)
\ac{eob} Hamiltonian (borrowed from the {\texttt{SEOBNRv4}} model~\cite{Bohe:2016gbl, Cotesta:2018fcv}), augmenting the spins equations of motions with
explicit non-circular terms. 
While general, this approach is computationally expensive, and it is not clear whether it is truly necessary to solve the full
spins equations of motion together with the \ac{eob} dynamics to obtain a faithful model for the precessing, eccentric regime.

In this paper, we present a simple yet efficient scheme to obtain waveforms from generic non-planar orbits,
that can be applied to any eccentric aligned-spins waveform model.
Section~\ref{sec:pn} is dedicated to a brief review of the \ac{pn} equations of motion for non-circularized precessing binaries.
Starting from the full 3\ac{pn} equations of motion, we apply successive approximations to gauge the importance of non-circular terms
in the evolutions of the spins.
Section~\ref{sec:nr} tests the intuitions gained in the \ac{pn} sector by identifying and inspecting the co-precessing (radiation) frame 
of a few chosen reference \ac{nr} simulations~\cite{Healy:2022wdn}. 
We show that the evolution in this frame resembles that of an aligned-spin system, as expected. We also compare two simulations
having same initial conditions but different eccentricities, corroborating the findings of Sec.~\ref{sec:pn}.
Section~\ref{sec:eob} presents the extension of the \TEOBResumSDali~model to include the evolution of the spins along non-circular orbits.
Putting together the insights gained from the \ac{pn} regime and the \ac{nr} simulations, we show that the scheme employed is
able to capture the main features of eccentric precessing waveforms.
Section~\ref{sec:validation} is dedicated to the validation of the model in the quasi-circular, precessing limit
against the same simulations considered in~\cite{Gamba:2021ydi}. We also present one phasing comparison against a precessing, 
eccentric simulation from the \RIT catalog.
Finally, Sec.~\ref{sec:conclusions} summarizes the main results and discusses the implications of this work as well as 
avenues for future developments.

Throughout the paper we use geometrized units, setting $G=c=1$. We denote the component masses of a binary system as $m_1, m_2$, 
and the total mass as $M = m_1 + m_2$; the mass fractions are $X_{1,2} = m_{1,2}/M$, the mass ratio is $q = m_1/m_2 \geq 1$, and the symmetric mass ratio is $\nu = q/(1+q)^2$, with
the reduced mass given by $\mu = M \nu$.
The spin vectors are $\Sa, \Sb$, and they are related to the dimensionless spins
$\bm{\chi}_{1,2}$ by $\Sa = m_1^2 \bm{\chi}_1, \Sb = m_2^2 \bm{\chi}_2$. The total spin
is given by $\mathbf{S} = \Sa + \Sb$, and the spin difference by $\mathbf{\Sigma} = \Sb/X_2 - \Sa/X_1$; the effective spin variable is 
$\mathbf{\chi}_\text{eff} = X_1 \chi_{1,z} + X_2 \chi_{2,z}$, while the orthogonal spin parameter is
$\chi_p = \max \left\{|\bm{\chi}_{1, \perp}|, \dfrac{4+3q}{4q^2+3q} |\bm{\chi}_{2, \perp}|\right\}$~\cite{Schmidt:2014iyl}.

\section{The PN sandbox}
\label{sec:pn}

While often unreliable from a quantitative point of view, \ac{pn} theory is a powerful tool to gain
qualitative insights on the dynamics of \acp{cbc}. In this section, we employ \ac{pn} equations of motion
to gauge the importance of non-circular terms in the evolution of the spins of a precessing binary, 
and more generically review the effects that non-quasicircular evolution has on the time-dependent rotation
that connects the co-precessing frame to the inertial one. Our approach is often pedagogical, aiming to
reinforce and extend the intuitions that have been acquired over the years during the development of quasi-spherical 
models for precessing binaries.

\subsection{Reference frames and equations of motion}
\label{subsec:reference_frames}

Consider a \ac{bbh} system with spins $\Sa, \Sb$ and total mass $M = m_1 + m_2$, moving in a non-planar orbit.
Following the notation employed in, e.g.,~\cite{Bohe:2012mr, Blanchet:2013haa},
we describe the system in center of mass coordinates, and introduce the relative position and velocity vectors
$\mbf{r}, \bm{v}$. 
We denote the unit vector of the relative position as $\mbf{n} = \mbf{r}/r$, and the unit
vector pointing along the (Newtonian) orbital angular momentum $\bm{L}$ as $\bm{\ell} = \mbf{n} \times \bm{v}/v $.
We choose an initial reference frame in which the $z$-axis is aligned with  $\bm{\ell}$, and the $x$-axis is aligned with $\mbf{n}$.
The $y$ axis is then chosen to complete a right-handed triad.

The equations of motion at 3\ac{pn} order in modified harmonic coordinates 
have the form~\cite{Bohe:2012mr, Blanchet:2013haa}:
\begin{subequations}
  \label{eq:eom_pn}
  \begin{align}
  \frac{\ud \bm{r}}{\ud t}&= \bm{v} \, , \\
  \frac{\ud \bm{v}}{\ud t}&=-\frac{G m}{r^2} \Bigl[
      \bigl(1+\mathcal{A}\bigr)\,\bm{n} + \mathcal{B}\,\bm{v} \Bigr] + \frac{\ud \bm{v}_{\rm S}}{\ud t} +
   \mathcal{O}\left( \frac{1}{c^7} \right)\,, \\
  \frac{\ud \mathbf{S}}{\ud t}&=\left(X_1\mathbf{\Omega}_1+
  X_2\mathbf{\Omega}_2\right)\mathbf{\times S}+
  \nu \left(\mathbf{\Omega_2-\mathbf{\Omega}_1}\right) \mathbf{\times \Sigma} \,
  , \\
  \frac{\ud \mathbf{\Sigma}}{\ud t}&=\left(X_2\mathbf{\Omega}_1+
  X_1\mathbf{\Omega}_2\right)\mathbf{\times \Sigma}+ 
  \left(\mathbf{\Omega_2-\mathbf{\Omega}_1}\right) \mathbf{\times S} \, ,
  \end{align}
\end{subequations}
with:
\begin{align}
  \label{OmegaCMstruct}
  \mathbf{\Omega}_i &= \bm{\ell} \left[
  \frac{1}{c^2}\alpha^{(i)}_\mathrm{1PN}
  +\frac{1}{c^4}\alpha^{(i)}_\mathrm{2PN}
  +\frac{1}{c^6}\alpha^{(i)}_\mathrm{3PN}
  + \mathcal{O}\left(\frac{1}{c^7}\right)\right] \,.
\end{align}
All coefficients listed in the equations above are functions of $\bm{r}, \bm{v}, \bm{S}, \bm{\Sigma}$ and of the symmetric mass ratio $\nu$ (or, alternatively,
the component masses $m_1$ and $m_2$), and can be read from Eqs.~(3.4a-3.7c) of \cite{Bohe:2012mr} 
and Eqs.~(355a - 355d), (356a - 356d) of \cite{Blanchet:2013haa}. 
Notably, the $\ud \bm{v}_{\rm S}/\ud t$ component of the
acceleration contains a term parallel to $\bm{\ell}$, which is responsible for the precession of the orbital plane.

Following~\cite{Akcay:2020qrj, Gamba:2021ydi}, we choose as our reference 
co-precessing frame $\{\mbf{x}', \mbf{y}', \mbf{z}'\}$ the one in which the $\mbf{z}'$ axis is aligned with the Newtonian
angular momentum $\bm{\ell}$ at all times. We parameterize the time-dependent rotation relating the 
$\{\mbf{x}, \mbf{y}, \mbf{z}\}$ and $\{\mbf{x}', \mbf{y}', \mbf{z}'\}$ frames as a sequence of three 
Euler angles $\alpha(t), \beta(t), \gamma(t)$:
\begin{subequations}
  \begin{align}
    \alpha &= \arctan\left(\frac{\ell^y}{\ell^x}\right) \, , \\
    \beta  &= \arccos\left(\ell^z\right) \, .
  \end{align}
\end{subequations}
The third angle $\gamma$ is obtained from $\alpha$ and $\beta$ by
\begin{equation}
\dot{\gamma} = \dot{\alpha} \cos{\beta} \, , \label{eq:gamma}
\end{equation}
where the overdot denotes differentiation with respect to time.
Note that, by construction, at the initial time $\beta(0) = 0$, while 
$\alpha(0)$ and $\gamma(0)$ are undefined according to the Eqs. above.
We resolve this ambiguity by setting $\alpha(0)$ according to 
App.A of~\cite{Gamba:2021ydi} and $\gamma(0) = \alpha(0)$.
This is not the only possible choice of co-precessing frame, as one could 
also specify the direction of the $\mbf{x}'$ axis at $t=0$ to be aligned 
with the initial $\mbf{n}$.

We solve the equations of motion Eqs.~\eqref{eq:eom_pn} numerically,
terminating the integration when the binary reaches the final peak of the orbital
frequency $\Omega = |\bm{v} \times \bm{r}|/r^2$, surpasses a maximum threshold time 
or a minimum radial separation of $r=4M$.
As we will also be interested in the evolution of systems along quasi-circular orbits,
in order to perform direct comparisons with the results obtained in the general case,
we implement an eccentricity-reduction scheme following~\cite{Tacik:2015tja, Tichy:2019ouu}.

We compute the dynamics of several systems with characteristics spanning the parameter space, 
varying the mass-ratio $q \in \left[1, 9\right]$, 
the component spins (with the initial values of the spin parameters $\chi_{\rm eff} \in \left[-0.9,0.9\right]$ and $\chi_p \in \left[0, 1\right]$),
and the orbit geometry. We consider both bound orbits (defined by the initial eccentricity, $e_0 \in \left[0.01, 0.9\right]$, and 
dimensionless semilatus rectum, $p_0$) and unbound configurations (scatterings, dynamical captures), 
where initial data is parametrized by the starting energy $\hat{E}_0 = E_0/\mu$ and (orbital) angular momentum $\hat{L}_{0} = L_0/\mu$
(varied from just above the separatrix to high values), at an initial defined separation $r_0 = 10\ 000M$.
 
We neglect, for simplicity, any terms in the PN expressions giving the initial velocities (see App.~\ref{app:PN_ics}) that
depend on the spins $\Sa, \Sb$. For unbound orbits, because of the very large initial separation, this leads to negligible differences between
the given initial energy and angular momentum and the values recovered from the dynamics using the spin-dependent expressions ($<10^{-5}$ for the former, 
$\lesssim 10^{-2}$ for the latter). For bound configurations, when the out-of-plane spin components are large and aligned with the orbital angular momentum, this approximation
can result in more important deviations in the eccentricity and semilatus rectum as estimated from the dynamics (a value of $e_0 = 0.9$ in input can result in 
a recovered initial eccentricity of $\sim 0.93$). However, since the eccentricity is in any case a non-gauge invariant parameter, 
we accept this (usually very) slight incongruence and treat the nominal value of $e_0$ as an approximate indicator of the degree of non-circularity in the orbit.

\subsection{Angular momentum vectors}
\label{subsec:angular_momentum}

\begin{figure}[t]
  \includegraphics[width=0.49\textwidth]{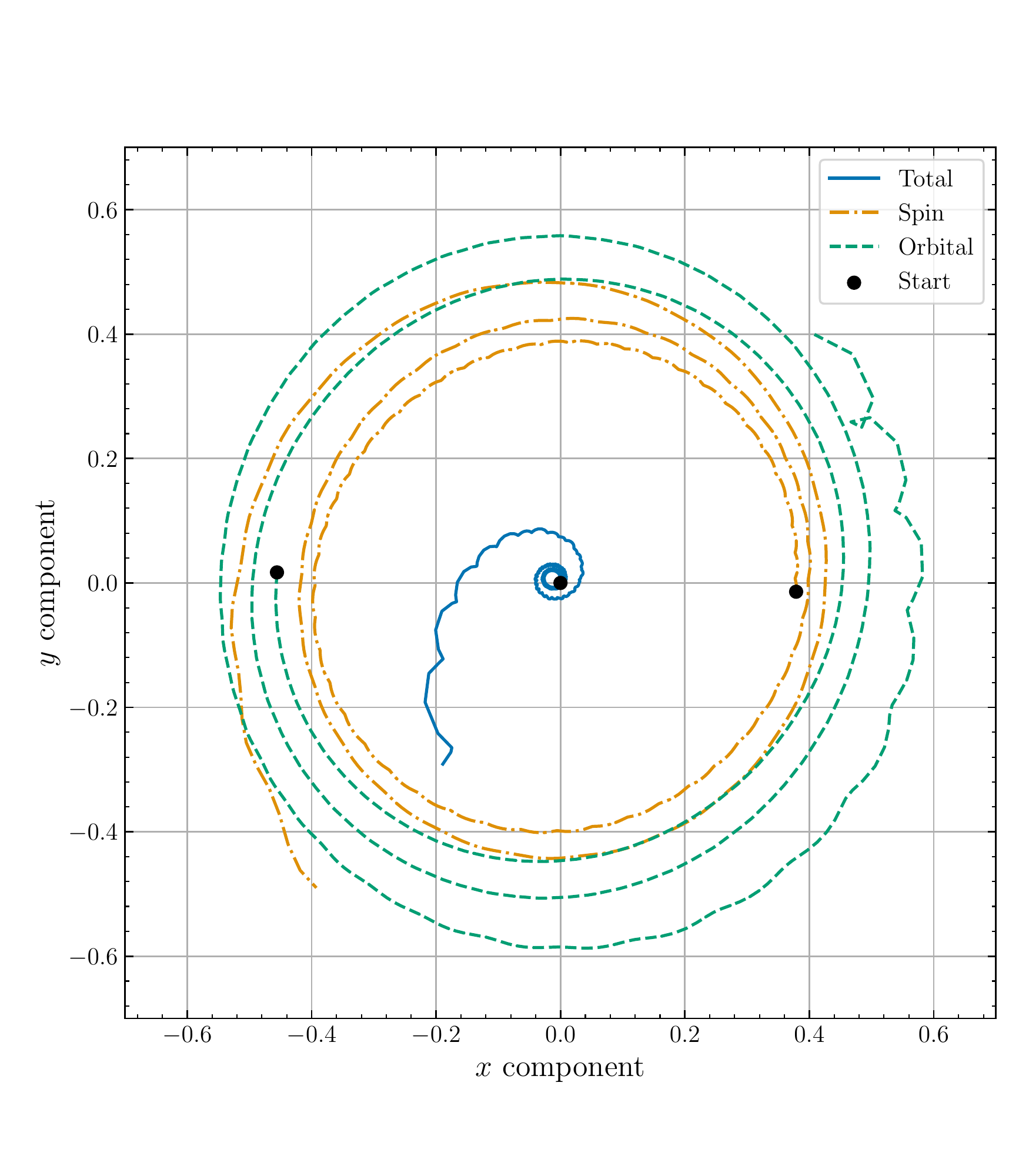}
  \caption{Tracks of the evolving angular momentum (total and orbital) and (total) spin unit vectors, projected onto the plane perpendicular to the initial
  value $\bm{J_0}$. The mass ratio is $q = 4$, the initial eccentricity is $e_0 = 0.7$, the semilatus rectum $p_0 = 30$, and the dimensionless spins 
  $\bm{\chi}_1 = \left(0, 0.8, 0.1\right), \bm{\chi}_2 = \left(0.4,0,0.2\right)$.
  \label{fig:J_vectors}}
\end{figure}

We begin by reviewing the main results concerning the dynamical properties of eccentric, precessing binaries in the PN approximation,
starting with a reminder of the effects of precession on the angular momentum vectors of the system.

If one neglects radiation reaction terms, the equations of motion for a generic precessing system admit 
two non-trivial conserved integrals in the center-of-mass frame: the energy and the total angular momentum vector, 
given by the sum of the orbital and spin contributions:
\begin{equation}
  \bm{J} = \bm{L} + \bm{S} \, .
\end{equation}
The evolution equations for the individual spins $\Sa, \Sb$ guarantee the conservation of their norm. Conversely, the modulus of the orbital angular 
momentum decays under radiation reaction, and with it $\bm{J}$. The resulting phenomenology for these vectors 
has been studied in detail in the literature (see e.g.~\cite{Apostolatos:1994mx,Racine:2008qv}), and our results are in line with 
the findings of such works irrespectively of the shape of the orbit.
At 3\ac{pn} order the total momentum $\bm{J}$ is characterized by slowly decreasing magnitude and approximately conserved direction (outside of
the last few orbits). In contrast, the orbital angular momentum $\bm{L}$ and the total spin $\bm{S}$ precess around $\bm{J}_0$ 
on cones with increasing aperture. Figure \ref{fig:J_vectors}, representative of all configurations studied\footnote{With the exception of some rather extreme ones 
with $q = 9$ and large, negative initial $z$-components in the spins. In this regime, the initial total spin and orbital angular momentum partially cancel out, but 
while the norm of $\bm{S}$ does not change much during the inspiral, $L$ decays due to radiation reaction, and this can lead to a peculiar \textit{increase} in the norm of $\bm{J}$,
as well as deviations from the phenomenology described in the text.}, 
shows the slowly outspiralling tracks of the unit vectors of $\bm{J}$, $\bm{L}$ and $\bm{S}$, projected onto a plane orthogonal to $\bm{J}_0$, for a system with mass ratio $q = 4$, 
large eccentricity $(e_0 = 0.7)$ and $\bm{\chi}_1 = \left(0, 0.8, 0.1\right), \bm{\chi}_2 = \left(0.4,0,0.2\right)$.
The long tail in the track of $\bm{J}$, where conservation is clearly broken, corresponds to just the final cycles of the inspiral.

\subsection{Phenomenolgy of the Euler angles}
\label{subsec:phenom_euler}

\begin{figure*}[t]
  \includegraphics[width=\textwidth]{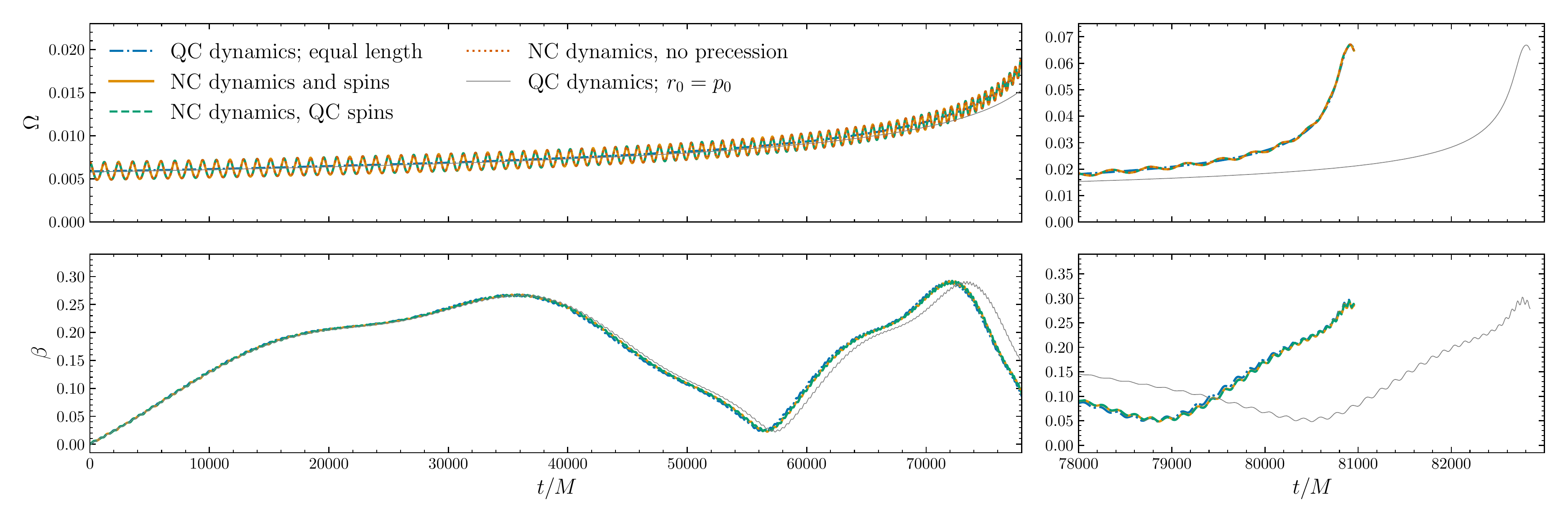}
  \caption{Evolution of the orbital frequency $\Omega$ (top) and the Euler angle $\beta$ (bottom) for an orbit with $q = 4$, 
  initial dimensionless spins $\bm{\chi}_{1} = \left(0, 0.2, 0.1\right)$ and $\bm{\chi}_2 = \left(0.4, 0, 0.2\right)$, 
  initial eccentricity $e_0 = 0.1$ and semilatus rectum $p_0 = 30$, using different prescriptions for the orbital and spin dynamics. 
  Also shown are two wholly QC orbits: one with initial separation equal to $p_0$; one with initial separation chosen so the length 
  of the orbit is the same as the NC case. Notice how the QC evolution of $\beta$ with suitable initial conditions almost exactly
  matches what is found in the NC orbits, with only small oscillating deviations on the orbital time scale.
  \label{fig:ecc_vs_circ_opt}}
\end{figure*}

We now move on to highlighting the relevant features of the evolution of the Euler angles, focusing on $\beta$, which is most directly
tied to the precession of the orbital plane.

\begin{figure*}[t]
  \includegraphics[width=\textwidth]{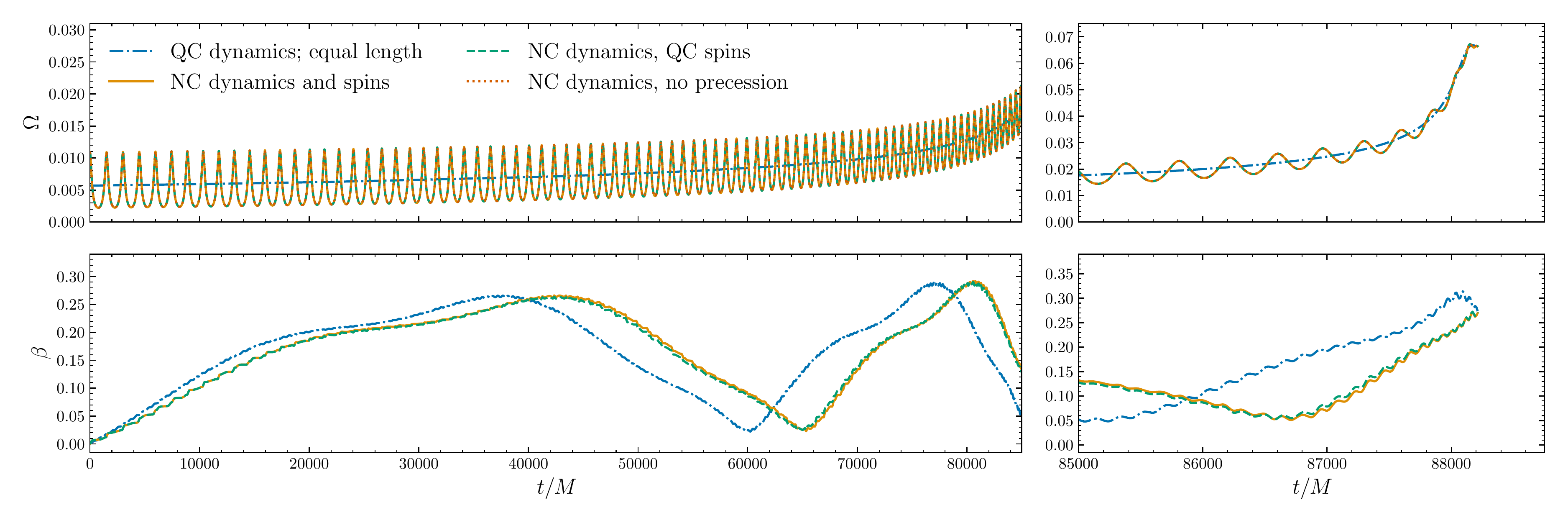}
  \includegraphics[width=\textwidth]{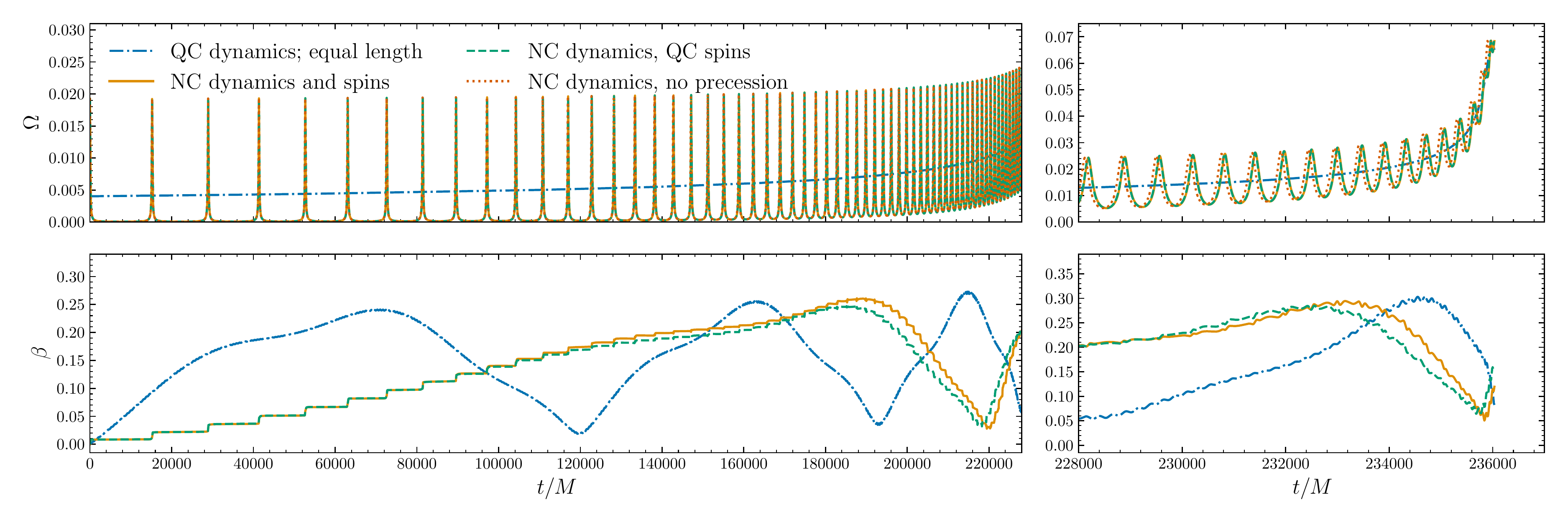}
  \caption{Evolution of the orbital frequency $\Omega$ and the Euler angle $\beta$ using different prescriptions for the orbital and spins dynamics, 
  for two eccentric orbits, both with $q = 4$, (initial) $\bm{\chi_{1}} = \left(0, 0.2, 0.1\right)$ and $\bm{\chi_2} = \left(0.4, 0, 0.2\right)$: $e_0 = 0.4$, $p_0 = 30$ (top), 
  and $e_0 = 0.9$, $p_0 = 30$ (bottom). In both cases an entirely QC orbit with same total duration is also shown. 
  As the eccentricity increases, the slow variation of $\beta$ on the precession timescale is no longer faithfully reproduced by the QC orbit; the difference between 
  the orbit with full NC spin and orbital evolution and that with QC spin dynamics remains notably small even at high eccentricities.
  \label{fig:ecc_vs_circ_48}}
\end{figure*}

Figure~\ref{fig:ecc_vs_circ_opt} shows the evolution of the orbital frequency $\Omega$ and of the Euler angle $\beta$ 
for one exemplary system with mass ratio $q = 4$, spins $\bm{\chi}_1 = \left(0,0.2,0.1\right), \bm{\chi}_2 = \left(0.4,0,0.2\right)$, initial eccentricity $e_0 = 0.1$ 
and semilatus rectum $p_0 = 30$ (see Sec.~\ref{subsec:nc_terms} for the meaning of the different curves).
As is the case for all systems we consider, $\beta$ starts from 0 (by construction), and over the course of the binary evolution undergoes 
a series of slow, but accelerating oscillations. It is immediately evident that the precession timescale 
characterizing these oscillations is much larger than the orbital period: the system completes $\simeq 55$ orbits during the first $\beta$ cycle in this example. 
This separation of timescales remains true for most of the configurations studied, as evidenced in Fig.~\ref{fig:ecc_vs_circ_48}, 
which shows two orbits with larger initial eccentricity ($e_0 = 0.4$ and $e_0 = 0.9$) but same initial spin vectors.
Effects on $\beta$ on the orbital timescale are present, at low eccentricities, exclusively in the form of small nutations around the overall, slower evolution.
As eccentricity grows and the orbits part from quasi-circularity, physical quantities such as the energy and the angular momentum vectors 
(and thus the Euler angles) undergo long stretches of stasis -- rather than slow, secular changes -- interspersed with 
short but intense bursts of activity, coinciding with periastron passages (see the bottom panel of Fig.~\ref{fig:ecc_vs_circ_48}).

While our focus has been on one single representative configuration of mass ratio and spins,
the main characteristic features of the evolution of $\beta$ do not change significantly when varying the initial conditions.
Generally speaking, larger eccentricities and/or spin components in the direction of the orbital angular momentum lead to fewer oscillations, 
the first very stretched out (this is especially true when the $z$-components of the spins are anti-aligned with the initial $\bm{\ell}$).
The maximum value of $\beta$ is only mildly influenced by the eccentricity, but depends strongly on the mass ratio 
and on the in-plane spin components (it is well known that larger $q$ and $\chi_p$ typically imply stronger precession).

\begin{figure*}[t]
  \includegraphics[width=\textwidth]{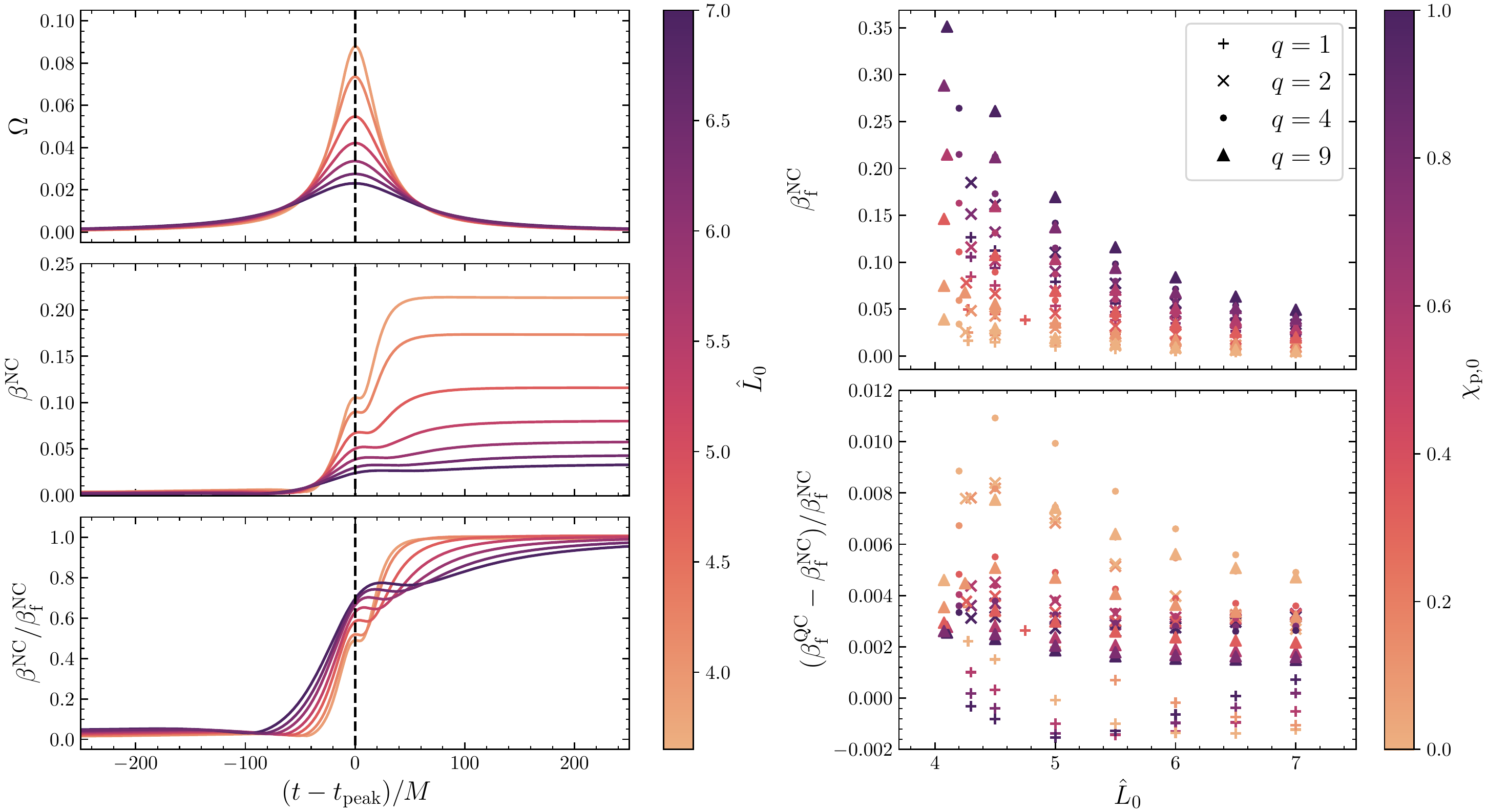}
  \caption{On the left, the orbital frequency $\Omega = |\bm{v} \times \bm{r}|/r^2$ (top) and Euler angle $\beta$ (absolute, middle, 
  and normalized by its asymptotic value, bottom) around the time of closest approach (corresponding to 0 on the horizontal axis, 
  marked by a vertical dashed line) for a series of hyperbolic encounters with mass-ratio $q = 4$, initial spins 
  $\bm{\chi}_1 = \left(0, 0.8, 0.1\right)$ and $\bm{\chi}_2 = \left(0.4,0,0.2\right)$ (so $\chi_{p,0} = 0.8$), initial energy $\hat{E}_0 = 0.02$ 
  and varying initial orbital angular momentum $\hat{L}_0$. Smaller values of $\hat{L}_0$ lead to closer
  encounters and to more pronounced and more sudden shifts in the orientation of the orbital plane.
  On the right: asymptotic value of $\beta$ for scatterings with energy $\hat{E}_0 = 0.02$, mass-ratio $q \in \left[1,9\right]$, 
  varying $\hat{L}_0$ and spins; comparison between $\beta_{\rm f}$ as calculated with the full non-circular model and 
  neglecting non-circular terms in the spins evolution. The in-plane initial spin components are varied in norm (but keeping their directions fixed), 
  while $\chi_{1z,0} = 0.1$ and $\chi_{2z,0} = 0.2$ for all cases, yielding $\chi_{\rm eff, 0} \in \left[0.11,0.15\right]$. 
  After $\hat{L}_0$, $\beta_{\rm f}$ is most strongly determined by the in-plane spin parameter, as can be expected; 
  a secondary effect is a general increase in the asymptotic value of the Euler angle with the mass-ratio.
  \label{fig:hyp_beta_omega}}
\end{figure*}

The behavior that characterizes highly non-circular orbits is maximally evident for scatterings and captures, 
where each encounter is accompanied by a sudden change in the direction of $\bm{\ell}$ and
of the binary properties (energy, absolute value of the orbital angular momentum, directions of the spins) which then remain 
constant until the next encounter (if there is one).
This is illustrated in Fig.~\ref{fig:hyp_beta_omega}, which shows the evolution of $\beta$ for a series of scattering events with 
fixed $q = 4$, spins $\bm{\chi}_1 = \left(0, 0.8, 0.1\right), \bm{\chi}_2 = \left(0.4, 0, 0.2\right)$, initial energy $\hat{E}_0 = 0.02$, and $\hat{L}_0$
growing from just above the threshold between bound and unbound orbits to $\hat{L}_0 = 7$.
The shift in the orbital plane is more pronounced for systems with a smaller distance of closest approach 
(and $\hat{L}_0$) as well as for systems with larger mass-ratio and/or orthogonal spin parameter $\chi_{\rm p}$\footnote{We mention in passing
that $\chi_{\rm p}$ is known~\cite{Thomas:2020uqj} to be an imperfect measure of the strength of spin precession. Configurations 
with initial in-plane spin components that are large in size, but oriented such that they cancel out in the total $\bm{S}_{\perp}$,
have large $\chi_{\rm p}$, but exhibit little to no precession if the mass ratio is close to 1. This is a consequence of the 
spin evolution equations in our PN study, and can also be seen in NR simulations as will be remarked later.} (see Fig.~\ref{fig:hyp_beta_omega}).
A curious feature is the inflection in the $\beta$ curve at periastron: initially suspected to be a spurious effect due to the
application of \ac{pn} expansions in the strong field regime for the closest encounters, it appears to be a feature
common to all cases considered, only becoming smoother and slower for larger $\hat{L}_0$.

\subsection{The importance of non-circular terms}
\label{subsec:nc_terms}

In order to better understand the importance of in-plane spins and non-circular terms in the system of 
Eqs.~\eqref{eq:eom_pn}, we solve the equations of motion and compare the
results obtained in four different scenarios:
\begin{itemize}
  \item[(i)] with complete non-cirular corrections fully accounted for in the evolution of the spins and the orbital dynamics 
  (orange in Fig.~\ref{fig:ecc_vs_circ_opt} and~\ref{fig:ecc_vs_circ_48});
  \item[(ii)] with non-circular corrections explicitly accounted for in $\ud \bm{v}/\ud t$, but not in $\Sadot, \Sbdot$ (green). 
  This is done by setting $\mathbf{\Omega}_{1,2}$ to its quasi-circular reduction $\mathbf{\Omega}_{1,2}^{\rm QC}$ in the evolution equations of $\Sadot, \Sbdot$, 
  as given by Eq.~(4.5) of Ref.~\cite{Bohe:2012mr}. Note that some generic-orbit effects are nonetheless present, as the parameter $x = \Omega^{2/3}$ 
  appearing in $\mathbf{\Omega}_{1,2}^{\rm QC}$ is inherited from the evolution of $\mathbf{r}$ and $\mathbf{v}$;
  \item[(iii)] with the dynamics of a quasi-spherical system (blue);
  \item[(iv)]  with the dynamics of a planar non-circularized system (dotted red).
\end{itemize}

Figures \ref{fig:ecc_vs_circ_opt} and \ref{fig:ecc_vs_circ_48} show the evolution of $\Omega(t)$ (top) 
and $\beta(t)$ (bottom) in the four scenarios above for the systems mentioned in the previous subsection.
A few things can be immediately observed:
(i)  first, the impact of the explicitly non-circular terms in $\Sadot, \Sbdot$ appears to be largely sub-dominant;
(ii) second, the planar, non-circularized evolution of $\Omega(t)$ provides a good approximation of the full evolution;
(iii) third, the shape of $\beta(t)$ (i.e, the number and height of $\beta$ peaks) in the case of the quasi-spherical evolution 
matches that of the eccentric evolution, but it is ``stretched'' over time, similar to the evolution of $\Omega(t)$. In fact, 
if the initial conditions for the quasi-spherical evolution are slightly varied to generate orbital dynamics of approximately the
same time length as the eccentric, precessing system, we find that $\beta$ is almost perfectly overlayed with the target one.
Points (i) and (ii) hold for every orbital and spin configuration considered, up to high eccentricity ($e_0 = 0.9$) and in-plane 
spin components ($\chi_p = 1.0$). The non-precessing evolution at most accumulates a phase difference with respect to the precessing ones
of one cycle in cases of very long, eccentric orbits\footnote{See also the bottom panel of Fig.~\ref{fig:hyp_beta_omega}, which highlights the small impact of the
non-circular terms in $\Sadot, \Sbdot$ on the asymptotic value of the Euler angle $\beta$ for scattering events}.
The validity of point (iii) is more limited, as can be seen in the figures. Increasing 
the eccentricity ($e_0 \gtrsim 0.2$), the evolutions of $\beta$ as computed using quasi-circular and non-circular dynamics become
less and less comparable, with different characteristic timescales, shapes and numbers of cycles.

Our analysis thus seems to indicate that, so long as an ``eccentric'' prescription for $\Omega(t)$ is employed, the quasi-circular expressions 
for the spins may be sufficient to describe the precession of the orbital angular momentum up to large values of eccentricity $e_0 \leq 0.9$.
Conversely, the worsening comparison between fully generic and quasi-circular evolutions as the eccentricity grows suggests that the
cumulative effect of the ``bursts'' observed at $e_0 \geq 0.6$ cannot be fully captured by an orbit-averaging procedure.
Indeed, Ref.~\cite{Fumagalli:2023hde} has shown that the multi-timescale approach cannot be straightforwardly applied to highly eccentric 
systems, even at large separations.
Nonetheless, it would be interesting to investigate whether it is possible to systematically map the non-circular dynamics to that of a 
circularized precessing system in the range $e_0 \in[0.2, 0.6]$, perhaps by focusing on matching orbit-averaged frequencies rather than
the time-length of the orbits themselves.

\subsection{Scatterings}
\label{subsec:scattering}

\begin{figure}[t]
  \includegraphics[width=0.49\textwidth]{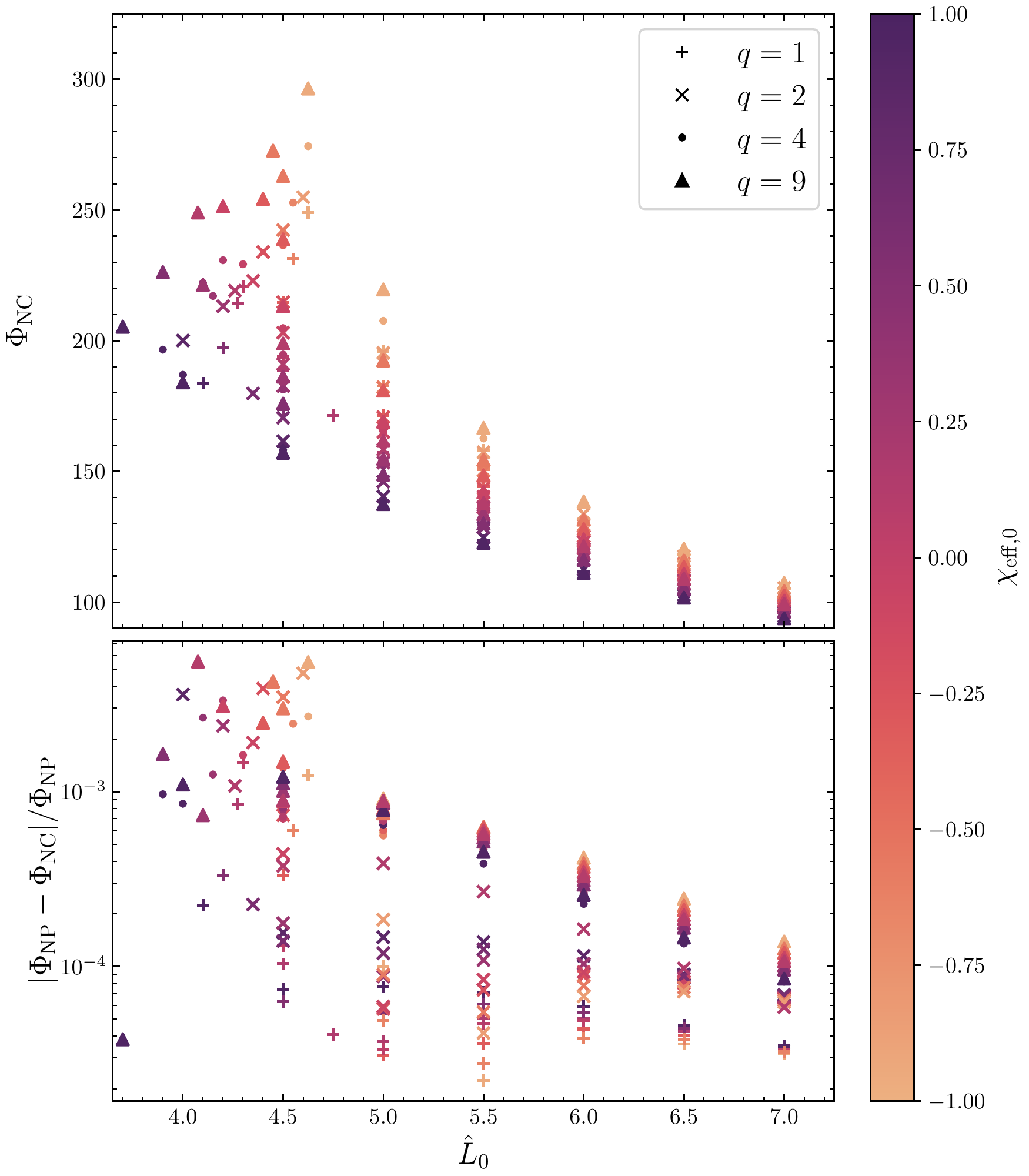}
  \caption{Top panel: the aziumuthal scattering angle $\Phi_{\rm NC}$, in degrees, as a function of the initial orbital angular momentum $\hat{L}_0$; color indicates 
  the starting value of the effective spin parameter $\chi_{\rm eff}$, while the marker corresponds to the mass-ratio $q$. The orthogonal spin parameter
  $\chi_p$ is fixed at 0.4 for this plot.
  Bottom panel: relative differences between $\Phi_{\rm NC}$ and the scattering angle $\Phi_{\rm NP}$ computed neglecting spin precession.
  \label{fig:hyp_scattering}}
\end{figure}
\begin{figure}[t]
  \includegraphics[width=0.49\textwidth]{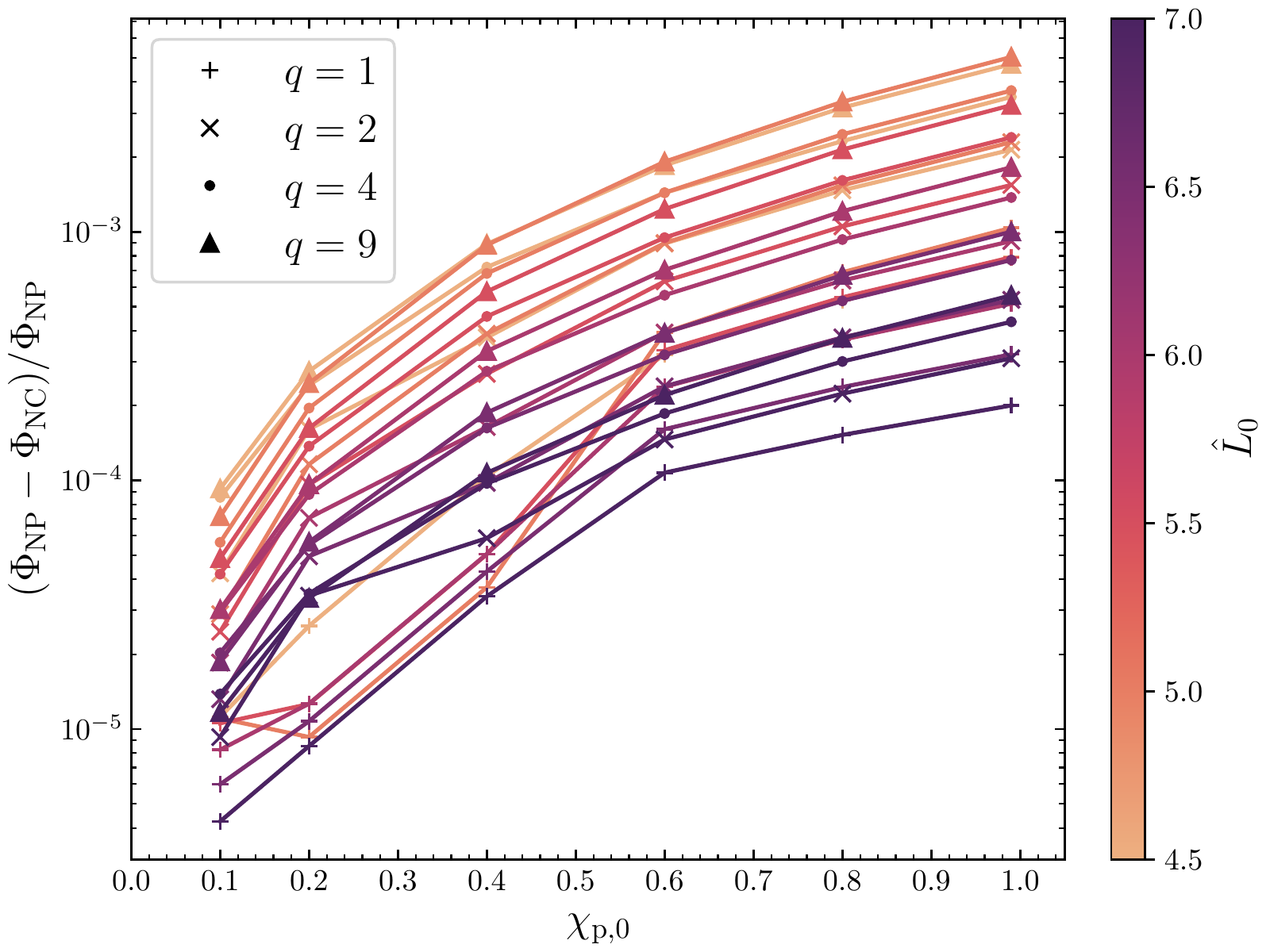}
  \caption{Relative difference in the azimuthal scattering angle between fully NC and non-precessing (NP) dynamics, for various mass ratios and initial angular momenta,
  as a function of the orthogonal spin parameter $\chi_{\rm p}$; the $z$-components of the initial dimensionless spins are fixed, $\chi_{1,z} = 0.1, \chi_{2,z} = 0.2$.
  Smaller $\Phi$ signifies weaker interaction at closest approach; these results suggest that, everything else held equal, increasing in-plane spins has
  a slightly repulsive effect on the dynamics.
  \label{fig:hyp_scattering_chip}}
\end{figure}

Given the importance of scatterings in the context of theoretical developments in \ac{gw} physics~\cite{Damour:2014afa, Hopper:2022rwo, Damour:2022ybd, Rettegno:2023ghr},
and in particular in the \ac{pm} expansion~\cite{Bern:2019nnu,Kalin:2020fhe,Bjerrum-Bohr:2021din,Bern:2021yeh,Dlapa:2021vgp,Damour:2020tta,DiVecchia:2021ndb,Cho:2021arx,DiVecchia:2021bdo,Herrmann:2021tct,Bini:2021gat,Bini:2021qvf,Manohar:2022dea,Dlapa:2022lmu}, 
we dedicate a brief discussion to scatterings of non-planar \acp{bbh}. 

We consider a series of hyperbolic encounters with mass-ratios $q \in [1,9]$, varying initial spins ($\chi_{\rm eff} \in \left[-0.9,0.9\right]$ and $\chi_p \in \left[0, 1\right]$), and initial energy $\hat{E}_0 = 0.02$.
Increasing the initial orbital angular momentum $\hat{L}_0$ from just above the separatrix between bound and unbound orbits to high values, 
we evolve the system from an initial separation $r_0 = 10\ 000M$, through the encounter, and out to the same final distance.

The outcome of a non-planar encounter cannot be adequately described by a single scattering angle. We instead introduce two separate angles:
$\Phi = \varphi_{\rm f} - \varphi_0 - \pi = \varphi_{\rm f} - \pi$, encoding the deflection in the original equatorial plane (perpendicular to $\bm{\ell}_0$) and defined as
the total variation in the azimuthal angular coordinate $\varphi$ tied to the initial inertial frame; 
and $\Theta = \theta_{\rm f} - \theta_0 = \theta_{\rm f} - \pi/2$, describing the out-of-plane component and similarly
defined in terms of the polar angle $\theta$.

Figure \ref{fig:hyp_scattering} displays the results for $\Phi$ as a function of $\hat{L}_0$, $q$ and of the initial value of the spin parameter $\chi_{\rm eff}$, 
for a collection of systems with $\chi_{\rm p,0} = 0.4$. After the obvious dependence on $\hat{L}_0$ (which determines the distance
of closest approach and impact parameter), we see that $\Phi$ is most correlated with the effective spin parameter $\chi_{\rm eff}$:
the scattering angle grows as $\chi_{\rm eff}$ decreases from positive (indicative of spin components initially aligned with the orbital angular momentum)
to negative values ($z$-components initially anti-aligned with $\bm{L}_0$). This effect is known, and can be intuitively understood as a consequence of the
spin-orbit interaction, which affects the effective radial potential~\cite{Rettegno:2023ghr}.
Overlayed with this is a positive correlation with the mass-ratio, although this seems to be a somewhat sub-leading effect 
(note however that the two variables are not independent: recall that $\chi_{\rm eff} = X_1 \chi_{1,z} + X_2 \chi_{2,z} = \left(q \chi_{1,z} + \chi_{2,z}\right)/\left(1+q\right)$).

The polar scattering angle $\Theta$ is tied to the asymptotic value of the Euler angle $\beta$, and they display similar dependence on the system parameters. 
In fact, $\Theta$ is geometrically bound to the interval $\left[-\beta,\beta\right]$; its exact value (and in particular its sign, which corresponds to deviation 
either above or below the original equatorial plane) depends on the orientation of the in-plane spin components.
In the configurations studied we found that $|\Theta|$ remains close to 0 in most cases, with deviations from the original equatorial plane 
rarely exceeding $5^\circ$, and at most reaching $15^\circ$ in cases of close encounters with high $\chi_{\rm p}$.

The orthogonal spin variable $\chi_{\rm p}$, encoding the strength of the precession effect, has a small impact on the orbital dynamics. 
Fixing the other parameters and increasing its magnitude from 0 to 1 leads to a decrease in $\Phi$ of the order of $\sim 1^\circ$, or 
$\lesssim 1\%$, in most cases (see Fig.~\ref{fig:hyp_scattering_chip}), with only the very closest encounters exhibiting stronger dependence on $\bm{S}_{\perp}$; 
this suggests that the in-plane spin components have a slightly repulsive effect in the binary interaction.

\section{The strong field regime}
\label{sec:nr}

\begin{figure*}
  \includegraphics[scale=0.7]{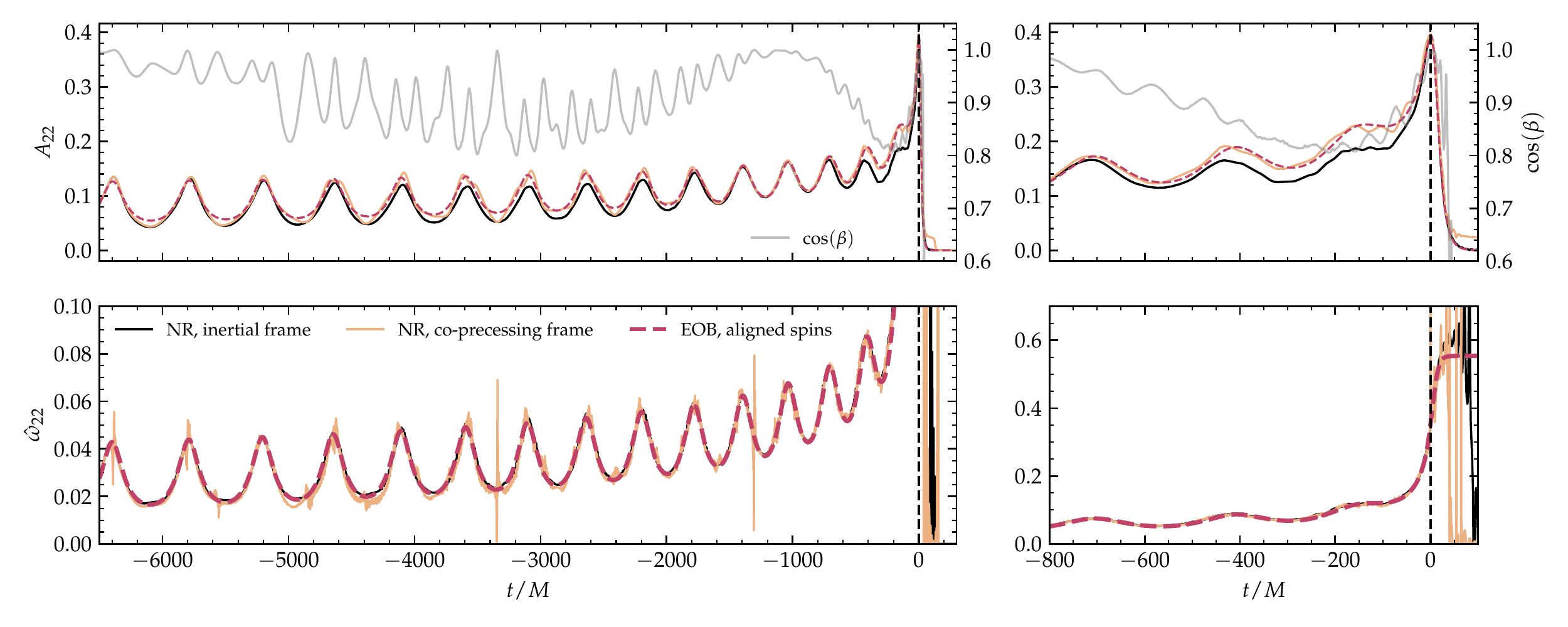}
  \caption{Comparison between amplitude and frequency evolutions in the inertial and co-precessing frames. Aligned spins 
  waveforms correctly capture the amplitude and frequency evolution of the co-precessing frame waveforms. Clear modulations, mimicking the
  oscillations in $\beta$, are visible in the inertial frame due to the precession of the spins.
  \label{fig:rit_eob_coprec}}
\end{figure*}

A comprehensive understanding of the strong field regime is pivotal for advancing the development of waveform models. 
Recently, simulations featuring both eccentricity and precession have started to be made publically available, 
with $7$ new simulations from the \MAYA~catalog and $115$ from the \RIT~catalog produced and shared over the past $2$ years~\cite{Healy:2022wdn,Ferguson:2023vta,Ferguson:2023mks}.
Unfortunately, most of these simulations exhibit a rather limited number of orbits before merger.
For instance, the longest \MAYA~simulation available lasts only about $1000 M$ (approximately $5$ orbits) before merger, 
while only $8$ simulations from the \RIT~catalog surpass this duration.
Furthermore, among these longer simulations, $3$ feature a zero value for the in-plane component of the total spin $\mathbf{S}_{\perp} = (S_x, S_y, 0) $.
This, combined with the fact that all of these systems are equal mass, implies that the orientation of the orbital
angular momentum throughout the inspiral does not evolve\footnote{Notably, this is true for about $30\%$ of the entirety of the \RIT~catalog of eccentric and precessing
simulations.}.
Consequently, the majority of existing simulations are unsuitable for studying the inspiral phase, with the few exceptions being limited to equal mass 
configurations and lacking multiple resolutions\footnote{Such simulations can nonetheless be employed for the study of the merger-ringdown phase, see e.g.~\cite{Carullo:2023kvj}.}.
In light of the limitations discussed, we concentrate on the two longest \RIT~simulations exhibiting 
clear precession effects, aiming to assess the validity of the physical intuition derived from the \ac{pn} sector. Clearly, this focus does 
not constitute a comprehensive exploration of the strong field regime, but rather serves as an initial step in this direction.

Past works~\cite{Schmidt:2010it, Schmidt:2012rh, Boyle:2011gg,OShaughnessy:2011pmr,Pekowsky:2013ska} have shown how it is possible to identify a co-precessing, 
non-inertial frame from simulations of quasi-spherical inspiralling binaries in which the modulations induced 
by the spins precession appear decoupled from the orbital dynamics. 
Here, we extend the analysis to non-spherical orbits, showing that similar conclusions appear to hold also for eccentric systems.
Given the lack of the full 3D information for the simulations that we consider, 
we extract the co-precessing frame following App.~A of \cite{Pekowsky:2013ska} directly from the 
waveform multipoles $h_{\ell m}$ or the Weyl scalar $\psi^4_{\ell m}$.
We identify the preferred radiation axis $\hat{\mathbf{V}}$ 
with the direction aligned with the principal direction of the $\langle \mathcal{L}_{(a} \mathcal{L}_{b)} \rangle$
tensor~\cite{OShaughnessy:2011pmr,Ochsner:2012dj}. 
After rotating the multipoles $h_{\ell m}$ and $\psi^4_{\ell m}$ to an inertial frame
aligned with the initial direction of the orbital angular momentum $\bm{\ell}$, we compute the Euler angles $\alpha, \beta, \gamma$ 
connecting the inertial ``source'' frame with the co-precessing frame (similar to what we did in the previous section), 
such that at each moment in time
\begin{equation}
  \hat{\mathbf{V}} = (\cos\alpha \sin\beta, \sin\alpha\sin\beta, \cos\beta) \, ,
\end{equation}
and the rotated multipoles read:
\begin{equation}
  w_{\ell m}^{\rm R} = \sum_{m'} D^{\ell}_{m m'}(R(\alpha, \beta, \gamma)^{-1}) w_{\ell m} \, ,
\end{equation}
where  $w_{\ell m}^{\rm R} = \{h_{\ell m}^{\rm R}, \psi^{4, \rm R}_{\ell m} \} $ and $R$ denotes the rotation
matrix associated with the Euler angles.

An example of $h_{2 2}$ in the radiation and inertial frames can be inspected from Fig.~\ref{fig:rit_eob_coprec} for the 
\texttt{RIT:eBBH:1632} simulation. This simulation is characterized by the following intrinsic parameters:
$q=1$, $\bm{\chi}_1 = (-0.7, 0, 0)$, $\bm{\chi}_2 = (-0.7, 0, 0)$ and initial eccentricity of $0.28$ at an initial radial separation of $r_0\sim 25 M$.
While subtle, the modulations due to precession in the inertial frame are clearly visible, especially in the mode's
amplitude close to the time of merger and around $3000 $M$ $ before this time. Predictably, these moments correspond to 
the times at which the $\beta$ angle is significantly different from zero, as can be seen from the top panel of the same 
figure. The frequency evolution of the $(2,2)$ mode, shown in the bottom panel, instead does not appear to be significantly 
affected by the precession of the spins, although large numerical error seems to be present.
We then compare the amplitude and frequency evolutions of the waveform in these two frames with those obtained with the aligned-spin \ac{eob} model of~\cite{Nagar:2024dzj},
that will be discussed in more detail in later sections (see Sec.~\ref{sec:eob}).
We fix the intrinsic parameters of the \ac{eob} model (mass ratio, $z$-component of the spins) to the initial values specified in the simulation's 
metadata, and choose initial values of orbit-averaged frequency, eccentricity and true anomaly such that the frequency peaks of the 
\ac{eob} waveform approximately match the ones of the \ac{nr} waveform, and the lengths of the waveforms are comparable.
Remarkably, the amplitude and frequency evolutions thus obtained match the ones observed in the co-precessing frame, indicating that
the simplification upon which the twisting procedure applied in the quasi-circular scenario is based holds also for eccentric systems.

\begin{figure*}[t]
  \includegraphics[scale=0.7]{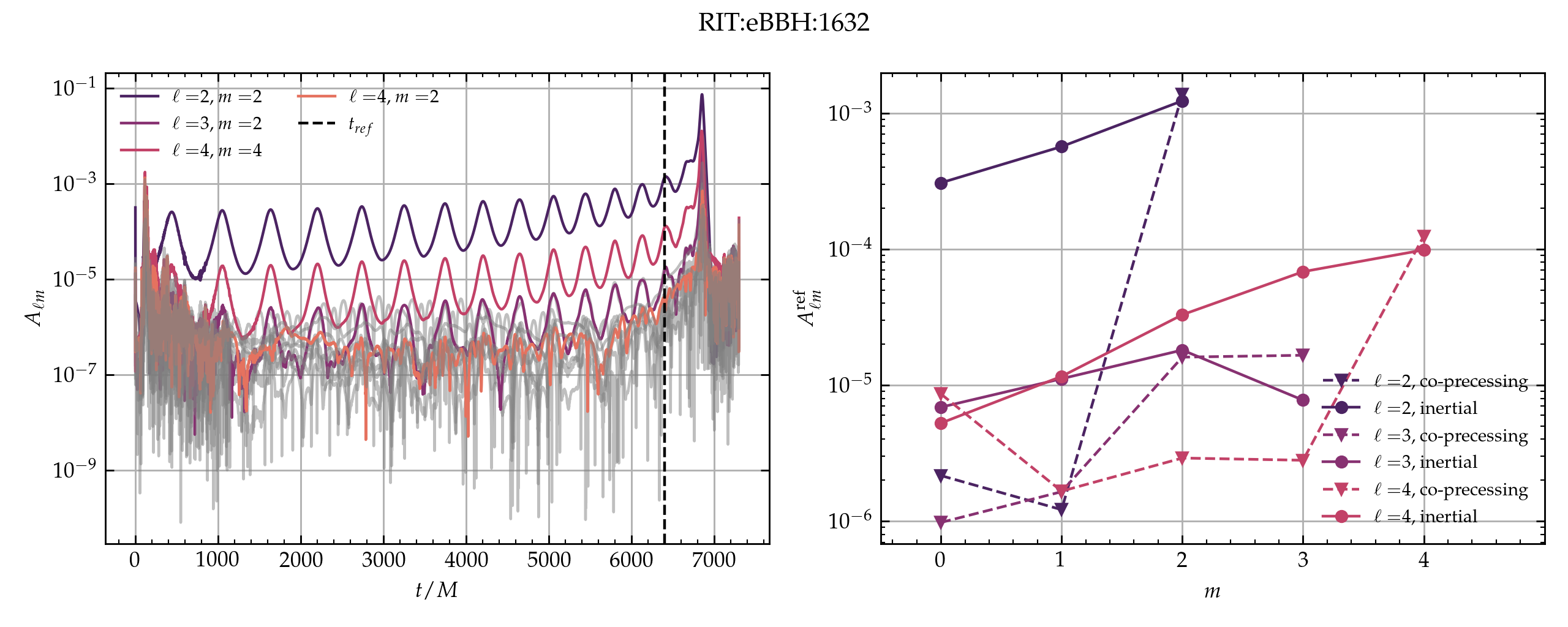}
  \caption{Evolution (left) and hierarchy (right) of $\psi^{4}_{\ell m}$ modes for the \texttt{RIT:eBBH:1632} simulation in the co-precessing and inertial frames.
  In the right panel, the amplitudes are evaluated ad a reference time $t_{\rm ref}$ corresponding to the last periastron before merger, and 
  shown as a function of $m$ for a fixed value of $\ell$, up to $\ell = 4$. The modes with $\ell > 2$, with the exception of the $(4,4)$ mode, 
  are affected by large numerical noise. Nonetheless, it is possible to appreciate that odd-$m$ modes in the co-precessing frame have typically 
  lower average amplitude than the even-$m$ ones. This is expected, as the system considered is an equal mass binary.
  \label{fig:RIT_amp_hier}}
\end{figure*}

Additional evidence that this is the case is provided by an inspection of the hierarchy of the waveform modes in the co-precessing and inertial
frames. We show the results of this analysis in Fig.~\ref{fig:RIT_amp_hier}, where we display the evolution of the amplitudes of the $\psi^4_{\ell m}$ modes
during the inspiral (left panel), and their values at a reference point corresponding to the last periastron before merger (right panel). 
The latter are shown as a function of $m$ for a fixed value of $\ell$, up to $\ell = 4$.
In spite of the large numerical noise that affects the modes with $\ell > 2$, with the exception of the (4,4) mode, it is nonetheless possible to appreciate 
that odd-$m$ modes in the co-precessing frame have typically lower average amplitude than the even-$m$ ones, as one would expect from an aligned-spin, 
equal mass simulation. This is especially visible for $m=1$ modes, whose amplitudes in the inertial frame are at least one order of magnitude larger than the ones in the co-precessing frame.
While not the case for the system consdered, we remind the reader that for more eccentric simulations ($e \geq 0.8$), e.g. \texttt{RIT:eBBH:1199} or \texttt{RIT:eBBH:1132}, 
the amplitude at merger of the $(2,0)$ mode is not negligible, but rather can be comparable to that of the $(2,2)$ mode, or larger. 
This is in contrast to the quasi-circular case, where the $(2,0)$ mode is approximately zero up until merger.

\begin{figure}
  \includegraphics[scale=0.49]{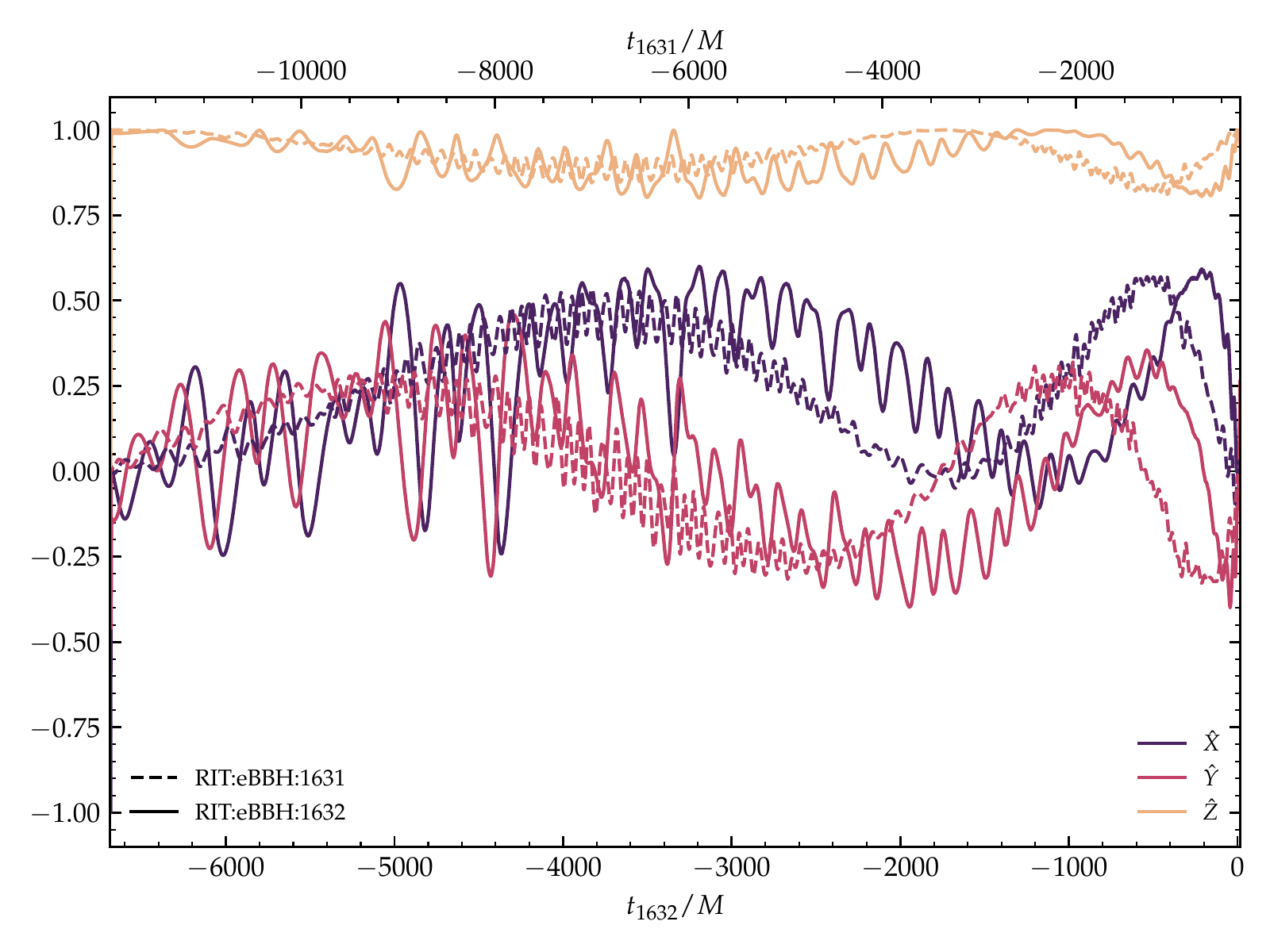}
  \caption{Comparison of the evolution of the radiation frame components $\hat{V}_x, \hat{V}_y, \hat{V}_z$ of the $\hat{\mathbf{V}}$ vector
  for the \texttt{RIT:eBBH:1631}~and \texttt{RIT:eBBH:1632}~simulations. The two simulations are characterized by the same intrinsic parameters, 
  but by different initial eccentricities. The Euler angles are weakly affected by the eccentricity of the orbit, and are morphologically 
  very similar in the two cases.
  \label{fig:rit_31_32}}
\end{figure}

We conclude this section by observing that the Euler angles extracted from the simulation considered above are morphologically very similar to the
ones computed from the \texttt{RIT:eBBH:1631} data. The latter is characterized by the same intrinsic parameters as \texttt{RIT:eBBh:1632}, 
but by a smaller initial eccentricity of $e \sim 0.19$ at approximately the same initial separation.
This fact is demonstrated in Fig.~\ref{fig:rit_31_32}, which shows the evolution of the components 
$\hat{V}_x, \hat{V}_y, \hat{V}_z$ of the radiation frame vector $\hat{\mathbf{V}}$ for the two simulations. 
The two evolutions are clearly characterized by different timescales (reprensented in the two $x$-axes of the plots), 
due to their different orbital eccentricity, but -- once appropriately rescaled -- appear rather close to one another.
This fact is consistent with the studies performed in the previous section, and suggests that the Euler angles are in general weakly 
affected by the eccentricity of the orbit even in the strong field regime, up to merger.

\section{An efficient description of eccentricity and precession}
\label{sec:eob}

After discussing the phenomenology of eccentric, precessing dynamics in both the \ac{pn} and \ac{nr} regimes,
it is finally time to move to the construction of a model that is able to quantitatively capture the waveforms
emitted by such systems, wielding the insights obtained through our previous studies.
In particular, we will first review the main features of the \ac{eob} model employed to obtain the co-precessing waveforms,
recall the simplifications that are made to obtain the quasi-spherical, precessing waveforms,
and finally discuss the generalization of the spin dynamics of Ref.~\cite{Akcay:2020qrj, Gamba:2021ydi} to the non-circular case.

\subsection{Co-precessing waveform model}
\label{subsec:co_precessing}
The baseline model that we employ for the description of the co-precessing motion
is the \TEOBResumSDali~approximant of~\cite{Nagar:2021gss,Nagar:2024dzj}.
Within the framework of this model, the dynamics of the system is obtained
from the \ac{eob} equations of motion implied by the Hamiltonian $H_{\rm EOB}$.
Assuming planar orbits, the latter is a function of the variables $\{r, p_{r*}, p_\varphi, \mathbf{S}, \mathbf{S}_*\}$, 
where $r$ is the radial separation, $p_{r*}, p_\varphi$ are the conjugate momenta to $r_* = \int dr (A/B)^{-1/2}, \varphi$, with $A,B$ being the EOB potentials,
and $\hat{S} = (S_1 + S_2)/M^2, \hat{S}_* =  (1/q S_1 + q S_2)/M^2$ 
are dimensionless combinations of the spins of the binary components.
Explicitly, the \ac{eob} Hamiltonian is given by:
\begin{equation}
  \hat{H}^{\rm EOB} = \frac{1}{\nu} \sqrt{1 + 2\nu (\hat{H}_{\rm eff} -1)} \,
\end{equation}
where $\hat{H}_{\rm eff}$ is the effective Hamiltonian:
\begin{equation}
  \hat{H}_{\rm eff}=\sqrt{A(1+p_\varphi^2 u_c^2 +Q)+p_{r_*}^2} + p_{\varphi}(G_{\hat{S}} \hat{S} + G_{\hat{S}*}\hat{S}_{*}) \, .
\end{equation}
The EOB potentials $A, D, Q$ are considered at local 5\ac{pn} order~\cite{Bini:2019nra}, and resummed according to \cite{Nagar:2021xnh}.
The gyro-gravitomagnetic functions $G_S, G_{S*}$, which encode spin-orbit contributions, are instead taken at \ac{nnlo}~\cite{Damour:2014sva,Nagar:2021xnh}, 
and inverse-resummed following usual \ac{eob} prescriptions~\cite{Nagar:2018zoe}. Even in spin effects are accounted for by the centrifugal radius 
$u_c$~\cite{Damour:2014sva,Nagar:2021xnh}, here considered up to \ac{nlo}.
Two NR-informed parameters, $a_6$ and $c_3$, complete the conservative sector of the model and ensure robustness
up to merger~\cite{Nagar:2024dzj}.
The motion of the system is then obtained by evolving the set of equations:
\begin{subequations}
  \label{eq:eob_eom}
  \begin{align}
    \frac{d r}{d {t}} &=\Bigl(\frac{A}{B}\Bigr)^{1/2} \frac{\partial \hat{H}^{\rm{EOB}}}{\partial p_{r*}}\left(r, p_{r*}, p_{\varphi}\right) \\ 
    \frac{d \varphi}{d {t}} &= \Omega = \frac{\partial \hat{H}^{\rm{EOB}}}{\partial p_{\varphi}}\left(r, p_{r*}, p_{\varphi}\right) \\ 
    \frac{d p_{r*}}{d {t}} &=\Bigl(\frac{A}{B}\Bigr)^{1/2}\Bigl(-\frac{\partial \hat{H}^{\rm{EOB}}}{\partial r}\left(r, p_{r*}, p_{\varphi}\right) + \hat{\mathcal{F}}_r \Bigr) \, ,\\ 
    \frac{d p_{\varphi}}{d {t}} &= \hat{\mathcal{F}}_{\varphi}\left(r, p_{r*}, p_{\varphi}\right) 
  \end{align}
\end{subequations}
which replace the equivalent equations for the \ac{pn} dynamics given by Eq.~\eqref{eq:eom_pn}. 
The radial and azimuthal back-reaction forces $\hat{\mathcal{F}}_\varphi$ and $\hat{\mathcal{F}}_r$ are
computed from the \ac{eob} waveform, and contain non-circular corrections via the leading order Newtonian prefactor $\hat{f}^{\rm nc}_\varphi$~\cite{Chiaramello:2020ehz}:
\begin{equation}
\hat{\mathcal{F}}_{\varphi}^{\rm EOB} = -\frac{32}{5}\nu r_{\omega}^4 \Omega^5 \hat{f}^{\rm nc}_{\varphi} \hat{f}(\Omega) \, .
\end{equation}
The model was recently shown to be more than $99\%$ faithful to both quasi-circular and eccentric~\cite{Nagar:2024dzj} 
\ac{nr} simulations up to merger and beyond. The model has also been tested against 15 non-spinning scattering simulations from Refs.~\cite{Damour:2014afa,Hopper:2022rwo}, 
21 spinning scattering systems from Ref.~\cite{Rettegno:2023ghr}, multiple dynamical captures simulations from Ref.~\cite{Andrade:2023trh, Gamba:2021gap} as well as a large 
number of test-mass Regge-Wheeler-Zerilli and Teukolsky waveforms~\cite{Albanesi:2021rby, Albanesi:2022ywx}. More details and an in-depth discussion of the model can be found in Ref.~\cite{Nagar:2024dzj}.

\subsection{Quasi-spherical precessing orbits}
\label{subsec:qc_precessing}

Most state-of-the-art \ac{eob} models for precessing, quasi-spherical \acp{bbh} do not directly solve 
a fully coupled \ac{pn} system of \acp{ode} describing the spins and orbital dynamics, such as Eq.~\eqref{eq:eom_pn}.
Instead, they employ a scheme which relies on a split between precessing and orbital evolutions, the former typically considered 
in \ac{pn} form, the latter from the planar, resummed \ac{eob} dynamics. 
This scheme, inspired by phenomenological models, was first introduced in the context of the \ac{eob} framework by 
Ref.~\cite{Akcay:2020qrj} and then further improved upon and refined in Ref.~\cite{Gamba:2021ydi,Ramos-Buades:2023ehm}.
The \acp{ode} system considered by the \TEOBResumS{}~family for the spins evolution is of the form:
\begin{subequations}
  \begin{align}
   \Sadot   & = f_1(\Omega, \eta, \Lhat, \Sa, \Sb) \, ,\\
   \Sbdot   & = f_2(\Omega, \eta, \Lhat, \Sa, \Sb) \, ,\\
   \dot{\bm \ell} & = f_L(\Omega, \eta, \Sadot, \Sbdot) \, ,\\
   \dot{\Omega} & = \dot{\Omega}_{\rm PN}(\Omega, \eta, \Lhat, \Sa, \Sb) \, ,
   \label{eq:LNdot}
   \end{align}
\end{subequations}
where $f_1, f_2, f_L$ and $\dot{\Omega}_{\rm PN}$ can be read from \cite{Akcay:2020qrj}.

Critically, independently evolving the spins and the orbital dynamics allows for a significant reduction in the computational cost of the model,
which -- in the quasi-circular limit -- can rely on analytical acceleration techniques such as the \ac{pa}~\cite{Nagar:2018gnk} and 
\ac{spa}~\cite{Gamba:2020ljo} to obtain the waveform.
The matching between the two evolutions of Eqs.~\eqref{eq:eob_eom} and Eqs.~\eqref{eq:LNdot} represents the most delicate point in the scheme, and is typically performed by
interpolating the Euler angles $\alpha, \beta, \gamma$ obtained from the precessing dynamics to the orbital (\ac{pn}) frequency, 
and then identifying $\Omega_{\rm PN}$ with the orbital \ac{eob} frequency $\dot{\varphi} = \partial \hat{H}^{\rm{EOB}}/\partial p_{\varphi}$ 
or the waveform frequency $\sim \omega_{22}/2$ to obtain the map to \ac{eob} time; see Sec. IIC of Ref.~\cite{Gamba:2021ydi} for more details.
%
Models based on this scheme were shown to be faithful to a large number of precessing, quasi-spherical \ac{nr} simulations from the \ac{sxs}
catalog. The \TEOBResumSGIOTTO~model in its first \ac{imr} precessing iteration, in particular, was validated against 99 \ac{nr} simulations in the \texttt{lvcnr} catalog,
and 20 additional simulations with mass ratios $q \leq 4$ and $\chi_p \leq 0.49$ with more than $70$ cycles.
Its median unfaithfulness against these sets was found to be $7 \times 10^{-3}$ and $5\times 10^{-3}$ respectively, for an inclination $\iota=\pi/3$~\cite{Gamba:2021ydi}.

\subsection{Generalized spins dynamics}
\label{subsec:gen_spin}
We now aim to extend the procedure summarized in the previous section to non-circular orbits.
There are a few obvious ways to do so. We list them below in order of growing complexity:
\begin{enumerate}

  \item[(i)] Use the evolution of $\Omega(t)$ given by \TEOBResumSDali~in place of $\dot{\Omega}_{\rm PN}$. 
  This immediately allows for the inclusion of eccentricity-related effects and -- since no interpolation or orbit averaging
  is required -- this strategy can be applied to all kinds of systems, including scatterings and captures.
  While no contributions of (explicit)  ``non circularity'' to $\Sadot, \Sbdot, \dot{\bm{\ell}}$ are considered, 
  the results of Sec.~\ref{sec:pn} indicate that this does not represent a significant issue for eccentric systems.
  A bigger drawback is represented by the fact that the evolution of the spins requires the EOB dynamics to be evolved first, 
  making it not straightforward to account for time-varying contributions to the orbital dynamics.

  \item[(ii)] Solve the quasi-circular PN spin precessing equations and use them to obtain the Euler angles
  $\alpha, \beta, \gamma$. Interpolate the angles in the frequency domain, and map them to an orbit-averaged EOB (orbital)
  frequency (see e.g. Sec. II D of \cite{Shaikh:2023ypz}). This method has the advantage of allowing for the inclusion of
  time-varying contributions to the orbital dynamics, so long as one is able to compute -- at each
  moment of the \ac{eob} dynamics evolution -- an orbit-averaged frequency.
  Once more, thanks to the difference in the orbital and precession timescales, this simple technique is expected to work well for mildly eccentric binaries.
  It however cannot be applied to e.g. scatterings or captures, and neglects all contributions of ``non circularity'' to the spins dynamics.

  \item[(iii)] Solve the quasi-circular PN spin precessing equations together with the generic \ac{eob} orbital dynamics.
  This method allows for the inclusion of time-varying spins contributions 
  while retaining both the ability to account for generic orbital configurations beyond eccentric systems and the theoretical simplifications that come with 
  employing a co-precessing orbital dynamics description\footnote{By this we mean that the \ac{eob} Hamiltonian and radiation reaction forces employed are those used for planar
  orbits, which are better studied than their non-planar counterparts}. 
  Its main limitation is represented by its increased computational cost with respect to the previous options, as it requires the solution of
  a coupled system of 12 \acp{ode}.

  \item[(iv)] Solve the spins dynamics equations of \cite{Klein:2021jtd,Schnittman:2004vq,Fumagalli:2023hde}, together with orbital dynamics explicitly parameterized in terms
  of (quasi-Keplerian) eccentricity, anomaly and orbit-averaged frequency. This method extends the previous one by including the effects of
  eccentricity on the spins dynamics, while still allowing for the inclusion of time-varying contributions to the orbital dynamics and being applicable to 
  larger values of eccentricity. However, it retains the limitation of not being applicable to all kinds of systems, as it assumes a parameterization of the
  orbital dynamics where eccentricity is explicitly present.
\end{enumerate}

For the time being, we choose to adopt the first strategy, as it is the simplest to implement and -- at the same time --
represents a good compromise between accuracy, computational cost and generality.

The chosen method is then implemented following these steps:
\begin{enumerate}
  \item[(i)] we first obtain the evolution of the system in the co-precessing frame, assuming that waveforms can be well approximated
  by those given by the \TEOBResumSDali~aligned-spins model. 
  \item[(ii)] From the co-precessing frame EOB dynamics we extract the orbital frequency evolution $\Omega(t)$ (or, alternatively, the waveform frequency 
  evolution  $\omega(t) \sim \omega_{22}(t)/2$\footnote{Notably, this relation is only approximately valid for low eccentricities.}).
  \item[(iii)] We employ this EOB evolution of the frequency to drive the evolution of the spins in the co-precessing frame, utilizing the 
  orbit-averaged PN expressions for $\Sadot, \Sbdot, \Lhatdot$ of \cite{Akcay:2020qrj}.
  \item[(iv)] With the evolution of the spins at hand, we compute the Euler angles and rotate the co-precessing waveforms to the inertial frame as 
  described in e.g.~\cite{Akcay:2020qrj, Gamba:2021ydi}.
  \item[(v)] We extend $\alpha, \beta, \gamma$ beyond merger by fixing them to their merger values.
\end{enumerate}

\begin{figure*}[t]
  \includegraphics[width=\textwidth]{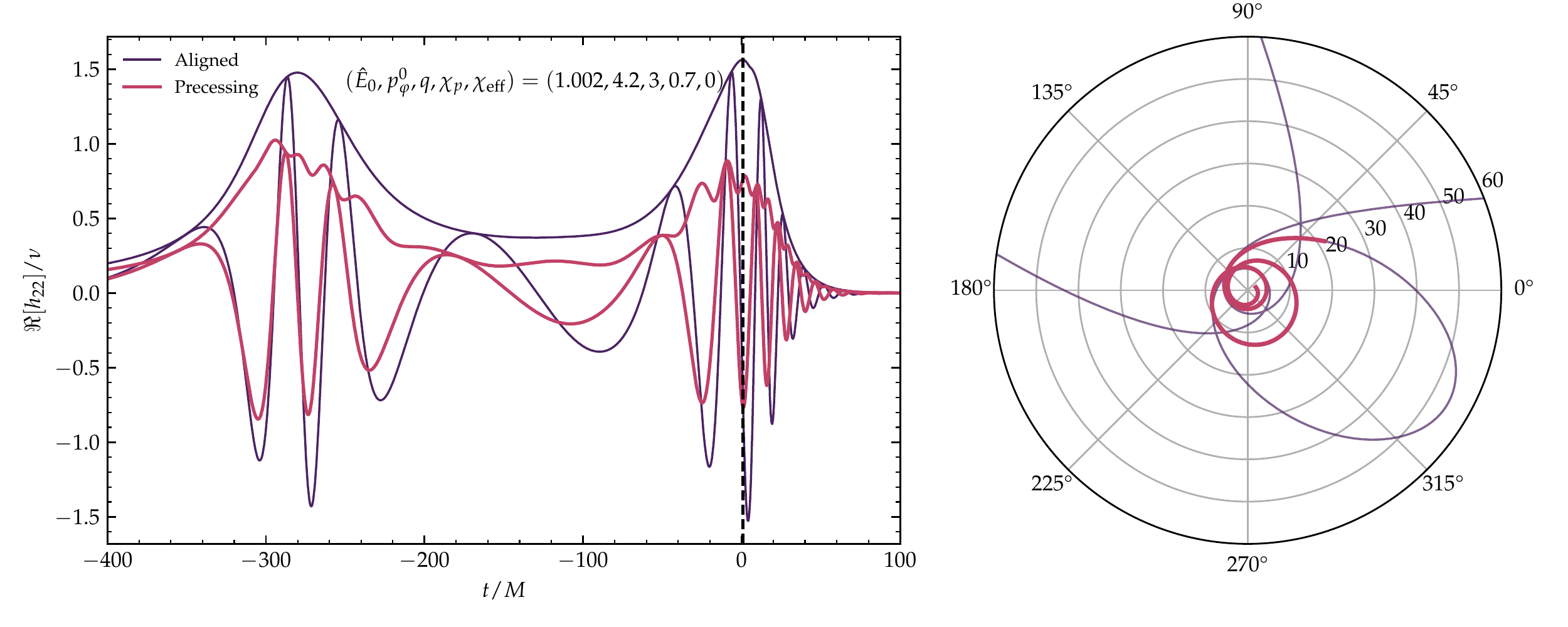}
  \caption{Quadrupolar waveform mode (left) and co-precessing orbit (right) for a dynamically captured system with mass ratio $q=3$, spins $\bm{\chi}_1 = (0.7, 0, 0)$, $\bm{\chi}_2 = (0, 0, 0)$
  and initial energy, angular momentum and EOB radial separation $1.002, 4.2$ and $10000$. We compare the precessing waveform to its aligned spin counterpart. The portion of the waveform
  shown in the left panel is highlighted in the orbit plot on the right with a dark red line.
  Noticeable modulations, due to the mixing between $\ell=2$ co-precessing modes, can be observed at each periastron passage and at the time of merger.
  \label{fig:capture}}
\end{figure*}

This scheme is computationally cheaper than a full evolution of the 3D \ac{eob} dynamics, 
with no additional \acp{ode} to solve and no need for a new Hamiltonian with respect to the quasi-spherical case.
It largely relies on the
assumption that non-circular terms in the spins dynamics are negligible and that the evolution of 
the orbital dynamics is not strongly affected by time-varying spins contributions.
Both assumptions were shown to be approximately true in Sec.~\ref{sec:pn}, and can be expected to hold up even
close to merger, where the system circularizes. The twisting procedure, too, does not appear to require modifications with respect to the 
quasi-circular case, as was empirically demonstrated in Sec.~\ref{sec:nr}. 
Beyond the mildly eccentric case, this scheme can be applied also to unbound systems, scatterings and captures (see Fig.~\ref{fig:capture}),
although the validity of the orbit-averaged expressions for the spins dynamics is uncertain in these regimes.

\section{Validation}
\label{sec:validation}
We conclude the results presented in this work by validating the model in the quasi-circular, precessing limit, comparing
its performance to that of the \TEOBResumSGIOTTO~model of Ref.~\cite{Gamba:2021ydi}.
We then move on to compare the model to one mildly eccentric, precessing waveform from the \RIT~database.
This comparison is performed in the time domain, focusing mainly on the the $(2,2)$ and $(2,1)$ modes.
Given the lack of availability of multiple resolutions, as well as the fact that most other eccentric 
and precessing waveforms are either not very eccentric or not very long, we leave a more in-depth validation 
of the model in this regime to future works, where we will also present new eccentric and precessing simulations 
of \acp{bbh}~\cite{Gamba:2024}.

\subsection{Quasi-circular, precessing limit}
Following the same procedure detailed in Sec.~III of Ref.~\cite{Gamba:2021ydi}, we compare the model presented in this work to 99
\ac{nr} simulations from the \texttt{lvcnr} catalog, spanning mass ratios $q \leq 6$, $\chi_p \leq 0.89$ and $\chi_{\rm eff} \in [-0.45, 0.65]$,
and 21 ``long'' simulations of \acp{bbh} with mass ratios $q \leq 4$ and spins $\chi_p \leq 0.49$ and $\chi_{\rm eff} \in [-0.2, 0.3]$.
We quantify the goodness of our model in terms of the sky-maximized unfaithfulness~\cite{Harry:2016ijz,Pratten:2020ceb,Harry:2017weg}, which is defined as:
\begin{equation}
  \bar{\mathcal{F}}_\text{SM}  = 1 - \max_{t_0^h, \varphi_0^h, \kappa^h, \xi_0} \frac{(s, h)}{\sqrt{(s, s)(h, h)}} \, ,
\end{equation}
where $s$ is the target (\ac{nr}) waveform, $h$ is the model waveform and the maximization is performed over reference time $t_0^h$,
reference phase $\varphi_0^h$, effective polarization angle $\kappa^h$ and over an initial rotation of the in-plane spins $\xi_0$.
This quantity is then \ac{snr} weighted, and averaged over the sky, the polarization and the initial phase of the target waveform:
we choose four different values for the effective polarization, $\kappa^s = \{0, \pi/4, \pi/2, 3\pi/4\}$, and 
six values for the target reference phase, $\varphi_{0}^s = \{ 0, 2\pi/5, 4\pi/5, 6\pi/5, 8\pi/5 \}$, to average over.
We fix the power spectral density of the detector to the Zero-Detuned High-Power advanced LIGO design sensitivity~\cite{Sn:advLIGO}, 
and compute mismatches from $20$ to $2048$ Hz using the $(\ell, |m|) = (2, 2), (2, 1), (3, 3), (4, 4)$ modes.

\begin{figure}[t]
  \includegraphics[scale=0.7]{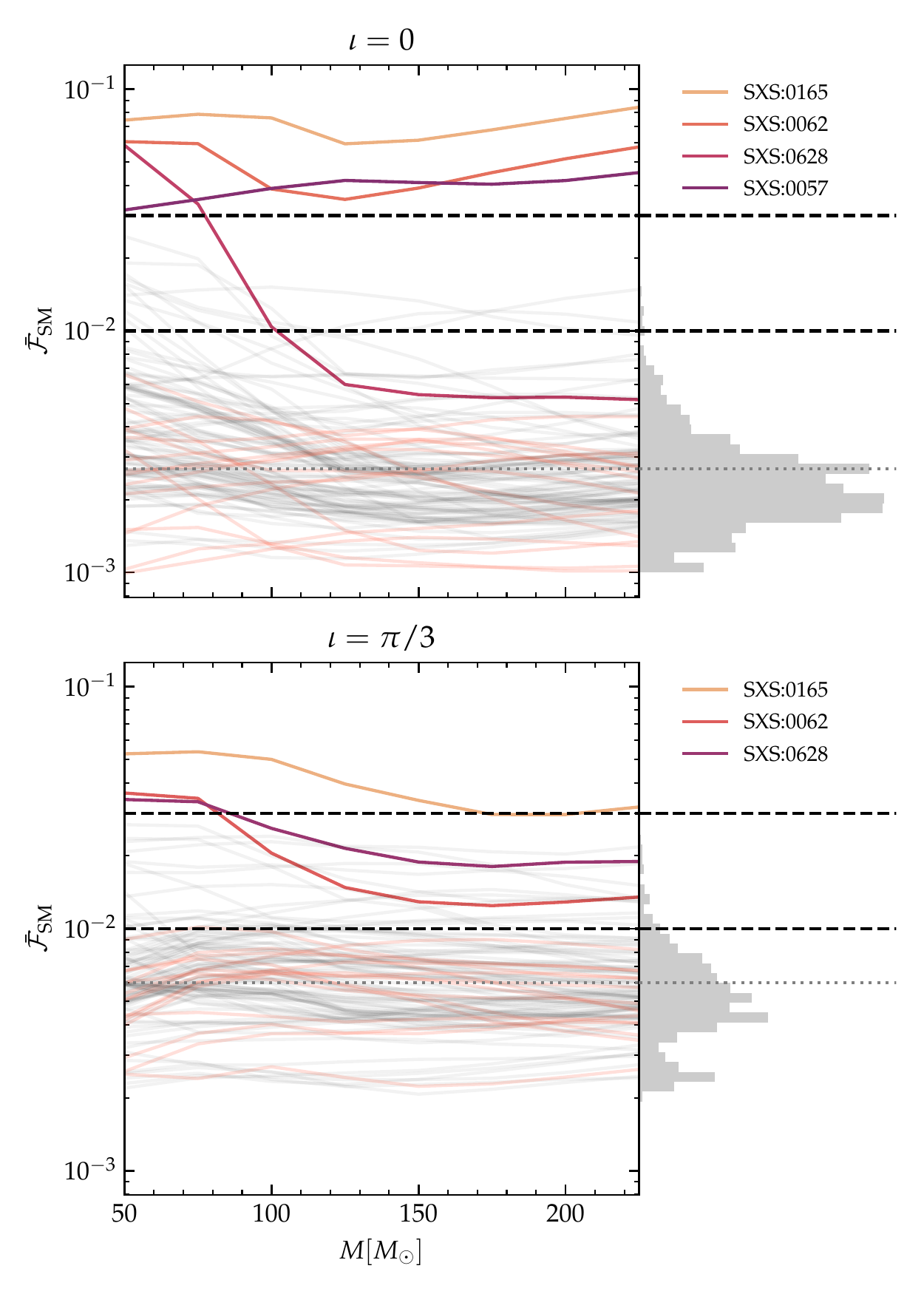}
  \caption{ Mismatches of the model for the $99$ \ac{nr} simulations of the \texttt{lvcnr} catalog (gray) and the 20 ``long'' simulations
  (red) discussed in the text, for different total masses and two different fixed values of inclination: 
  $\iota=0$ (top) and $\iota=\pi/3$ (bottom). We highlight with colored lines all simulations which at any total mass value cross the $3\%$ threshold. Between inclinations,
  the more problematic simulations are the same, but the overall performance of the model worsens for increasing $\iota$, as expected. We find median unfaithfulnesses of 
  $0.003^{+0.009}_{-0.001}$ and  $0.006^{+0.010}_{-0.003}$ for $\iota=0$ and $\iota=\pi/3$ respectively.
  \label{fig:validation_qc}}
\end{figure}

Results are displayed in Fig.~\ref{fig:validation_qc}, where we show the mismatches of the model for the simulations considered as 
a function of the total mass of the system for two target inclinations $\iota=0$ and $\iota=\pi/3$. 
Global distributions of the mismatches are also shown as histograms in the right panel of the same figure.
We find that for the majority of the simulations the model is able to maintain an unfaithfulness below $3\%$, with the notable exceptions
of the \texttt{SXS:BBH:0165}, \texttt{SXS:BBH:0062}, \texttt{SXS:BBH:0628} and \texttt{SXS:BBH:0057} simulations.
These systems are characterized by either very asymmetric mass ratios ($q \geq 5$), strong precession ($\chi_p > 0.7$)
or both, and are known to be challenging for \ac{gw} models in general (see Tab.~2 of~\cite{MacUilliam:2024oif}).
The global unfaithfulness found for the set considered with $\iota=0$ is $0.003^{+0.009}_{-0.001}$, where the we employ the standard notation of
quoting the median and the $90\%$ confidence interval of the distribution.
As inclination is increased from $\iota=0$ to $\iota=\pi/3$, the model becomes overall less faithful, especially for the merger-ringdown portion of the waveform.
Once more, this is not surprising, given that for more face-on systems the importance of higher modes increases, and they are both (i) more affected by the
precession of the spins and (ii) less well-modelled in the aligned-spin limit.
In this case, the global unfaithfulness is $0.006^{+0.010}_{-0.003}$, with the same notation as before.
Overall, we find the performance of the model in the quasi-circular precessing limit to be comparable to that of other state 
of the art models, by indirectly comparing to the results of Ref.~\cite{Gamba:2021ydi,MacUilliam:2024oif}.

\subsection{RIT:eBBH:1632}

\begin{figure*}
  \includegraphics[scale=0.7]{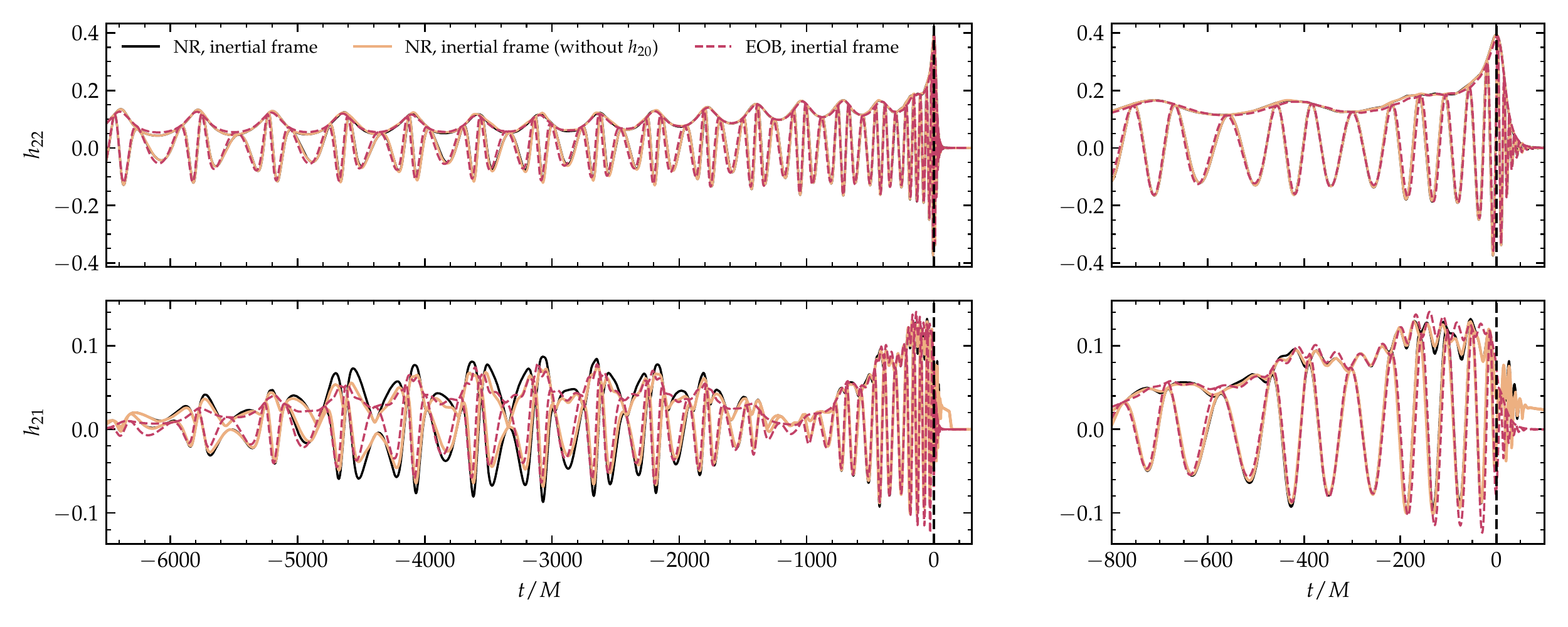}
  \caption{Comparison between RIT NR waveform multipoles $h_{\ell m}$ = (2,2) and (2,1) (black and orange lines) and the EOB model (red line) presented in this work.
  While the (2,2) EOB and NR modes do not display large precession-induced modulations, the (2,1) mode is clearly affected by the precession of the spins.
  For this mode, it is possible to observe large EOB/NR differences during the early inspiral ($t\sim-5500$ to $t\sim-3000$). 
  It is not clear whether these differences are due to real inaccuracies of the EOB model, or rather can be reconduced to unphysical features of the simulation itself.
  Overall, our EOB model is able to capture the main features of the NR data in terms of both amplitude and phase evolution up to merger and beyond.
  \label{fig:eobnr_rit_1632}}
\end{figure*}

The performance of the model in the eccentric, precessing regime is tested via a time-domain comparison against the \RIT~simulation
\texttt{RIT:eBBH:1632}, already discussed in Sec.~\ref{sec:nr} when considering the behavior of radiation-frame waveforms.
Directly comparing waveforms in the inertial frame is not straightforward, as precessing waveforms are characterized by one additional
degree of freedom with respect to the aligned-spin case: an initial angle $\xi$ that determines the orientation of the in-plane spins.
Therefore, comparing waveforms in the inertial frame would require devising a method to align them while varying three different parameters at the same time: 
initial eccentricity, initial anomaly and $\xi$ (assuming a fixed initial orbit-averaged frequency).
This can in princple be performed by, e.g., minimizing mismatches over a certain frequency range via either multi-dimensional
numerical minimization or a simpler grid search~\cite{Ramos-Buades:2023ehm}. Given the complex functional dependence of the waveform on the parameters,
these methods are not expected to be particularly efficient, and typically require a large number of waveform evaluations to be performed to succeed.
Therefore, rather than employing such a procedure, we choose to focus on the (2,2) mode and first find the eccentricity and anomaly $e_0$ and $\zeta_0$
that align the \ac{eob} waveform with the \ac{nr} one in the co-precessing frame. Then -- keeping these fixed -- we vary $\xi$ to obtain the 
$\Delta t_{22}$, $\Delta \phi_{22}$ values that provide the best alignment in the inertial frame.
The initial conditions and shifts so obtained are used to also align the $(2,1)$ mode, recalling that $\Delta \phi_{\ell m} = m/2 \Delta \phi_{22}$.

\begin{figure}
  \includegraphics[scale=0.6]{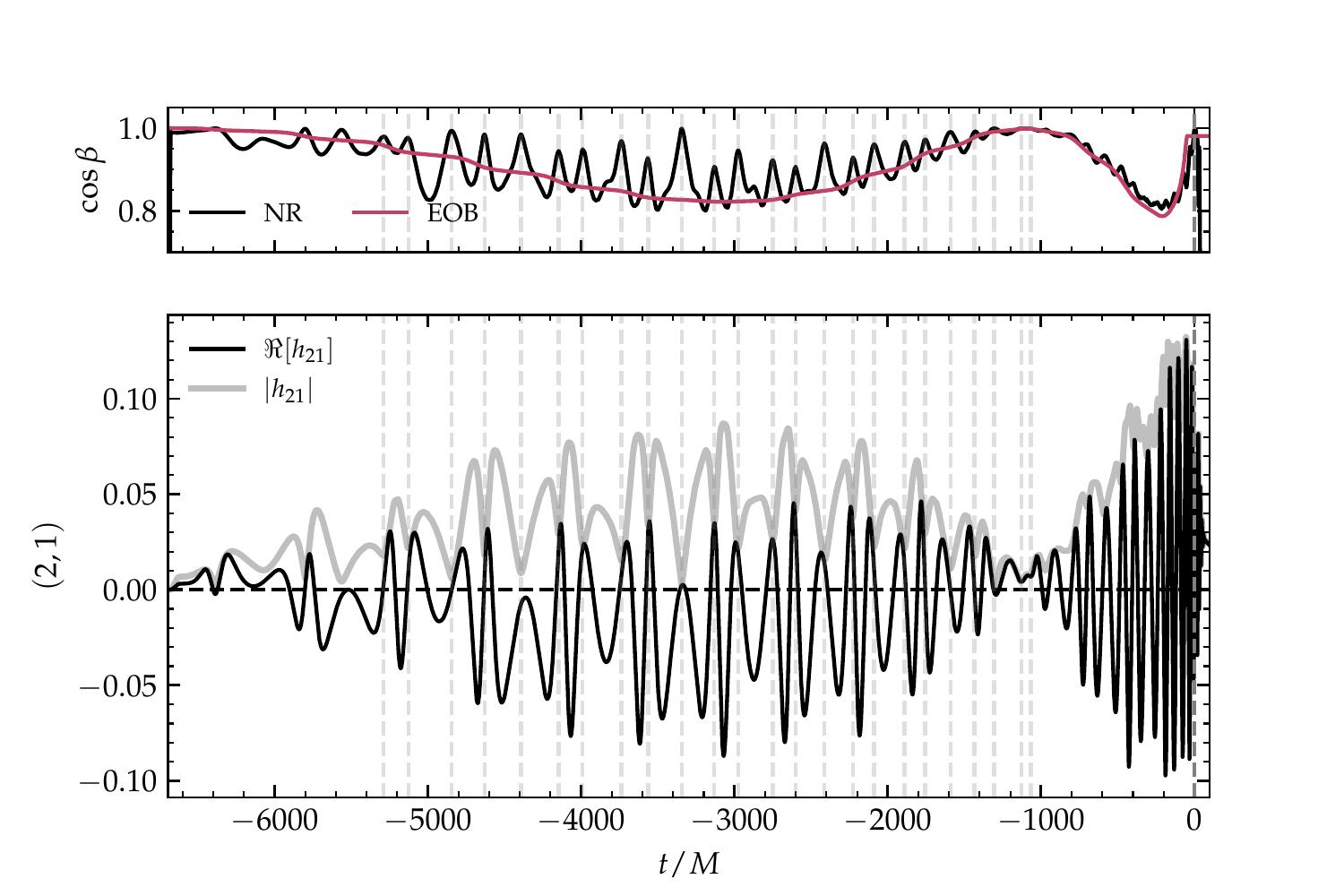}
  \caption{Top: evolution of the Euler angle $\beta$ extracted from the RIT:eBBH:1632 simulation (black) and the \ac{eob} prediction (red). 
  Bottom: real part (black) and amplitude (gray) of the inertial-frame (2,1) mode of the same \ac{nr}~simulation. 
  We highlight with dashed vertical lines the minima of the amplitude of the (2,1) mode between $-5500 M$ and $-1000 M$, 
  which are not present in the \ac{eob} prediction: they correspond to the times when the $\beta$ angle extracted from the \ac{nr} simulation is closest to $0$.
  These spikes in $\beta$ do not, however, clearly correspond to periastron or apastron passages, or other reference points in the orbit.
  A noticeable ``drift'' can be observed from the real part of the NR waveform, which does not oscillate around $y=0$. 
  Recalling that the \ac{nr} Euler angles are extracted by identifying the radiation frame from the waveform modes, 
  it apparent that this drift is closely related to the behavior of the \ac{nr} $\beta$. What remains unclear is whether it is due to
  a real physical effect, which induces oscillations in $\beta$ reflected in the inertial-frame waveform, or rather to inaccuracies in the \ac{nr} data itself,
  which are then reflected in the calculation of Euler angles.
  \label{fig:eobnr_rit_1632_detail}}
\end{figure}

The results of this procedure are displayed in Fig.~\ref{fig:eobnr_rit_1632}.
The model correctly reproduces the amplitude peaks of the $(2,2)$ mode, as well as the less pronounced modulations due to
the precession of the orbital plane. The $(2,1)$ mode in the inertial frame is computed entirely from the $(2, |2|)$ co-precessing mode,
and therefore it is more informative -- in principle -- regarding the performance of the model.
We find that the \ac{eob} prediction is qualitatively consistent with the \ac{nr} data, especially close to merger where 
the model approximates the amplitude modulations with remarkable accuracy ($t\sim -400$ to $-200$), and overall captures the envelope
of the waveform amplitude.
Gauging the quantitative performance of the model across the inspiral is more challenging, as unusual features appear in the \ac{nr} waveform.
Indeed, between the times of $t\sim -5500$ and $t\sim -3000$ the (2,1) \ac{nr} mode displays zeroes in the amplitude, which are not present in the \ac{eob} prediction, as well as
an overall ``drift'' (see Fig.~\ref{fig:eobnr_rit_1632_detail}).
This effect was initially thought to be related to the presence of a non-zero (2,0) coprecessing mode. However, after computing the \ac{nr} radiation-frame modes, removing the (2,0)
and rotating back to the inertial frame, the presence of such zeroes does not appear to be significantly affected.
The physical reality of this feature is therefore uncertain, and given that the merger-ringdown portion of the waveform appears significantly affected by
errors in the simulation itself, the differences observed in the inspiral may be due to limitations in the \ac{nr} data rather than in the \ac{eob} model.

\section{Conclusions}
\label{sec:conclusions}

In this work we discussed the phenomenology of eccentric, precessing \acp{bbh} in both the \ac{pn} and \ac{nr} regimes.
After a brief review of the \ac{pn} equations of motion for these kind of systems, we studied the morphology of
the Euler angles $\alpha, \beta, \gamma$ that connect the co-precessing, $\bm{\ell}$-aligned frame with the inertial frame
for various kind of non-circular binaries (bound systems, eccentric and quasi-circular, scatterings, captures), highlighting common features
and differences. In view of the development of our non-circular, precessing model, we have assessed the importance of 
the explicitly non-circular contributions to the spin dynamics of the system, finding that they are largely negliglible, even up to high values of $e \sim 0.9$.
This result is in line with previous studies~\cite{Ireland:2019tao,Fumagalli:2023hde}, and suggests that our model can be expected to be accurate up to such eccentricities.
We have also briefly discussed the impact of precession on the scattering angle of low-energy \ac{bbh} systems, finding again that spin precession does not significantly affect
aspects of the orbital dynamics. Indeed, it contributes $\lesssim 1\%$ to the azimuthal scattering angle, which displays a much stronger dependence on the mass ratio and 
the out-of-plane spin components, while generating usually only a modest deviation from the original orbital plane.

We then performed a preliminary study of eccentric, precessing numerical relativity simulations. We highlighted that the
radiation-frame waveforms obtained by finding the direction of the tensor $\langle \mathcal{L}_{(a} \mathcal{L}_{b)} \rangle$ are well approximated by aligned spins waveforms,
and that the hierarchy of the modes in the co-precessing frame is consistent with that of aligned-spin systems. This is in line with the
expectations from the quasi-circular limit, and indicates that the twisting procedure routinely employed for these scenarios can be straightforwardly
extended to the eccentric case. 
We also showed that, consistently with the \ac{pn} results, the details of the orbit do not significantly affect the radiation-frame
evolution, as one would expect from the separation of timescales between orbit, precession and back-reaction.

Finally, we presented a model that is able to quantitatively describe the $(2,2)$ mode of the waveforms emitted 
by non-circularized systems.  We discussed the different avenues that can be easily followed to extend any time domain eccentric
waveform model to the precessing case, and  chose to adopt the simplest one, which is based on the orbit-averaged expressions for the spin dynamics
of Ref.~\cite{Akcay:2020qrj}. We then presented the first results of this model, showing that it is faithful to \ac{nr} simulations in the
quasi-circular limit, and performing one EOB/NR comparison for the largest eccentricity simulation (with precession) in the \RIT~database 
with more than $10$ orbital cycles.
While the model appears to be accurate up to merger and beyond in the $(2,2)$ mode, the behavior of the $(2,1)$ mode is only qualitatively captured
during the inspiral. It is unclear whether this is due to limitations of the \ac{eob} model or inaccuracies of the \ac{nr} data itself.

As the community slowly undertakes the endeavor of producing new highly accurate, eccentric, precessing \ac{nr} simulations,
the \ac{eob} model presented in this work should be considered as a predictive tool, a-posteriori confirmed (or falsified) by \ac{nr}. 
Along these lines, in future works we will perform a more in-depth \ac{nr} validation of the model within this challenging regime. 
This will allow us to clearly identify its limitations and define its current range of applicability, indicating avenues for future improvements.
At the same time, we also aim to incorporate the description of the $(2,0)$ mode, known to be significant for highly eccentric systems, 
and -- more ambitiously -- to build a full \ac{imr} model accounting for non-circularity and precession also beyond merger.
In doing so, we anticipate refining our understanding of eccentric, non-planar coalescing \acp{cbc}, ultimately 
enriching our comprehension of these complex systems.

\acknowledgments
The authors warmly thank the anonymous referee for their insightful comments.
In particular, point (iii) of the list in Sec.~\ref{subsec:gen_spin} stems from the referee's suggestion, which the 
authors merely expanded upon and for which they do not claim any intellectual credit.
The authors also wish to thank G.~Pratten, G.~Carullo, A.~Nagar, S.~Bernuzzi, D.~Radice and B.~Sathyaprakash for useful discussions and feedback.
RG and DC additonally thank A.~``Nali'' Scarrone, D.~Rettore and A.~Mango for inspiring them throughout the development of this work.
Finally, the authors wish to thank the \MAYA, \RIT~and \ac{sxs} collaborations for sharing their simulations and mantaining their public
databases.
RG acknowledges support by the Deutsche Forschungsgemeinschaft (DFG) under Grant No.
406116891 within the Research Training Group RTG 2522/1 and from NSF Grant PHY-2020275
(Network for Neutrinos, Nuclear Astrophysics, and Symmetries (N3AS)).
DC acknowledges support from the Italian Ministry of University and Research (MUR) via the PRIN 2022ZHYFA2, {\it GRavitational wavEform models for coalescing compAct binaries
with eccenTricity} (GREAT).
SN acknowledges support from DAAD and KVPY for funding.
\TEOBResumS{} is publicly available at \url{https://bitbucket.org/eob_ihes/teobresums}. The version
employed in this work is tagged via the arXiv submission number of the paper itself.

\appendix
\section{\ac{pn} initial conditions}
\label{app:PN_ics}
Section~\ref{sec:pn} of this work relies on the integration of the \ac{pn} equations of motion 
(Eqs.~\ref{eq:eom_pn}) for different kinds of orbital dynamics: quasi-circular, eccentric
and hyperbolic (or capture). Orbits corresponding to these three main faimilies are realized by choosing
appropriate initial conditions, i.e. values of the initial radial separation and velocity.
In this appendix we provide expressions for computing the initial conditions for eccentric and hyperbolic orbits
in terms of more convenient sets of parameters: eccentricity and semi-latus rectum for the former, and energy and angular momentum for the
latter. As already mentioned in the main text, we neglect spin-orbit terms in the expressions for the energy and angular momentum that we use to set up the initial conditions
(they are included in the equations of motion); their impact is expected to be small, and this approximation does not invalidate the conclusions of our study of the PN dynamics.

In the expressions in this appendix we include explicit powers of $1/c$ to keep track of PN order.
\subsection{Eccentric orbits}
We define the eccentricity $e$ and semi-latus rectum $p$ by their relation to the periastron and apastron radii $r_{p,a}$, as in Newtonian gravity:
\begin{equation}
  r_p = \frac{p}{1+e},  \qquad r_a = \dfrac{p}{1-e}.
\end{equation}
To calculate the velocities $v_{p,a}$ at the extremes of the orbit we use the 3\ac{pn} expressions for the energy and angular momentum as functions of 
$r, v$ and $\dot{r}$ given in Eq. 23 of~\cite{Memmesheimer:2004cv}: we require that both $\hat{E} = E/\mu$ and $\hat{L} = L/\mu$ have the same value when evaluated at
periastron and apastron, and solve the resulting system of equations for the \ac{pn}-expanded $v_{p,a} \left(e,p\right)$:
\begin{align}
  \hat{E} \left(r_p, v_p, \dot{r} = 0\right) &= \hat{E} \left(r_a, v_a, \dot{r} = 0\right) \\
  \hat{L} \left(r_p, v_p, \dot{r} = 0\right) &= \hat{L} \left(r_a, v_a, \dot{r} = 0\right).
\end{align}
We show here the result for the periastron velocity $v_p \left(e,p\right)$, where we start our eccentric orbits:
\begin{widetext}
\begin{equation}
\begin{split}
 v_{p} = \dfrac{1+e}{\sqrt{p}} &\Biggl\{1  + \dfrac{1}{2 p c^2}\left[e^2+8 e (\nu -1)-\nu -3\right] + \frac{1}{8 p^2 c^4}\Bigl[e^4 \left(-24 \nu ^2+8 \nu +3\right)+ 4 e^3 \left(6 \nu ^2-7 \nu -4\right) \\
           &+ e^2 \Bigl(48 \nu ^2-135 \nu +58\Bigr) + 8 e \left(9 \nu ^2-6 \nu +8\right)-41 \nu ^2+111 \nu +15\Bigr] \\
           &+ \frac{1}{p^3 c^6} \Biggl[e^6 \left(\frac{61 \nu ^3}{8}-\frac{91 \nu ^2}{8}+3 \nu +\frac{5}{16}\right) +e^5 \left(-\frac{69 \nu ^3}{4}+\frac{125 \nu ^2}{4}-\frac{113 \nu }{8}-\frac{3}{2}\right)\\
           &+e^4 \left(-\frac{137 \nu^3}{8}+\frac{203 \nu ^2}{8}+\frac{\left(4305 \pi ^2-123848\right) \nu }{13440}+\frac{47}{16}\right) +e^3 \left(53 \nu ^3-\frac{531 \nu ^2}{4}+\frac{\left(391108-4305 \pi ^2\right) \nu }{3360}-8\right)\\
           &+e^2 \left(\frac{59 \nu ^3}{8}-\frac{77 \nu ^2}{4}+\frac{\left(51608-1435 \pi ^2\right) \nu }{2240}-\frac{313}{16}\right) +e \left(-\frac{13 \nu ^3}{4}+\frac{279 \nu ^2}{4}+\frac{\left(-29696-12915 \pi^2\right) \nu }{3360}-\frac{29}{2}\right)\\
           &-\frac{79 \nu ^3}{16}+\frac{141 \nu ^2}{8}+\frac{\left(4305 \pi ^2-855976\right) \nu }{13440}-\frac{35}{16}\Biggr] \Biggr\}
\end{split}
\end{equation}
\end{widetext}

\subsection{Hyperbolic orbits}
For hyperbolic orbits, we choose the initial values of the energy $\hat{E}_0$ and orbital angular momentum $\hat{L}_0$, as well as the starting orbital separation $r$.
We then invert the PN-expanded expressions of the energy and angular momentum (Eq. 23 of~\cite{Memmesheimer:2004cv}) to find the radial and azimuthal components of the velocity vector:
\begin{widetext}
  \begin{align}
    v_r^2 &= 2 \Ez-\frac{\Jz^2}{r^2}+\frac{2}{r} + \dfrac{1}{c^2}  \Biggl[\Ez^2 (9 \nu -3)+\frac{\Ez \Jz^2}{r^2} (2-6 \nu )+\dfrac{1}{r^2}(5 \nu -10)+\frac{\Ez}{r} (14 \nu -12)+\frac{\Jz^2}{r^3} (8-7 \nu)\Biggr] \nonumber \\
        &+ \dfrac{1}{c^4} \Biggl[\Ez^3 \left(16 \nu ^2-19 \nu +4\right)+\frac{\Ez^2 \Jz^2}{r^2} \left(-15 \nu ^2+15 \nu -3\right)+\dfrac{\Ez}{r^2} \left(36\nu ^2-127 \nu +36\right) + \frac{\Ez^2}{r} \left(42 \nu ^2-96 \nu +18\right)\nonumber \\
        &+\frac{\Ez \Jz^2}{r^3} \left(-55 \nu ^2+73 \nu-16\right)+\dfrac{1}{r^3} \left(10 \nu ^2-\frac{81 \nu }{2}+26\right) + \frac{\Jz^4}{r^5} \left(\frac{15 \nu }{4}-\frac{\nu ^2}{4}\right)+\frac{\Jz^2}{r^4}\left(-38 \nu ^2+\frac{267 \nu }{4}-33\right)\Biggr]\nonumber \\
        & +\dfrac{1}{c^6} \Biggl[\Ez^4 \left(14 \nu ^3-\frac{211 \nu ^2}{4}+32 \nu -5\right)+\frac{\Ez^3}{r} \left(70 \nu ^3-342 \nu ^2+186 \nu -24\right)+\frac{\Ez \Jz^2}{r^3} \left(-192 \nu ^3+\frac{651 \nu ^2}{2}-168 \nu +24\right) \nonumber \\
        & +\dfrac{\Ez}{r^3} \left(74 \nu ^3-\frac{911 \nu ^2}{2}+\frac{41 \pi ^2 \nu }{8}+\frac{165353 \nu }{210}-76\right)+\frac{\Ez^3 \Jz^2}{r^2} \left(-20 \nu ^3+50 \nu ^2-27 \nu +4\right)+\dfrac{\Ez^2}{r^2} \Bigl(114 \nu ^3-659 \nu ^2 \nonumber \\
        &+\frac{1738 \nu }{3}-54\Bigr) +\frac{\Ez \Jz^2}{r^4} \left(-366 \nu ^3+\frac{1439 \nu ^2}{2}-\frac{2463 \nu }{4}+66\right)+\dfrac{1}{r^4} \left(16 \nu ^3-\frac{403 \nu ^2}{4}+\frac{41 \pi ^2 \nu}{8}+\frac{60341 \nu }{210}-50\right)\nonumber \\
        &+\frac{\Ez \Jz^4}{r^5} \left(-\frac{27 \nu ^3}{4}+\frac{241 \nu ^2}{4}-\frac{71 \nu }{4}\right) +\dfrac{\Jz^2}{r^5} \left(-186 \nu ^3+392 \nu ^2-\frac{41 \pi ^2 \nu }{32}-\frac{36929 \nu }{70}+96\right)\nonumber \\
        &+\frac{\Jz^6}{r^7} \left(\frac{13 \nu^3}{8}-\frac{73 \nu ^2}{8}+\frac{23 \nu }{8}\right)+\frac{\Jz^4}{r^6} \left(-6 \nu ^3+\frac{667 \nu ^2}{8}-\frac{11 \nu}{8}\right)\Biggr]
  \end{align}

  \begin{align}
      v_{\varphi} &= \dfrac{\hat{L}_0}{r} \Biggl\{1 + \dfrac{1}{c^2} \left[\Ez (-1 + 3 \nu) + \dfrac{-4 + 4 \nu}{r}\right] \nonumber \\
                  &+ \dfrac{1}{c^4} \Biggl[\dfrac{\Jz^2}{r^3} \left( -\dfrac{3}{2} \nu - \nu^2 \right) + \Ez^2 \left( 1 - \frac{9 \nu}{2} + 3 \nu^2 \right) + \dfrac{1}{r^2} \left(9 - \dfrac{67}{4} \nu + 17 \nu^2\right) + \dfrac{\Ez}{r} (4 - 24 \nu + 20 \nu^2) \Biggr] \nonumber \\
                  &+ \dfrac{1}{c^6} \Biggl[\dfrac{\Jz^2}{r^4} \left( \frac{11 \nu}{12} - \frac{75 \nu^2}{4} - 15 \nu^3 \right) + \dfrac{\Jz^4}{r^5} \left( -\frac{9 \nu}{8} + 3 \nu^2 + \frac{3 \nu^3}{4} \right) + \Ez^3 \left( -1 + 6 \nu - \frac{17 \nu^2}{2} + \nu^3 \right) \nonumber \\
                  &+ \frac{\Ez^2}{r} \left( -4 + 34 \nu - 76 \nu^2 + 42 \nu^3 \right) + \dfrac{1}{r^3} \left(-16 + \frac{74677 \nu}{840} - \frac{41 \pi^2 \nu}{32} - \frac{157 \nu^2}{2} + 74 \nu^3\right) \nonumber \\
                  &+ \dfrac{\Ez \Jz^2}{r^3} (4 \nu - 12 \nu^2 - 6 \nu^3) + \dfrac{\Ez}{r^2} \left(-9 + \frac{1615 \nu}{12} - \frac{699 \nu^2}{4} + 117 \nu^3 \right)  \Biggr] \Biggr\}
    \end{align}
\end{widetext}

%

\begin{thebibliography}{128}%
\makeatletter
\providecommand \@ifxundefined [1]{%
 \@ifx{#1\undefined}
}%
\providecommand \@ifnum [1]{%
 \ifnum #1\expandafter \@firstoftwo
 \else \expandafter \@secondoftwo
 \fi
}%
\providecommand \@ifx [1]{%
 \ifx #1\expandafter \@firstoftwo
 \else \expandafter \@secondoftwo
 \fi
}%
\providecommand \natexlab [1]{#1}%
\providecommand \enquote  [1]{``#1''}%
\providecommand \bibnamefont  [1]{#1}%
\providecommand \bibfnamefont [1]{#1}%
\providecommand \citenamefont [1]{#1}%
\providecommand \href@noop [0]{\@secondoftwo}%
\providecommand \href [0]{\begingroup \@sanitize@url \@href}%
\providecommand \@href[1]{\@@startlink{#1}\@@href}%
\providecommand \@@href[1]{\endgroup#1\@@endlink}%
\providecommand \@sanitize@url [0]{\catcode `\\12\catcode `\$12\catcode
  `\&12\catcode `\#12\catcode `\^12\catcode `\_12\catcode `\%12\relax}%
\providecommand \@@startlink[1]{}%
\providecommand \@@endlink[0]{}%
\providecommand \url  [0]{\begingroup\@sanitize@url \@url }%
\providecommand \@url [1]{\endgroup\@href {#1}{\urlprefix }}%
\providecommand \urlprefix  [0]{URL }%
\providecommand \Eprint [0]{\href }%
\providecommand \doibase [0]{http://dx.doi.org/}%
\providecommand \selectlanguage [0]{\@gobble}%
\providecommand \bibinfo  [0]{\@secondoftwo}%
\providecommand \bibfield  [0]{\@secondoftwo}%
\providecommand \translation [1]{[#1]}%
\providecommand \BibitemOpen [0]{}%
\providecommand \bibitemStop [0]{}%
\providecommand \bibitemNoStop [0]{.\EOS\space}%
\providecommand \EOS [0]{\spacefactor3000\relax}%
\providecommand \BibitemShut  [1]{\csname bibitem#1\endcsname}%
\let\auto@bib@innerbib\@empty
\bibitem [{\citenamefont {Abbott}\ \emph
  {et~al.}(2017{\natexlab{a}})\citenamefont {Abbott} \emph
  {et~al.}}]{Abbott:2017xzu}%
  \BibitemOpen
  \bibfield  {author} {\bibinfo {author} {\bibfnamefont {B.~P.}\ \bibnamefont
  {Abbott}} \emph {et~al.} (\bibinfo {collaboration} {LIGO Scientific,
  VINROUGE, Las Cumbres Observatory, DLT40, Virgo, 1M2H, MASTER}),\ }\bibfield
  {title} {\enquote {\bibinfo {title} {{A gravitational-wave standard siren
  measurement of the Hubble constant}},}\ }\href {\doibase 10.1038/nature24471}
  {\bibfield  {journal} {\bibinfo  {journal} {Nature}\ } (\bibinfo {year}
  {2017}{\natexlab{a}}),\ 10.1038/nature24471},\ \Eprint
  {http://arxiv.org/abs/1710.05835} {arXiv:1710.05835 [astro-ph.CO]}
  \BibitemShut {NoStop}%
\bibitem [{\citenamefont {Abbott}\ \emph
  {et~al.}(2017{\natexlab{b}})\citenamefont {Abbott} \emph
  {et~al.}}]{TheLIGOScientific:2017qsa}%
  \BibitemOpen
  \bibfield  {author} {\bibinfo {author} {\bibfnamefont {Benjamin~P.}\
  \bibnamefont {Abbott}} \emph {et~al.} (\bibinfo {collaboration} {Virgo, LIGO
  Scientific}),\ }\bibfield  {title} {\enquote {\bibinfo {title} {{GW170817:
  Observation of Gravitational Waves from a Binary Neutron Star Inspiral}},}\
  }\href {\doibase 10.1103/PhysRevLett.119.161101} {\bibfield  {journal}
  {\bibinfo  {journal} {Phys. Rev. Lett.}\ }\textbf {\bibinfo {volume} {119}},\
  \bibinfo {pages} {161101} (\bibinfo {year} {2017}{\natexlab{b}})},\ \Eprint
  {http://arxiv.org/abs/1710.05832} {arXiv:1710.05832 [gr-qc]} \BibitemShut
  {NoStop}%
\bibitem [{\citenamefont {Abbott}\ \emph {et~al.}(2018)\citenamefont {Abbott}
  \emph {et~al.}}]{Abbott:2018exr}%
  \BibitemOpen
  \bibfield  {author} {\bibinfo {author} {\bibfnamefont {B.~P.}\ \bibnamefont
  {Abbott}} \emph {et~al.} (\bibinfo {collaboration} {LIGO Scientific,
  Virgo}),\ }\bibfield  {title} {\enquote {\bibinfo {title} {{GW170817:
  Measurements of neutron star radii and equation of state}},}\ }\href
  {\doibase 10.1103/PhysRevLett.121.161101} {\bibfield  {journal} {\bibinfo
  {journal} {Phys. Rev. Lett.}\ }\textbf {\bibinfo {volume} {121}},\ \bibinfo
  {pages} {161101} (\bibinfo {year} {2018})},\ \Eprint
  {http://arxiv.org/abs/1805.11581} {arXiv:1805.11581 [gr-qc]} \BibitemShut
  {NoStop}%
\bibitem [{\citenamefont {Annala}\ \emph {et~al.}(2018)\citenamefont {Annala},
  \citenamefont {Gorda}, \citenamefont {Kurkela},\ and\ \citenamefont
  {Vuorinen}}]{Annala:2017llu}%
  \BibitemOpen
  \bibfield  {author} {\bibinfo {author} {\bibfnamefont {Eemeli}\ \bibnamefont
  {Annala}}, \bibinfo {author} {\bibfnamefont {Tyler}\ \bibnamefont {Gorda}},
  \bibinfo {author} {\bibfnamefont {Aleksi}\ \bibnamefont {Kurkela}}, \ and\
  \bibinfo {author} {\bibfnamefont {Aleksi}\ \bibnamefont {Vuorinen}},\
  }\bibfield  {title} {\enquote {\bibinfo {title} {{Gravitational-wave
  constraints on the neutron-star-matter Equation of State}},}\ }\href
  {\doibase 10.1103/PhysRevLett.120.172703} {\bibfield  {journal} {\bibinfo
  {journal} {Phys. Rev. Lett.}\ }\textbf {\bibinfo {volume} {120}},\ \bibinfo
  {pages} {172703} (\bibinfo {year} {2018})},\ \Eprint
  {http://arxiv.org/abs/1711.02644} {arXiv:1711.02644 [astro-ph.HE]}
  \BibitemShut {NoStop}%
\bibitem [{\citenamefont {Radice}\ \emph {et~al.}(2018)\citenamefont {Radice},
  \citenamefont {Perego}, \citenamefont {Hotokezaka}, \citenamefont {Fromm},
  \citenamefont {Bernuzzi},\ and\ \citenamefont {Roberts}}]{Radice:2018pdn}%
  \BibitemOpen
  \bibfield  {author} {\bibinfo {author} {\bibfnamefont {David}\ \bibnamefont
  {Radice}}, \bibinfo {author} {\bibfnamefont {Albino}\ \bibnamefont {Perego}},
  \bibinfo {author} {\bibfnamefont {Kenta}\ \bibnamefont {Hotokezaka}},
  \bibinfo {author} {\bibfnamefont {Steven~A.}\ \bibnamefont {Fromm}}, \bibinfo
  {author} {\bibfnamefont {Sebastiano}\ \bibnamefont {Bernuzzi}}, \ and\
  \bibinfo {author} {\bibfnamefont {Luke~F.}\ \bibnamefont {Roberts}},\
  }\bibfield  {title} {\enquote {\bibinfo {title} {{Binary Neutron Star
  Mergers: Mass Ejection, Electromagnetic Counterparts and Nucleosynthesis}},}\
  }\href {\doibase 10.3847/1538-4357/aaf054} {\bibfield  {journal} {\bibinfo
  {journal} {Astrophys. J.}\ }\textbf {\bibinfo {volume} {869}},\ \bibinfo
  {pages} {130} (\bibinfo {year} {2018})},\ \Eprint
  {http://arxiv.org/abs/1809.11161} {arXiv:1809.11161 [astro-ph.HE]}
  \BibitemShut {NoStop}%
\bibitem [{\citenamefont {De}\ \emph {et~al.}(2018)\citenamefont {De},
  \citenamefont {Finstad}, \citenamefont {Lattimer}, \citenamefont {Brown},
  \citenamefont {Berger},\ and\ \citenamefont {Biwer}}]{De:2018uhw}%
  \BibitemOpen
  \bibfield  {author} {\bibinfo {author} {\bibfnamefont {Soumi}\ \bibnamefont
  {De}}, \bibinfo {author} {\bibfnamefont {Daniel}\ \bibnamefont {Finstad}},
  \bibinfo {author} {\bibfnamefont {James~M.}\ \bibnamefont {Lattimer}},
  \bibinfo {author} {\bibfnamefont {Duncan~A.}\ \bibnamefont {Brown}}, \bibinfo
  {author} {\bibfnamefont {Edo}\ \bibnamefont {Berger}}, \ and\ \bibinfo
  {author} {\bibfnamefont {Christopher~M.}\ \bibnamefont {Biwer}},\ }\bibfield
  {title} {\enquote {\bibinfo {title} {{Tidal Deformabilities and Radii of
  Neutron Stars from the Observation of GW170817}},}\ }\href {\doibase
  10.1103/PhysRevLett.121.259902, 10.1103/PhysRevLett.121.091102} {\bibfield
  {journal} {\bibinfo  {journal} {Phys. Rev. Lett.}\ }\textbf {\bibinfo
  {volume} {121}},\ \bibinfo {pages} {091102} (\bibinfo {year} {2018})},\
  \bibinfo {note} {[Erratum: Phys. Rev. Lett.121,no.25,259902(2018)]},\ \Eprint
  {http://arxiv.org/abs/1804.08583} {arXiv:1804.08583 [astro-ph.HE]}
  \BibitemShut {NoStop}%
\bibitem [{\citenamefont {Fattoyev}\ \emph {et~al.}(2018)\citenamefont
  {Fattoyev}, \citenamefont {Piekarewicz},\ and\ \citenamefont
  {Horowitz}}]{Fattoyev:2017jql}%
  \BibitemOpen
  \bibfield  {author} {\bibinfo {author} {\bibfnamefont {F.~J.}\ \bibnamefont
  {Fattoyev}}, \bibinfo {author} {\bibfnamefont {J.}~\bibnamefont
  {Piekarewicz}}, \ and\ \bibinfo {author} {\bibfnamefont {C.~J.}\ \bibnamefont
  {Horowitz}},\ }\bibfield  {title} {\enquote {\bibinfo {title} {{Neutron skins
  and neutron stars in the multi-messenger era}},}\ }\href {\doibase
  10.1103/PhysRevLett.120.172702} {\bibfield  {journal} {\bibinfo  {journal}
  {Phys. Rev. Lett.}\ }\textbf {\bibinfo {volume} {120}},\ \bibinfo {pages}
  {172702} (\bibinfo {year} {2018})},\ \Eprint
  {http://arxiv.org/abs/1711.06615} {arXiv:1711.06615 [nucl-th]} \BibitemShut
  {NoStop}%
\bibitem [{\citenamefont {Most}\ \emph {et~al.}(2018)\citenamefont {Most},
  \citenamefont {Weih}, \citenamefont {Rezzolla},\ and\ \citenamefont
  {Schaffner-Bielich}}]{Most:2018hfd}%
  \BibitemOpen
  \bibfield  {author} {\bibinfo {author} {\bibfnamefont {Elias~R.}\
  \bibnamefont {Most}}, \bibinfo {author} {\bibfnamefont {Lukas~R.}\
  \bibnamefont {Weih}}, \bibinfo {author} {\bibfnamefont {Luciano}\
  \bibnamefont {Rezzolla}}, \ and\ \bibinfo {author} {\bibfnamefont {Jürgen}\
  \bibnamefont {Schaffner-Bielich}},\ }\bibfield  {title} {\enquote {\bibinfo
  {title} {{New constraints on radii and tidal deformabilities of neutron stars
  from GW170817}},}\ }\href {\doibase 10.1103/PhysRevLett.120.261103}
  {\bibfield  {journal} {\bibinfo  {journal} {Phys. Rev. Lett.}\ }\textbf
  {\bibinfo {volume} {120}},\ \bibinfo {pages} {261103} (\bibinfo {year}
  {2018})},\ \Eprint {http://arxiv.org/abs/1803.00549} {arXiv:1803.00549
  [gr-qc]} \BibitemShut {NoStop}%
\bibitem [{\citenamefont {Raithel}\ \emph {et~al.}(2018)\citenamefont
  {Raithel}, \citenamefont {\"Ozel},\ and\ \citenamefont
  {Psaltis}}]{Raithel:2018ncd}%
  \BibitemOpen
  \bibfield  {author} {\bibinfo {author} {\bibfnamefont {Carolyn}\ \bibnamefont
  {Raithel}}, \bibinfo {author} {\bibfnamefont {Feryal}\ \bibnamefont
  {\"Ozel}}, \ and\ \bibinfo {author} {\bibfnamefont {Dimitrios}\ \bibnamefont
  {Psaltis}},\ }\bibfield  {title} {\enquote {\bibinfo {title} {{Tidal
  deformability from GW170817 as a direct probe of the neutron star radius}},}\
  }\href {\doibase 10.3847/2041-8213/aabcbf} {\bibfield  {journal} {\bibinfo
  {journal} {Astrophys. J. Lett.}\ }\textbf {\bibinfo {volume} {857}},\
  \bibinfo {pages} {L23} (\bibinfo {year} {2018})},\ \Eprint
  {http://arxiv.org/abs/1803.07687} {arXiv:1803.07687 [astro-ph.HE]}
  \BibitemShut {NoStop}%
\bibitem [{\citenamefont {Tews}\ \emph {et~al.}(2018)\citenamefont {Tews},
  \citenamefont {Margueron},\ and\ \citenamefont {Reddy}}]{Tews:2018chv}%
  \BibitemOpen
  \bibfield  {author} {\bibinfo {author} {\bibfnamefont {I.}~\bibnamefont
  {Tews}}, \bibinfo {author} {\bibfnamefont {J.}~\bibnamefont {Margueron}}, \
  and\ \bibinfo {author} {\bibfnamefont {S.}~\bibnamefont {Reddy}},\ }\bibfield
   {title} {\enquote {\bibinfo {title} {{Critical examination of constraints on
  the equation of state of dense matter obtained from GW170817}},}\ }\href
  {\doibase 10.1103/PhysRevC.98.045804} {\bibfield  {journal} {\bibinfo
  {journal} {Phys. Rev.}\ }\textbf {\bibinfo {volume} {C98}},\ \bibinfo {pages}
  {045804} (\bibinfo {year} {2018})},\ \Eprint
  {http://arxiv.org/abs/1804.02783} {arXiv:1804.02783 [nucl-th]} \BibitemShut
  {NoStop}%
\bibitem [{\citenamefont {Meidam}\ \emph {et~al.}(2014)\citenamefont {Meidam},
  \citenamefont {Agathos}, \citenamefont {Van Den~Broeck}, \citenamefont
  {Veitch},\ and\ \citenamefont {Sathyaprakash}}]{Meidam:2014jpa}%
  \BibitemOpen
  \bibfield  {author} {\bibinfo {author} {\bibfnamefont {J.}~\bibnamefont
  {Meidam}}, \bibinfo {author} {\bibfnamefont {M.}~\bibnamefont {Agathos}},
  \bibinfo {author} {\bibfnamefont {C.}~\bibnamefont {Van Den~Broeck}},
  \bibinfo {author} {\bibfnamefont {J.}~\bibnamefont {Veitch}}, \ and\ \bibinfo
  {author} {\bibfnamefont {B.~S.}\ \bibnamefont {Sathyaprakash}},\ }\bibfield
  {title} {\enquote {\bibinfo {title} {{Testing the no-hair theorem with black
  hole ringdowns using TIGER}},}\ }\href {\doibase 10.1103/PhysRevD.90.064009}
  {\bibfield  {journal} {\bibinfo  {journal} {Phys. Rev.}\ }\textbf {\bibinfo
  {volume} {D90}},\ \bibinfo {pages} {064009} (\bibinfo {year} {2014})},\
  \Eprint {http://arxiv.org/abs/1406.3201} {arXiv:1406.3201 [gr-qc]}
  \BibitemShut {NoStop}%
\bibitem [{\citenamefont {Abbott}\ \emph {et~al.}(2016)\citenamefont {Abbott}
  \emph {et~al.}}]{TheLIGOScientific:2016src}%
  \BibitemOpen
  \bibfield  {author} {\bibinfo {author} {\bibfnamefont {B.~P.}\ \bibnamefont
  {Abbott}} \emph {et~al.} (\bibinfo {collaboration} {LIGO Scientific,
  Virgo}),\ }\bibfield  {title} {\enquote {\bibinfo {title} {{Tests of general
  relativity with GW150914}},}\ }\href {\doibase
  10.1103/PhysRevLett.116.221101, 10.1103/PhysRevLett.121.129902} {\bibfield
  {journal} {\bibinfo  {journal} {Phys. Rev. Lett.}\ }\textbf {\bibinfo
  {volume} {116}},\ \bibinfo {pages} {221101} (\bibinfo {year} {2016})},\
  \bibinfo {note} {[Erratum: Phys. Rev. Lett.121,no.12,129902(2018)]},\ \Eprint
  {http://arxiv.org/abs/1602.03841} {arXiv:1602.03841 [gr-qc]} \BibitemShut
  {NoStop}%
\bibitem [{\citenamefont {Abbott}\ \emph {et~al.}(2019)\citenamefont {Abbott}
  \emph {et~al.}}]{LIGOScientific:2018dkp}%
  \BibitemOpen
  \bibfield  {author} {\bibinfo {author} {\bibfnamefont {B.~P.}\ \bibnamefont
  {Abbott}} \emph {et~al.} (\bibinfo {collaboration} {LIGO Scientific,
  Virgo}),\ }\bibfield  {title} {\enquote {\bibinfo {title} {{Tests of General
  Relativity with GW170817}},}\ }\href {\doibase
  10.1103/PhysRevLett.123.011102} {\bibfield  {journal} {\bibinfo  {journal}
  {Phys. Rev. Lett.}\ }\textbf {\bibinfo {volume} {123}},\ \bibinfo {pages}
  {011102} (\bibinfo {year} {2019})},\ \Eprint
  {http://arxiv.org/abs/1811.00364} {arXiv:1811.00364 [gr-qc]} \BibitemShut
  {NoStop}%
\bibitem [{\citenamefont {Abbott}\ \emph
  {et~al.}(2021{\natexlab{a}})\citenamefont {Abbott} \emph
  {et~al.}}]{LIGOScientific:2020ibl}%
  \BibitemOpen
  \bibfield  {author} {\bibinfo {author} {\bibfnamefont {R.}~\bibnamefont
  {Abbott}} \emph {et~al.} (\bibinfo {collaboration} {LIGO Scientific,
  Virgo}),\ }\bibfield  {title} {\enquote {\bibinfo {title} {{GWTC-2: Compact
  Binary Coalescences Observed by LIGO and Virgo During the First Half of the
  Third Observing Run}},}\ }\href {\doibase 10.1103/PhysRevX.11.021053}
  {\bibfield  {journal} {\bibinfo  {journal} {Phys. Rev. X}\ }\textbf {\bibinfo
  {volume} {11}},\ \bibinfo {pages} {021053} (\bibinfo {year}
  {2021}{\natexlab{a}})},\ \Eprint {http://arxiv.org/abs/2010.14527}
  {arXiv:2010.14527 [gr-qc]} \BibitemShut {NoStop}%
\bibitem [{\citenamefont {Abbott}\ \emph
  {et~al.}(2020{\natexlab{a}})\citenamefont {Abbott} \emph
  {et~al.}}]{LIGOScientific:2020stg}%
  \BibitemOpen
  \bibfield  {author} {\bibinfo {author} {\bibfnamefont {R.}~\bibnamefont
  {Abbott}} \emph {et~al.} (\bibinfo {collaboration} {LIGO Scientific,
  Virgo}),\ }\bibfield  {title} {\enquote {\bibinfo {title} {{GW190412:
  Observation of a Binary-Black-Hole Coalescence with Asymmetric Masses}},}\
  }\href {\doibase 10.1103/PhysRevD.102.043015} {\bibfield  {journal} {\bibinfo
   {journal} {Phys. Rev. D}\ }\textbf {\bibinfo {volume} {102}},\ \bibinfo
  {pages} {043015} (\bibinfo {year} {2020}{\natexlab{a}})},\ \Eprint
  {http://arxiv.org/abs/2004.08342} {arXiv:2004.08342 [astro-ph.HE]}
  \BibitemShut {NoStop}%
\bibitem [{\citenamefont {Abbott}\ \emph {et~al.}(2023)\citenamefont {Abbott}
  \emph {et~al.}}]{KAGRA:2021vkt}%
  \BibitemOpen
  \bibfield  {author} {\bibinfo {author} {\bibfnamefont {R.}~\bibnamefont
  {Abbott}} \emph {et~al.} (\bibinfo {collaboration} {KAGRA, VIRGO, LIGO
  Scientific}),\ }\bibfield  {title} {\enquote {\bibinfo {title} {{GWTC-3:
  Compact Binary Coalescences Observed by LIGO and Virgo during the Second Part
  of the Third Observing Run}},}\ }\href {\doibase 10.1103/PhysRevX.13.041039}
  {\bibfield  {journal} {\bibinfo  {journal} {Phys. Rev. X}\ }\textbf {\bibinfo
  {volume} {13}},\ \bibinfo {pages} {041039} (\bibinfo {year} {2023})},\
  \Eprint {http://arxiv.org/abs/2111.03606} {arXiv:2111.03606 [gr-qc]}
  \BibitemShut {NoStop}%
\bibitem [{\citenamefont {Varma}\ \emph {et~al.}(2022)\citenamefont {Varma},
  \citenamefont {Biscoveanu}, \citenamefont {Islam}, \citenamefont {Shaik},
  \citenamefont {Haster}, \citenamefont {Isi}, \citenamefont {Farr},
  \citenamefont {Field},\ and\ \citenamefont {Vitale}}]{Varma:2022pld}%
  \BibitemOpen
  \bibfield  {author} {\bibinfo {author} {\bibfnamefont {Vijay}\ \bibnamefont
  {Varma}}, \bibinfo {author} {\bibfnamefont {Sylvia}\ \bibnamefont
  {Biscoveanu}}, \bibinfo {author} {\bibfnamefont {Tousif}\ \bibnamefont
  {Islam}}, \bibinfo {author} {\bibfnamefont {Feroz~H.}\ \bibnamefont {Shaik}},
  \bibinfo {author} {\bibfnamefont {Carl-Johan}\ \bibnamefont {Haster}},
  \bibinfo {author} {\bibfnamefont {Maximiliano}\ \bibnamefont {Isi}}, \bibinfo
  {author} {\bibfnamefont {Will~M.}\ \bibnamefont {Farr}}, \bibinfo {author}
  {\bibfnamefont {Scott~E.}\ \bibnamefont {Field}}, \ and\ \bibinfo {author}
  {\bibfnamefont {Salvatore}\ \bibnamefont {Vitale}},\ }\bibfield  {title}
  {\enquote {\bibinfo {title} {{Evidence of Large Recoil Velocity from a Black
  Hole Merger Signal}},}\ }\href {\doibase 10.1103/PhysRevLett.128.191102}
  {\bibfield  {journal} {\bibinfo  {journal} {Phys. Rev. Lett.}\ }\textbf
  {\bibinfo {volume} {128}},\ \bibinfo {pages} {191102} (\bibinfo {year}
  {2022})},\ \Eprint {http://arxiv.org/abs/2201.01302} {arXiv:2201.01302
  [astro-ph.HE]} \BibitemShut {NoStop}%
\bibitem [{\citenamefont {Abbott}\ \emph
  {et~al.}(2020{\natexlab{b}})\citenamefont {Abbott} \emph
  {et~al.}}]{Abbott:2020tfl}%
  \BibitemOpen
  \bibfield  {author} {\bibinfo {author} {\bibfnamefont {R.}~\bibnamefont
  {Abbott}} \emph {et~al.} (\bibinfo {collaboration} {LIGO Scientific,
  Virgo}),\ }\bibfield  {title} {\enquote {\bibinfo {title} {{GW190521: A
  Binary Black Hole Merger with a Total Mass of 150\,\,M\ensuremath{\odot}}},}\
  }\href {\doibase 10.1103/PhysRevLett.125.101102} {\bibfield  {journal}
  {\bibinfo  {journal} {Phys. Rev. Lett.}\ }\textbf {\bibinfo {volume} {125}},\
  \bibinfo {pages} {101102} (\bibinfo {year} {2020}{\natexlab{b}})},\ \Eprint
  {http://arxiv.org/abs/2009.01075} {arXiv:2009.01075 [gr-qc]} \BibitemShut
  {NoStop}%
\bibitem [{\citenamefont {Abbott}\ \emph
  {et~al.}(2020{\natexlab{c}})\citenamefont {Abbott} \emph
  {et~al.}}]{Abbott:2020mjq}%
  \BibitemOpen
  \bibfield  {author} {\bibinfo {author} {\bibfnamefont {R.}~\bibnamefont
  {Abbott}} \emph {et~al.} (\bibinfo {collaboration} {LIGO Scientific,
  Virgo}),\ }\bibfield  {title} {\enquote {\bibinfo {title} {{Properties and
  astrophysical implications of the 150 Msun binary black hole merger
  GW190521}},}\ }\href {\doibase 10.3847/2041-8213/aba493} {\bibfield
  {journal} {\bibinfo  {journal} {Astrophys. J. Lett.}\ }\textbf {\bibinfo
  {volume} {900}},\ \bibinfo {pages} {L13} (\bibinfo {year}
  {2020}{\natexlab{c}})},\ \Eprint {http://arxiv.org/abs/2009.01190}
  {arXiv:2009.01190 [astro-ph.HE]} \BibitemShut {NoStop}%
\bibitem [{\citenamefont {Abbott}\ \emph
  {et~al.}(2017{\natexlab{c}})\citenamefont {Abbott} \emph
  {et~al.}}]{LIGOScientific:2017ync}%
  \BibitemOpen
  \bibfield  {author} {\bibinfo {author} {\bibfnamefont {B.~P.}\ \bibnamefont
  {Abbott}} \emph {et~al.} (\bibinfo {collaboration} {LIGO Scientific, Virgo,
  Fermi GBM, INTEGRAL, IceCube, AstroSat Cadmium Zinc Telluride Imager Team,
  IPN, Insight-Hxmt, ANTARES, Swift, AGILE Team, 1M2H Team, Dark Energy Camera
  GW-EM, DES, DLT40, GRAWITA, Fermi-LAT, ATCA, ASKAP, Las Cumbres Observatory
  Group, OzGrav, DWF (Deeper Wider Faster Program), AST3, CAASTRO, VINROUGE,
  MASTER, J-GEM, GROWTH, JAGWAR, CaltechNRAO, TTU-NRAO, NuSTAR, Pan-STARRS,
  MAXI Team, TZAC Consortium, KU, Nordic Optical Telescope, ePESSTO, GROND,
  Texas Tech University, SALT Group, TOROS, BOOTES, MWA, CALET, IKI-GW
  Follow-up, H.E.S.S., LOFAR, LWA, HAWC, Pierre Auger, ALMA, Euro VLBI Team, Pi
  of Sky, Chandra Team at McGill University, DFN, ATLAS Telescopes, High Time
  Resolution Universe Survey, RIMAS, RATIR, SKA South Africa/MeerKAT}),\
  }\bibfield  {title} {\enquote {\bibinfo {title} {{Multi-messenger
  Observations of a Binary Neutron Star Merger}},}\ }\href {\doibase
  10.3847/2041-8213/aa91c9} {\bibfield  {journal} {\bibinfo  {journal}
  {Astrophys. J. Lett.}\ }\textbf {\bibinfo {volume} {848}},\ \bibinfo {pages}
  {L12} (\bibinfo {year} {2017}{\natexlab{c}})},\ \Eprint
  {http://arxiv.org/abs/1710.05833} {arXiv:1710.05833 [astro-ph.HE]}
  \BibitemShut {NoStop}%
\bibitem [{\citenamefont {Abbott}\ \emph
  {et~al.}(2020{\natexlab{d}})\citenamefont {Abbott} \emph
  {et~al.}}]{LIGOScientific:2020aai}%
  \BibitemOpen
  \bibfield  {author} {\bibinfo {author} {\bibfnamefont {B.~P.}\ \bibnamefont
  {Abbott}} \emph {et~al.} (\bibinfo {collaboration} {LIGO Scientific,
  Virgo}),\ }\bibfield  {title} {\enquote {\bibinfo {title} {{GW190425:
  Observation of a Compact Binary Coalescence with Total Mass $\sim 3.4
  M_{\odot}$}},}\ }\href {\doibase 10.3847/2041-8213/ab75f5} {\bibfield
  {journal} {\bibinfo  {journal} {Astrophys. J. Lett.}\ }\textbf {\bibinfo
  {volume} {892}},\ \bibinfo {pages} {L3} (\bibinfo {year}
  {2020}{\natexlab{d}})},\ \Eprint {http://arxiv.org/abs/2001.01761}
  {arXiv:2001.01761 [astro-ph.HE]} \BibitemShut {NoStop}%
\bibitem [{\citenamefont {Abbott}\ \emph
  {et~al.}(2021{\natexlab{b}})\citenamefont {Abbott} \emph
  {et~al.}}]{LIGOScientific:2021qlt}%
  \BibitemOpen
  \bibfield  {author} {\bibinfo {author} {\bibfnamefont {R.}~\bibnamefont
  {Abbott}} \emph {et~al.} (\bibinfo {collaboration} {LIGO Scientific, KAGRA,
  VIRGO}),\ }\bibfield  {title} {\enquote {\bibinfo {title} {{Observation of
  Gravitational Waves from Two Neutron Star\textendash{}Black Hole
  Coalescences}},}\ }\href {\doibase 10.3847/2041-8213/ac082e} {\bibfield
  {journal} {\bibinfo  {journal} {Astrophys. J. Lett.}\ }\textbf {\bibinfo
  {volume} {915}},\ \bibinfo {pages} {L5} (\bibinfo {year}
  {2021}{\natexlab{b}})},\ \Eprint {http://arxiv.org/abs/2106.15163}
  {arXiv:2106.15163 [astro-ph.HE]} \BibitemShut {NoStop}%
\bibitem [{\citenamefont {Gayathri}\ \emph {et~al.}(2022)\citenamefont
  {Gayathri}, \citenamefont {Healy}, \citenamefont {Lange}, \citenamefont
  {O'Brien}, \citenamefont {Szczepanczyk}, \citenamefont {Bartos},
  \citenamefont {Campanelli}, \citenamefont {Klimenko}, \citenamefont
  {Lousto},\ and\ \citenamefont {O'Shaughnessy}}]{Gayathri:2020coq}%
  \BibitemOpen
  \bibfield  {author} {\bibinfo {author} {\bibfnamefont {V.}~\bibnamefont
  {Gayathri}}, \bibinfo {author} {\bibfnamefont {J.}~\bibnamefont {Healy}},
  \bibinfo {author} {\bibfnamefont {J.}~\bibnamefont {Lange}}, \bibinfo
  {author} {\bibfnamefont {B.}~\bibnamefont {O'Brien}}, \bibinfo {author}
  {\bibfnamefont {M.}~\bibnamefont {Szczepanczyk}}, \bibinfo {author}
  {\bibfnamefont {Imre}\ \bibnamefont {Bartos}}, \bibinfo {author}
  {\bibfnamefont {M.}~\bibnamefont {Campanelli}}, \bibinfo {author}
  {\bibfnamefont {S.}~\bibnamefont {Klimenko}}, \bibinfo {author}
  {\bibfnamefont {C.~O.}\ \bibnamefont {Lousto}}, \ and\ \bibinfo {author}
  {\bibfnamefont {R.}~\bibnamefont {O'Shaughnessy}},\ }\bibfield  {title}
  {\enquote {\bibinfo {title} {{Eccentricity estimate for black hole mergers
  with numerical relativity simulations}},}\ }\href {\doibase
  10.1038/s41550-021-01568-w} {\bibfield  {journal} {\bibinfo  {journal}
  {Nature Astron.}\ }\textbf {\bibinfo {volume} {6}},\ \bibinfo {pages}
  {344--349} (\bibinfo {year} {2022})},\ \Eprint
  {http://arxiv.org/abs/2009.05461} {arXiv:2009.05461 [astro-ph.HE]}
  \BibitemShut {NoStop}%
\bibitem [{\citenamefont {Romero-Shaw}\ \emph {et~al.}(2020)\citenamefont
  {Romero-Shaw}, \citenamefont {Lasky}, \citenamefont {Thrane},\ and\
  \citenamefont {Bustillo}}]{Romero-Shaw:2020thy}%
  \BibitemOpen
  \bibfield  {author} {\bibinfo {author} {\bibfnamefont {Isobel~M.}\
  \bibnamefont {Romero-Shaw}}, \bibinfo {author} {\bibfnamefont {Paul~D.}\
  \bibnamefont {Lasky}}, \bibinfo {author} {\bibfnamefont {Eric}\ \bibnamefont
  {Thrane}}, \ and\ \bibinfo {author} {\bibfnamefont {Juan~Calderon}\
  \bibnamefont {Bustillo}},\ }\bibfield  {title} {\enquote {\bibinfo {title}
  {{GW190521: orbital eccentricity and signatures of dynamical formation in a
  binary black hole merger signal}},}\ }\href {\doibase
  10.3847/2041-8213/abbe26} {\bibfield  {journal} {\bibinfo  {journal}
  {Astrophys. J. Lett.}\ }\textbf {\bibinfo {volume} {903}},\ \bibinfo {pages}
  {L5} (\bibinfo {year} {2020})},\ \Eprint {http://arxiv.org/abs/2009.04771}
  {arXiv:2009.04771 [astro-ph.HE]} \BibitemShut {NoStop}%
\bibitem [{\citenamefont {Bustillo}\ \emph {et~al.}(2021)\citenamefont
  {Bustillo}, \citenamefont {Sanchis-Gual}, \citenamefont {Torres-Forn\'e},\
  and\ \citenamefont {Font}}]{CalderonBustillo:2020odh}%
  \BibitemOpen
  \bibfield  {author} {\bibinfo {author} {\bibfnamefont {Juan~Calder\'on}\
  \bibnamefont {Bustillo}}, \bibinfo {author} {\bibfnamefont {Nicolas}\
  \bibnamefont {Sanchis-Gual}}, \bibinfo {author} {\bibfnamefont {Alejandro}\
  \bibnamefont {Torres-Forn\'e}}, \ and\ \bibinfo {author} {\bibfnamefont
  {Jos\'e~A.}\ \bibnamefont {Font}},\ }\bibfield  {title} {\enquote {\bibinfo
  {title} {{Confusing Head-On Collisions with Precessing Intermediate-Mass
  Binary Black Hole Mergers}},}\ }\href {\doibase
  10.1103/PhysRevLett.126.201101} {\bibfield  {journal} {\bibinfo  {journal}
  {Phys. Rev. Lett.}\ }\textbf {\bibinfo {volume} {126}},\ \bibinfo {pages}
  {201101} (\bibinfo {year} {2021})},\ \Eprint
  {http://arxiv.org/abs/2009.01066} {arXiv:2009.01066 [gr-qc]} \BibitemShut
  {NoStop}%
\bibitem [{\citenamefont {Shibata}\ \emph {et~al.}(2021)\citenamefont
  {Shibata}, \citenamefont {Kiuchi}, \citenamefont {Fujibayashi},\ and\
  \citenamefont {Sekiguchi}}]{Shibata:2021sau}%
  \BibitemOpen
  \bibfield  {author} {\bibinfo {author} {\bibfnamefont {Masaru}\ \bibnamefont
  {Shibata}}, \bibinfo {author} {\bibfnamefont {Kenta}\ \bibnamefont {Kiuchi}},
  \bibinfo {author} {\bibfnamefont {Sho}\ \bibnamefont {Fujibayashi}}, \ and\
  \bibinfo {author} {\bibfnamefont {Yuichiro}\ \bibnamefont {Sekiguchi}},\
  }\bibfield  {title} {\enquote {\bibinfo {title} {{Alternative possibility of
  GW190521: Gravitational waves from high-mass black hole-disk systems}},}\
  }\href {\doibase 10.1103/PhysRevD.103.063037} {\bibfield  {journal} {\bibinfo
   {journal} {Phys. Rev. D}\ }\textbf {\bibinfo {volume} {103}},\ \bibinfo
  {pages} {063037} (\bibinfo {year} {2021})},\ \Eprint
  {http://arxiv.org/abs/2101.05440} {arXiv:2101.05440 [astro-ph.HE]}
  \BibitemShut {NoStop}%
\bibitem [{\citenamefont {Nitz}\ and\ \citenamefont
  {Capano}(2021)}]{Nitz:2020mga}%
  \BibitemOpen
  \bibfield  {author} {\bibinfo {author} {\bibfnamefont {Alexander~H.}\
  \bibnamefont {Nitz}}\ and\ \bibinfo {author} {\bibfnamefont {Collin~D.}\
  \bibnamefont {Capano}},\ }\bibfield  {title} {\enquote {\bibinfo {title}
  {{GW190521 may be an intermediate mass ratio inspiral}},}\ }\href {\doibase
  10.3847/2041-8213/abccc5} {\bibfield  {journal} {\bibinfo  {journal}
  {Astrophys. J. Lett.}\ }\textbf {\bibinfo {volume} {907}},\ \bibinfo {pages}
  {L9} (\bibinfo {year} {2021})},\ \Eprint {http://arxiv.org/abs/2010.12558}
  {arXiv:2010.12558 [astro-ph.HE]} \BibitemShut {NoStop}%
\bibitem [{\citenamefont {Gamba}\ \emph {et~al.}(2023)\citenamefont {Gamba},
  \citenamefont {Breschi}, \citenamefont {Carullo}, \citenamefont {Albanesi},
  \citenamefont {Rettegno}, \citenamefont {Bernuzzi},\ and\ \citenamefont
  {Nagar}}]{Gamba:2021gap}%
  \BibitemOpen
  \bibfield  {author} {\bibinfo {author} {\bibfnamefont {Rossella}\
  \bibnamefont {Gamba}}, \bibinfo {author} {\bibfnamefont {Matteo}\
  \bibnamefont {Breschi}}, \bibinfo {author} {\bibfnamefont {Gregorio}\
  \bibnamefont {Carullo}}, \bibinfo {author} {\bibfnamefont {Simone}\
  \bibnamefont {Albanesi}}, \bibinfo {author} {\bibfnamefont {Piero}\
  \bibnamefont {Rettegno}}, \bibinfo {author} {\bibfnamefont {Sebastiano}\
  \bibnamefont {Bernuzzi}}, \ and\ \bibinfo {author} {\bibfnamefont
  {Alessandro}\ \bibnamefont {Nagar}},\ }\bibfield  {title} {\enquote {\bibinfo
  {title} {{GW190521 as a dynamical capture of two nonspinning black holes}},}\
  }\href {\doibase 10.1038/s41550-022-01813-w} {\bibfield  {journal} {\bibinfo
  {journal} {Nature Astron.}\ }\textbf {\bibinfo {volume} {7}},\ \bibinfo
  {pages} {11--17} (\bibinfo {year} {2023})},\ \Eprint
  {http://arxiv.org/abs/2106.05575} {arXiv:2106.05575 [gr-qc]} \BibitemShut
  {NoStop}%
\bibitem [{\citenamefont {Romero-Shaw}\ \emph {et~al.}(2023)\citenamefont
  {Romero-Shaw}, \citenamefont {Gerosa},\ and\ \citenamefont
  {Loutrel}}]{Romero-Shaw:2022fbf}%
  \BibitemOpen
  \bibfield  {author} {\bibinfo {author} {\bibfnamefont {Isobel~M.}\
  \bibnamefont {Romero-Shaw}}, \bibinfo {author} {\bibfnamefont {Davide}\
  \bibnamefont {Gerosa}}, \ and\ \bibinfo {author} {\bibfnamefont {Nicholas}\
  \bibnamefont {Loutrel}},\ }\bibfield  {title} {\enquote {\bibinfo {title}
  {{Eccentricity or spin precession? Distinguishing subdominant effects in
  gravitational-wave data}},}\ }\href {\doibase 10.1093/mnras/stad031}
  {\bibfield  {journal} {\bibinfo  {journal} {Mon. Not. Roy. Astron. Soc.}\
  }\textbf {\bibinfo {volume} {519}},\ \bibinfo {pages} {5352--5357} (\bibinfo
  {year} {2023})},\ \Eprint {http://arxiv.org/abs/2211.07528} {arXiv:2211.07528
  [astro-ph.HE]} \BibitemShut {NoStop}%
\bibitem [{\citenamefont {Chandra}\ \emph {et~al.}(2023)\citenamefont
  {Chandra}, \citenamefont {Pai}, \citenamefont {Leong},\ and\ \citenamefont
  {Calder\'on~Bustillo}}]{Chandra:2023nge}%
  \BibitemOpen
  \bibfield  {author} {\bibinfo {author} {\bibfnamefont {Koustav}\ \bibnamefont
  {Chandra}}, \bibinfo {author} {\bibfnamefont {Archana}\ \bibnamefont {Pai}},
  \bibinfo {author} {\bibfnamefont {Samson H.~W.}\ \bibnamefont {Leong}}, \
  and\ \bibinfo {author} {\bibfnamefont {Juan}\ \bibnamefont
  {Calder\'on~Bustillo}},\ }\bibfield  {title} {\enquote {\bibinfo {title}
  {{Impact of Bayesian Priors on the Inferred Masses of Quasi-Circular
  Intermediate-Mass Black Hole Binaries}},}\ }\href@noop {} {\  (\bibinfo
  {year} {2023})},\ \Eprint {http://arxiv.org/abs/2309.01683} {arXiv:2309.01683
  [gr-qc]} \BibitemShut {NoStop}%
\bibitem [{\citenamefont {Apostolatos}\ \emph {et~al.}(1994)\citenamefont
  {Apostolatos}, \citenamefont {Cutler}, \citenamefont {Sussman},\ and\
  \citenamefont {Thorne}}]{Apostolatos:1994mx}%
  \BibitemOpen
  \bibfield  {author} {\bibinfo {author} {\bibfnamefont {Theocharis~A.}\
  \bibnamefont {Apostolatos}}, \bibinfo {author} {\bibfnamefont {Curt}\
  \bibnamefont {Cutler}}, \bibinfo {author} {\bibfnamefont {Gerald~J.}\
  \bibnamefont {Sussman}}, \ and\ \bibinfo {author} {\bibfnamefont {Kip~S.}\
  \bibnamefont {Thorne}},\ }\bibfield  {title} {\enquote {\bibinfo {title}
  {{Spin induced orbital precession and its modulation of the gravitational
  wave forms from merging binaries}},}\ }\href {\doibase
  10.1103/PhysRevD.49.6274} {\bibfield  {journal} {\bibinfo  {journal} {Phys.
  Rev.}\ }\textbf {\bibinfo {volume} {D49}},\ \bibinfo {pages} {6274--6297}
  (\bibinfo {year} {1994})}\BibitemShut {NoStop}%
\bibitem [{\citenamefont {Buonanno}\ \emph {et~al.}(2003)\citenamefont
  {Buonanno}, \citenamefont {Chen},\ and\ \citenamefont
  {Vallisneri}}]{Buonanno:2002fy}%
  \BibitemOpen
  \bibfield  {author} {\bibinfo {author} {\bibfnamefont {Alessandra}\
  \bibnamefont {Buonanno}}, \bibinfo {author} {\bibfnamefont {Yan-bei}\
  \bibnamefont {Chen}}, \ and\ \bibinfo {author} {\bibfnamefont {Michele}\
  \bibnamefont {Vallisneri}},\ }\bibfield  {title} {\enquote {\bibinfo {title}
  {{Detecting gravitational waves from precessing binaries of spinning compact
  objects: Adiabatic limit}},}\ }\href {\doibase 10.1103/PhysRevD.67.104025,
  10.1103/PhysRevD.74.029904} {\bibfield  {journal} {\bibinfo  {journal} {Phys.
  Rev.}\ }\textbf {\bibinfo {volume} {D67}},\ \bibinfo {pages} {104025}
  (\bibinfo {year} {2003})},\ \bibinfo {note} {[Erratum: Phys.
  Rev.D74,029904(2006)]},\ \Eprint {http://arxiv.org/abs/gr-qc/0211087}
  {arXiv:gr-qc/0211087 [gr-qc]} \BibitemShut {NoStop}%
\bibitem [{\citenamefont {Schmidt}\ \emph {et~al.}(2011)\citenamefont
  {Schmidt}, \citenamefont {Hannam}, \citenamefont {Husa},\ and\ \citenamefont
  {Ajith}}]{Schmidt:2010it}%
  \BibitemOpen
  \bibfield  {author} {\bibinfo {author} {\bibfnamefont {Patricia}\
  \bibnamefont {Schmidt}}, \bibinfo {author} {\bibfnamefont {Mark}\
  \bibnamefont {Hannam}}, \bibinfo {author} {\bibfnamefont {Sascha}\
  \bibnamefont {Husa}}, \ and\ \bibinfo {author} {\bibfnamefont
  {P.}~\bibnamefont {Ajith}},\ }\bibfield  {title} {\enquote {\bibinfo {title}
  {{Tracking the precession of compact binaries from their gravitational-wave
  signal}},}\ }\href {\doibase 10.1103/PhysRevD.84.024046} {\bibfield
  {journal} {\bibinfo  {journal} {Phys. Rev.}\ }\textbf {\bibinfo {volume}
  {D84}},\ \bibinfo {pages} {024046} (\bibinfo {year} {2011})},\ \Eprint
  {http://arxiv.org/abs/1012.2879} {arXiv:1012.2879 [gr-qc]} \BibitemShut
  {NoStop}%
\bibitem [{\citenamefont {Schmidt}\ \emph {et~al.}(2012)\citenamefont
  {Schmidt}, \citenamefont {Hannam},\ and\ \citenamefont
  {Husa}}]{Schmidt:2012rh}%
  \BibitemOpen
  \bibfield  {author} {\bibinfo {author} {\bibfnamefont {Patricia}\
  \bibnamefont {Schmidt}}, \bibinfo {author} {\bibfnamefont {Mark}\
  \bibnamefont {Hannam}}, \ and\ \bibinfo {author} {\bibfnamefont {Sascha}\
  \bibnamefont {Husa}},\ }\bibfield  {title} {\enquote {\bibinfo {title}
  {{Towards models of gravitational waveforms from generic binaries: A simple
  approximate mapping between precessing and non-precessing inspiral
  signals}},}\ }\href {\doibase 10.1103/PhysRevD.86.104063} {\bibfield
  {journal} {\bibinfo  {journal} {Phys. Rev.}\ }\textbf {\bibinfo {volume}
  {D86}},\ \bibinfo {pages} {104063} (\bibinfo {year} {2012})},\ \Eprint
  {http://arxiv.org/abs/1207.3088} {arXiv:1207.3088 [gr-qc]} \BibitemShut
  {NoStop}%
\bibitem [{\citenamefont {Boyle}\ \emph {et~al.}(2011)\citenamefont {Boyle},
  \citenamefont {Owen},\ and\ \citenamefont {Pfeiffer}}]{Boyle:2011gg}%
  \BibitemOpen
  \bibfield  {author} {\bibinfo {author} {\bibfnamefont {Michael}\ \bibnamefont
  {Boyle}}, \bibinfo {author} {\bibfnamefont {Robert}\ \bibnamefont {Owen}}, \
  and\ \bibinfo {author} {\bibfnamefont {Harald~P.}\ \bibnamefont {Pfeiffer}},\
  }\bibfield  {title} {\enquote {\bibinfo {title} {{A geometric approach to the
  precession of compact binaries}},}\ }\href {\doibase
  10.1103/PhysRevD.84.124011} {\bibfield  {journal} {\bibinfo  {journal} {Phys.
  Rev.}\ }\textbf {\bibinfo {volume} {D84}},\ \bibinfo {pages} {124011}
  (\bibinfo {year} {2011})},\ \Eprint {http://arxiv.org/abs/1110.2965}
  {arXiv:1110.2965 [gr-qc]} \BibitemShut {NoStop}%
\bibitem [{\citenamefont {O'Shaughnessy}\ \emph {et~al.}(2011)\citenamefont
  {O'Shaughnessy}, \citenamefont {Vaishnav}, \citenamefont {Healy},
  \citenamefont {Meeks},\ and\ \citenamefont
  {Shoemaker}}]{OShaughnessy:2011pmr}%
  \BibitemOpen
  \bibfield  {author} {\bibinfo {author} {\bibfnamefont {R.}~\bibnamefont
  {O'Shaughnessy}}, \bibinfo {author} {\bibfnamefont {B.}~\bibnamefont
  {Vaishnav}}, \bibinfo {author} {\bibfnamefont {J.}~\bibnamefont {Healy}},
  \bibinfo {author} {\bibfnamefont {Z.}~\bibnamefont {Meeks}}, \ and\ \bibinfo
  {author} {\bibfnamefont {D.}~\bibnamefont {Shoemaker}},\ }\bibfield  {title}
  {\enquote {\bibinfo {title} {{Efficient asymptotic frame selection for binary
  black hole spacetimes using asymptotic radiation}},}\ }\href {\doibase
  10.1103/PhysRevD.84.124002} {\bibfield  {journal} {\bibinfo  {journal} {Phys.
  Rev.}\ }\textbf {\bibinfo {volume} {D84}},\ \bibinfo {pages} {124002}
  (\bibinfo {year} {2011})},\ \Eprint {http://arxiv.org/abs/1109.5224}
  {arXiv:1109.5224 [gr-qc]} \BibitemShut {NoStop}%
\bibitem [{\citenamefont {Pan}\ \emph {et~al.}(2014)\citenamefont {Pan},
  \citenamefont {Buonanno}, \citenamefont {Taracchini}, \citenamefont {Kidder},
  \citenamefont {Mroue} \emph {et~al.}}]{Pan:2013rra}%
  \BibitemOpen
  \bibfield  {author} {\bibinfo {author} {\bibfnamefont {Yi}~\bibnamefont
  {Pan}}, \bibinfo {author} {\bibfnamefont {Alessandra}\ \bibnamefont
  {Buonanno}}, \bibinfo {author} {\bibfnamefont {Andrea}\ \bibnamefont
  {Taracchini}}, \bibinfo {author} {\bibfnamefont {Lawrence~E.}\ \bibnamefont
  {Kidder}}, \bibinfo {author} {\bibfnamefont {Abdul~H.}\ \bibnamefont
  {Mroue}},  \emph {et~al.},\ }\bibfield  {title} {\enquote {\bibinfo {title}
  {{Inspiral-merger-ringdown waveforms of spinning, precessing black-hole
  binaries in the effective-one-body formalism}},}\ }\href {\doibase
  10.1103/PhysRevD.89.084006} {\bibfield  {journal} {\bibinfo  {journal}
  {Phys.Rev.}\ }\textbf {\bibinfo {volume} {D89}},\ \bibinfo {pages} {084006}
  (\bibinfo {year} {2014})},\ \Eprint {http://arxiv.org/abs/1307.6232}
  {arXiv:1307.6232 [gr-qc]} \BibitemShut {NoStop}%
\bibitem [{\citenamefont {Thompson}\ \emph {et~al.}(2024)\citenamefont
  {Thompson}, \citenamefont {Hamilton}, \citenamefont {London}, \citenamefont
  {Ghosh}, \citenamefont {Kolitsidou}, \citenamefont {Hoy},\ and\ \citenamefont
  {Hannam}}]{Thompson:2023ase}%
  \BibitemOpen
  \bibfield  {author} {\bibinfo {author} {\bibfnamefont {Jonathan~E.}\
  \bibnamefont {Thompson}}, \bibinfo {author} {\bibfnamefont {Eleanor}\
  \bibnamefont {Hamilton}}, \bibinfo {author} {\bibfnamefont {Lionel}\
  \bibnamefont {London}}, \bibinfo {author} {\bibfnamefont {Shrobana}\
  \bibnamefont {Ghosh}}, \bibinfo {author} {\bibfnamefont {Panagiota}\
  \bibnamefont {Kolitsidou}}, \bibinfo {author} {\bibfnamefont {Charlie}\
  \bibnamefont {Hoy}}, \ and\ \bibinfo {author} {\bibfnamefont {Mark}\
  \bibnamefont {Hannam}},\ }\bibfield  {title} {\enquote {\bibinfo {title}
  {{Phenomenological gravitational-wave model for precessing black-hole
  binaries with higher multipoles and asymmetries}},}\ }\href {\doibase
  10.1103/PhysRevD.109.063012} {\bibfield  {journal} {\bibinfo  {journal}
  {Phys. Rev. D}\ }\textbf {\bibinfo {volume} {109}},\ \bibinfo {pages}
  {063012} (\bibinfo {year} {2024})},\ \Eprint
  {http://arxiv.org/abs/2312.10025} {arXiv:2312.10025 [gr-qc]} \BibitemShut
  {NoStop}%
\bibitem [{\citenamefont {Ghosh}\ \emph {et~al.}(2024)\citenamefont {Ghosh},
  \citenamefont {Kolitsidou},\ and\ \citenamefont {Hannam}}]{Ghosh:2023mhc}%
  \BibitemOpen
  \bibfield  {author} {\bibinfo {author} {\bibfnamefont {Shrobana}\
  \bibnamefont {Ghosh}}, \bibinfo {author} {\bibfnamefont {Panagiota}\
  \bibnamefont {Kolitsidou}}, \ and\ \bibinfo {author} {\bibfnamefont {Mark}\
  \bibnamefont {Hannam}},\ }\bibfield  {title} {\enquote {\bibinfo {title}
  {{First frequency-domain phenomenological model of the multipole asymmetry in
  gravitational-wave signals from binary-black-hole coalescence}},}\ }\href
  {\doibase 10.1103/PhysRevD.109.024061} {\bibfield  {journal} {\bibinfo
  {journal} {Phys. Rev. D}\ }\textbf {\bibinfo {volume} {109}},\ \bibinfo
  {pages} {024061} (\bibinfo {year} {2024})},\ \Eprint
  {http://arxiv.org/abs/2310.16980} {arXiv:2310.16980 [gr-qc]} \BibitemShut
  {NoStop}%
\bibitem [{\citenamefont {Hamilton}\ \emph {et~al.}(2023)\citenamefont
  {Hamilton}, \citenamefont {London},\ and\ \citenamefont
  {Hannam}}]{Hamilton:2023znn}%
  \BibitemOpen
  \bibfield  {author} {\bibinfo {author} {\bibfnamefont {Eleanor}\ \bibnamefont
  {Hamilton}}, \bibinfo {author} {\bibfnamefont {Lionel}\ \bibnamefont
  {London}}, \ and\ \bibinfo {author} {\bibfnamefont {Mark}\ \bibnamefont
  {Hannam}},\ }\bibfield  {title} {\enquote {\bibinfo {title} {{Ringdown
  frequencies in black holes formed from precessing black-hole binaries}},}\
  }\href {\doibase 10.1103/PhysRevD.107.104035} {\bibfield  {journal} {\bibinfo
   {journal} {Phys. Rev. D}\ }\textbf {\bibinfo {volume} {107}},\ \bibinfo
  {pages} {104035} (\bibinfo {year} {2023})},\ \Eprint
  {http://arxiv.org/abs/2301.06558} {arXiv:2301.06558 [gr-qc]} \BibitemShut
  {NoStop}%
\bibitem [{\citenamefont {London}\ \emph {et~al.}(2018)\citenamefont {London},
  \citenamefont {Khan}, \citenamefont {Fauchon-Jones}, \citenamefont {Forteza},
  \citenamefont {Hannam}, \citenamefont {Husa}, \citenamefont {Kalaghatgi},
  \citenamefont {Ohme},\ and\ \citenamefont {Pannarale}}]{London:2017bcn}%
  \BibitemOpen
  \bibfield  {author} {\bibinfo {author} {\bibfnamefont {Lionel}\ \bibnamefont
  {London}}, \bibinfo {author} {\bibfnamefont {Sebastian}\ \bibnamefont
  {Khan}}, \bibinfo {author} {\bibfnamefont {Edward}\ \bibnamefont
  {Fauchon-Jones}}, \bibinfo {author} {\bibfnamefont {Xisco~Jim\'enez}\
  \bibnamefont {Forteza}}, \bibinfo {author} {\bibfnamefont {Mark}\
  \bibnamefont {Hannam}}, \bibinfo {author} {\bibfnamefont {Sascha}\
  \bibnamefont {Husa}}, \bibinfo {author} {\bibfnamefont {Chinmay}\
  \bibnamefont {Kalaghatgi}}, \bibinfo {author} {\bibfnamefont {Frank}\
  \bibnamefont {Ohme}}, \ and\ \bibinfo {author} {\bibfnamefont {Francesco}\
  \bibnamefont {Pannarale}},\ }\bibfield  {title} {\enquote {\bibinfo {title}
  {{First higher-multipole model of gravitational waves from spinning and
  coalescing black-hole binaries}},}\ }\href {\doibase
  10.1103/PhysRevLett.120.161102} {\bibfield  {journal} {\bibinfo  {journal}
  {Phys. Rev. Lett.}\ }\textbf {\bibinfo {volume} {120}},\ \bibinfo {pages}
  {161102} (\bibinfo {year} {2018})},\ \Eprint
  {http://arxiv.org/abs/1708.00404} {arXiv:1708.00404 [gr-qc]} \BibitemShut
  {NoStop}%
\bibitem [{\citenamefont {Garc\'\i{}a-Quir\'os}\ \emph
  {et~al.}(2020)\citenamefont {Garc\'\i{}a-Quir\'os}, \citenamefont {Colleoni},
  \citenamefont {Husa}, \citenamefont {Estell\'es}, \citenamefont {Pratten},
  \citenamefont {Ramos-Buades}, \citenamefont {Mateu-Lucena},\ and\
  \citenamefont {Jaume}}]{Garcia-Quiros:2020qpx}%
  \BibitemOpen
  \bibfield  {author} {\bibinfo {author} {\bibfnamefont {Cecilio}\ \bibnamefont
  {Garc\'\i{}a-Quir\'os}}, \bibinfo {author} {\bibfnamefont {Marta}\
  \bibnamefont {Colleoni}}, \bibinfo {author} {\bibfnamefont {Sascha}\
  \bibnamefont {Husa}}, \bibinfo {author} {\bibfnamefont {H\'ector}\
  \bibnamefont {Estell\'es}}, \bibinfo {author} {\bibfnamefont {Geraint}\
  \bibnamefont {Pratten}}, \bibinfo {author} {\bibfnamefont {Antoni}\
  \bibnamefont {Ramos-Buades}}, \bibinfo {author} {\bibfnamefont {Maite}\
  \bibnamefont {Mateu-Lucena}}, \ and\ \bibinfo {author} {\bibfnamefont
  {Rafel}\ \bibnamefont {Jaume}},\ }\bibfield  {title} {\enquote {\bibinfo
  {title} {{Multimode frequency-domain model for the gravitational wave signal
  from nonprecessing black-hole binaries}},}\ }\href {\doibase
  10.1103/PhysRevD.102.064002} {\bibfield  {journal} {\bibinfo  {journal}
  {Phys. Rev. D}\ }\textbf {\bibinfo {volume} {102}},\ \bibinfo {pages}
  {064002} (\bibinfo {year} {2020})},\ \Eprint
  {http://arxiv.org/abs/2001.10914} {arXiv:2001.10914 [gr-qc]} \BibitemShut
  {NoStop}%
\bibitem [{\citenamefont {Khan}\ \emph {et~al.}(2020)\citenamefont {Khan},
  \citenamefont {Ohme}, \citenamefont {Chatziioannou},\ and\ \citenamefont
  {Hannam}}]{Khan:2019kot}%
  \BibitemOpen
  \bibfield  {author} {\bibinfo {author} {\bibfnamefont {Sebastian}\
  \bibnamefont {Khan}}, \bibinfo {author} {\bibfnamefont {Frank}\ \bibnamefont
  {Ohme}}, \bibinfo {author} {\bibfnamefont {Katerina}\ \bibnamefont
  {Chatziioannou}}, \ and\ \bibinfo {author} {\bibfnamefont {Mark}\
  \bibnamefont {Hannam}},\ }\bibfield  {title} {\enquote {\bibinfo {title}
  {{Including higher order multipoles in gravitational-wave models for
  precessing binary black holes}},}\ }\href {\doibase
  10.1103/PhysRevD.101.024056} {\bibfield  {journal} {\bibinfo  {journal}
  {Phys. Rev. D}\ }\textbf {\bibinfo {volume} {101}},\ \bibinfo {pages}
  {024056} (\bibinfo {year} {2020})},\ \Eprint
  {http://arxiv.org/abs/1911.06050} {arXiv:1911.06050 [gr-qc]} \BibitemShut
  {NoStop}%
\bibitem [{\citenamefont {Hannam}\ \emph {et~al.}(2014)\citenamefont {Hannam},
  \citenamefont {Schmidt}, \citenamefont {Boh\'e}, \citenamefont {Haegel},
  \citenamefont {Husa}, \citenamefont {Ohme}, \citenamefont {Pratten},\ and\
  \citenamefont {P\"urrer}}]{Hannam:2013oca}%
  \BibitemOpen
  \bibfield  {author} {\bibinfo {author} {\bibfnamefont {Mark}\ \bibnamefont
  {Hannam}}, \bibinfo {author} {\bibfnamefont {Patricia}\ \bibnamefont
  {Schmidt}}, \bibinfo {author} {\bibfnamefont {Alejandro}\ \bibnamefont
  {Boh\'e}}, \bibinfo {author} {\bibfnamefont {Le\"\i{}la}\ \bibnamefont
  {Haegel}}, \bibinfo {author} {\bibfnamefont {Sascha}\ \bibnamefont {Husa}},
  \bibinfo {author} {\bibfnamefont {Frank}\ \bibnamefont {Ohme}}, \bibinfo
  {author} {\bibfnamefont {Geraint}\ \bibnamefont {Pratten}}, \ and\ \bibinfo
  {author} {\bibfnamefont {Michael}\ \bibnamefont {P\"urrer}},\ }\bibfield
  {title} {\enquote {\bibinfo {title} {{Simple Model of Complete Precessing
  Black-Hole-Binary Gravitational Waveforms}},}\ }\href {\doibase
  10.1103/PhysRevLett.113.151101} {\bibfield  {journal} {\bibinfo  {journal}
  {Phys. Rev. Lett.}\ }\textbf {\bibinfo {volume} {113}},\ \bibinfo {pages}
  {151101} (\bibinfo {year} {2014})},\ \Eprint {http://arxiv.org/abs/1308.3271}
  {arXiv:1308.3271 [gr-qc]} \BibitemShut {NoStop}%
\bibitem [{\citenamefont {Schmidt}\ \emph {et~al.}(2015)\citenamefont
  {Schmidt}, \citenamefont {Ohme},\ and\ \citenamefont
  {Hannam}}]{Schmidt:2014iyl}%
  \BibitemOpen
  \bibfield  {author} {\bibinfo {author} {\bibfnamefont {Patricia}\
  \bibnamefont {Schmidt}}, \bibinfo {author} {\bibfnamefont {Frank}\
  \bibnamefont {Ohme}}, \ and\ \bibinfo {author} {\bibfnamefont {Mark}\
  \bibnamefont {Hannam}},\ }\bibfield  {title} {\enquote {\bibinfo {title}
  {{Towards models of gravitational waveforms from generic binaries II:
  Modelling precession effects with a single effective precession
  parameter}},}\ }\href {\doibase 10.1103/PhysRevD.91.024043} {\bibfield
  {journal} {\bibinfo  {journal} {Phys. Rev.}\ }\textbf {\bibinfo {volume}
  {D91}},\ \bibinfo {pages} {024043} (\bibinfo {year} {2015})},\ \Eprint
  {http://arxiv.org/abs/1408.1810} {arXiv:1408.1810 [gr-qc]} \BibitemShut
  {NoStop}%
\bibitem [{\citenamefont {Khan}\ \emph {et~al.}(2019)\citenamefont {Khan},
  \citenamefont {Chatziioannou}, \citenamefont {Hannam},\ and\ \citenamefont
  {Ohme}}]{Khan:2018fmp}%
  \BibitemOpen
  \bibfield  {author} {\bibinfo {author} {\bibfnamefont {Sebastian}\
  \bibnamefont {Khan}}, \bibinfo {author} {\bibfnamefont {Katerina}\
  \bibnamefont {Chatziioannou}}, \bibinfo {author} {\bibfnamefont {Mark}\
  \bibnamefont {Hannam}}, \ and\ \bibinfo {author} {\bibfnamefont {Frank}\
  \bibnamefont {Ohme}},\ }\bibfield  {title} {\enquote {\bibinfo {title}
  {{Phenomenological model for the gravitational-wave signal from precessing
  binary black holes with two-spin effects}},}\ }\href {\doibase
  10.1103/PhysRevD.100.024059} {\bibfield  {journal} {\bibinfo  {journal}
  {Phys. Rev.}\ }\textbf {\bibinfo {volume} {D100}},\ \bibinfo {pages} {024059}
  (\bibinfo {year} {2019})},\ \Eprint {http://arxiv.org/abs/1809.10113}
  {arXiv:1809.10113 [gr-qc]} \BibitemShut {NoStop}%
\bibitem [{\citenamefont {Pratten}\ \emph {et~al.}(2021)\citenamefont {Pratten}
  \emph {et~al.}}]{Pratten:2020ceb}%
  \BibitemOpen
  \bibfield  {author} {\bibinfo {author} {\bibfnamefont {Geraint}\ \bibnamefont
  {Pratten}} \emph {et~al.},\ }\bibfield  {title} {\enquote {\bibinfo {title}
  {{Computationally efficient models for the dominant and subdominant harmonic
  modes of precessing binary black holes}},}\ }\href {\doibase
  10.1103/PhysRevD.103.104056} {\bibfield  {journal} {\bibinfo  {journal}
  {Phys. Rev. D}\ }\textbf {\bibinfo {volume} {103}},\ \bibinfo {pages}
  {104056} (\bibinfo {year} {2021})},\ \Eprint
  {http://arxiv.org/abs/2004.06503} {arXiv:2004.06503 [gr-qc]} \BibitemShut
  {NoStop}%
\bibitem [{\citenamefont {Ossokine}\ \emph {et~al.}(2020)\citenamefont
  {Ossokine} \emph {et~al.}}]{Ossokine:2020kjp}%
  \BibitemOpen
  \bibfield  {author} {\bibinfo {author} {\bibfnamefont {Serguei}\ \bibnamefont
  {Ossokine}} \emph {et~al.},\ }\bibfield  {title} {\enquote {\bibinfo {title}
  {{Multipolar Effective-One-Body Waveforms for Precessing Binary Black Holes:
  Construction and Validation}},}\ }\href {\doibase
  10.1103/PhysRevD.102.044055} {\bibfield  {journal} {\bibinfo  {journal}
  {Phys. Rev. D}\ }\textbf {\bibinfo {volume} {102}},\ \bibinfo {pages}
  {044055} (\bibinfo {year} {2020})},\ \Eprint
  {http://arxiv.org/abs/2004.09442} {arXiv:2004.09442 [gr-qc]} \BibitemShut
  {NoStop}%
\bibitem [{\citenamefont {Akcay}\ \emph {et~al.}(2021)\citenamefont {Akcay},
  \citenamefont {Gamba},\ and\ \citenamefont {Bernuzzi}}]{Akcay:2020qrj}%
  \BibitemOpen
  \bibfield  {author} {\bibinfo {author} {\bibfnamefont {Sarp}\ \bibnamefont
  {Akcay}}, \bibinfo {author} {\bibfnamefont {Rossella}\ \bibnamefont {Gamba}},
  \ and\ \bibinfo {author} {\bibfnamefont {Sebastiano}\ \bibnamefont
  {Bernuzzi}},\ }\bibfield  {title} {\enquote {\bibinfo {title} {{A hybrid
  post-Newtonian -- effective-one-body scheme for spin-precessing
  compact-binary waveforms}},}\ }\href {\doibase 10.1103/PhysRevD.103.024014}
  {\bibfield  {journal} {\bibinfo  {journal} {Phys. Rev. D}\ }\textbf {\bibinfo
  {volume} {103}},\ \bibinfo {pages} {024014} (\bibinfo {year} {2021})},\
  \Eprint {http://arxiv.org/abs/2005.05338} {arXiv:2005.05338 [gr-qc]}
  \BibitemShut {NoStop}%
\bibitem [{\citenamefont {Ramos-Buades}\ \emph {et~al.}(2023)\citenamefont
  {Ramos-Buades}, \citenamefont {Buonanno}, \citenamefont {Estell\'es},
  \citenamefont {Khalil}, \citenamefont {Mihaylov}, \citenamefont {Ossokine},
  \citenamefont {Pompili},\ and\ \citenamefont
  {Shiferaw}}]{Ramos-Buades:2023ehm}%
  \BibitemOpen
  \bibfield  {author} {\bibinfo {author} {\bibfnamefont {Antoni}\ \bibnamefont
  {Ramos-Buades}}, \bibinfo {author} {\bibfnamefont {Alessandra}\ \bibnamefont
  {Buonanno}}, \bibinfo {author} {\bibfnamefont {H\'ector}\ \bibnamefont
  {Estell\'es}}, \bibinfo {author} {\bibfnamefont {Mohammed}\ \bibnamefont
  {Khalil}}, \bibinfo {author} {\bibfnamefont {Deyan~P.}\ \bibnamefont
  {Mihaylov}}, \bibinfo {author} {\bibfnamefont {Serguei}\ \bibnamefont
  {Ossokine}}, \bibinfo {author} {\bibfnamefont {Lorenzo}\ \bibnamefont
  {Pompili}}, \ and\ \bibinfo {author} {\bibfnamefont {Mahlet}\ \bibnamefont
  {Shiferaw}},\ }\bibfield  {title} {\enquote {\bibinfo {title} {{Next
  generation of accurate and efficient multipolar precessing-spin
  effective-one-body waveforms for binary black holes}},}\ }\href {\doibase
  10.1103/PhysRevD.108.124037} {\bibfield  {journal} {\bibinfo  {journal}
  {Phys. Rev. D}\ }\textbf {\bibinfo {volume} {108}},\ \bibinfo {pages}
  {124037} (\bibinfo {year} {2023})},\ \Eprint
  {http://arxiv.org/abs/2303.18046} {arXiv:2303.18046 [gr-qc]} \BibitemShut
  {NoStop}%
\bibitem [{\citenamefont {Buonanno}\ and\ \citenamefont
  {Damour}(1999)}]{Buonanno:1998gg}%
  \BibitemOpen
  \bibfield  {author} {\bibinfo {author} {\bibfnamefont {A.}~\bibnamefont
  {Buonanno}}\ and\ \bibinfo {author} {\bibfnamefont {T.}~\bibnamefont
  {Damour}},\ }\bibfield  {title} {\enquote {\bibinfo {title} {{Effective
  one-body approach to general relativistic two-body dynamics}},}\ }\href
  {\doibase 10.1103/PhysRevD.59.084006} {\bibfield  {journal} {\bibinfo
  {journal} {Phys. Rev.}\ }\textbf {\bibinfo {volume} {D59}},\ \bibinfo {pages}
  {084006} (\bibinfo {year} {1999})},\ \Eprint
  {http://arxiv.org/abs/gr-qc/9811091} {arXiv:gr-qc/9811091} \BibitemShut
  {NoStop}%
\bibitem [{\citenamefont {Buonanno}\ and\ \citenamefont
  {Damour}(2000)}]{Buonanno:2000ef}%
  \BibitemOpen
  \bibfield  {author} {\bibinfo {author} {\bibfnamefont {Alessandra}\
  \bibnamefont {Buonanno}}\ and\ \bibinfo {author} {\bibfnamefont {Thibault}\
  \bibnamefont {Damour}},\ }\bibfield  {title} {\enquote {\bibinfo {title}
  {{Transition from inspiral to plunge in binary black hole coalescences}},}\
  }\href {\doibase 10.1103/PhysRevD.62.064015} {\bibfield  {journal} {\bibinfo
  {journal} {Phys. Rev.}\ }\textbf {\bibinfo {volume} {D62}},\ \bibinfo {pages}
  {064015} (\bibinfo {year} {2000})},\ \Eprint
  {http://arxiv.org/abs/gr-qc/0001013} {arXiv:gr-qc/0001013} \BibitemShut
  {NoStop}%
\bibitem [{\citenamefont {Damour}\ \emph {et~al.}(2000)\citenamefont {Damour},
  \citenamefont {Jaranowski},\ and\ \citenamefont {Schaefer}}]{Damour:2000we}%
  \BibitemOpen
  \bibfield  {author} {\bibinfo {author} {\bibfnamefont {Thibault}\
  \bibnamefont {Damour}}, \bibinfo {author} {\bibfnamefont {Piotr}\
  \bibnamefont {Jaranowski}}, \ and\ \bibinfo {author} {\bibfnamefont
  {Gerhard}\ \bibnamefont {Schaefer}},\ }\bibfield  {title} {\enquote {\bibinfo
  {title} {{On the determination of the last stable orbit for circular general
  relativistic binaries at the third postNewtonian approximation}},}\ }\href
  {\doibase 10.1103/PhysRevD.62.084011} {\bibfield  {journal} {\bibinfo
  {journal} {Phys. Rev.}\ }\textbf {\bibinfo {volume} {D62}},\ \bibinfo {pages}
  {084011} (\bibinfo {year} {2000})},\ \Eprint
  {http://arxiv.org/abs/gr-qc/0005034} {arXiv:gr-qc/0005034 [gr-qc]}
  \BibitemShut {NoStop}%
\bibitem [{\citenamefont {Damour}(2001)}]{Damour:2001tu}%
  \BibitemOpen
  \bibfield  {author} {\bibinfo {author} {\bibfnamefont {Thibault}\
  \bibnamefont {Damour}},\ }\bibfield  {title} {\enquote {\bibinfo {title}
  {{Coalescence of two spinning black holes: An effective one- body
  approach}},}\ }\href {\doibase 10.1103/PhysRevD.64.124013} {\bibfield
  {journal} {\bibinfo  {journal} {Phys. Rev.}\ }\textbf {\bibinfo {volume}
  {D64}},\ \bibinfo {pages} {124013} (\bibinfo {year} {2001})},\ \Eprint
  {http://arxiv.org/abs/gr-qc/0103018} {arXiv:gr-qc/0103018} \BibitemShut
  {NoStop}%
\bibitem [{\citenamefont {Damour}\ \emph {et~al.}(2008)\citenamefont {Damour},
  \citenamefont {Jaranowski},\ and\ \citenamefont
  {Sch{\"a}fer}}]{Damour:2008qf}%
  \BibitemOpen
  \bibfield  {author} {\bibinfo {author} {\bibfnamefont {Thibault}\
  \bibnamefont {Damour}}, \bibinfo {author} {\bibfnamefont {Piotr}\
  \bibnamefont {Jaranowski}}, \ and\ \bibinfo {author} {\bibfnamefont
  {Gerhard}\ \bibnamefont {Sch{\"a}fer}},\ }\bibfield  {title} {\enquote
  {\bibinfo {title} {{Effective one body approach to the dynamics of two
  spinning black holes with next-to-leading order spin-orbit coupling}},}\
  }\href {\doibase 10.1103/PhysRevD.78.024009} {\bibfield  {journal} {\bibinfo
  {journal} {Phys.Rev.}\ }\textbf {\bibinfo {volume} {D78}},\ \bibinfo {pages}
  {024009} (\bibinfo {year} {2008})},\ \Eprint {http://arxiv.org/abs/0803.0915}
  {arXiv:0803.0915 [gr-qc]} \BibitemShut {NoStop}%
\bibitem [{\citenamefont {Nagar}(2011)}]{Nagar:2011fx}%
  \BibitemOpen
  \bibfield  {author} {\bibinfo {author} {\bibfnamefont {Alessandro}\
  \bibnamefont {Nagar}},\ }\bibfield  {title} {\enquote {\bibinfo {title}
  {{Effective one body Hamiltonian of two spinning black-holes with
  next-to-next-to-leading order spin-orbit coupling}},}\ }\href {\doibase
  10.1103/PhysRevD.84.084028} {\bibfield  {journal} {\bibinfo  {journal}
  {Phys.Rev.}\ }\textbf {\bibinfo {volume} {D84}},\ \bibinfo {pages} {084028}
  (\bibinfo {year} {2011})},\ \Eprint {http://arxiv.org/abs/1106.4349}
  {arXiv:1106.4349 [gr-qc]} \BibitemShut {NoStop}%
\bibitem [{\citenamefont {Damour}\ \emph {et~al.}(2015)\citenamefont {Damour},
  \citenamefont {Jaranowski},\ and\ \citenamefont
  {Sch\"afer}}]{Damour:2015isa}%
  \BibitemOpen
  \bibfield  {author} {\bibinfo {author} {\bibfnamefont {Thibault}\
  \bibnamefont {Damour}}, \bibinfo {author} {\bibfnamefont {Piotr}\
  \bibnamefont {Jaranowski}}, \ and\ \bibinfo {author} {\bibfnamefont
  {Gerhard}\ \bibnamefont {Sch\"afer}},\ }\bibfield  {title} {\enquote
  {\bibinfo {title} {{Fourth post-Newtonian effective one-body dynamics}},}\
  }\href {\doibase 10.1103/PhysRevD.91.084024} {\bibfield  {journal} {\bibinfo
  {journal} {Phys. Rev. D}\ }\textbf {\bibinfo {volume} {91}},\ \bibinfo
  {pages} {084024} (\bibinfo {year} {2015})},\ \Eprint
  {http://arxiv.org/abs/1502.07245} {arXiv:1502.07245 [gr-qc]} \BibitemShut
  {NoStop}%
\bibitem [{\citenamefont {Hinder}\ \emph {et~al.}(2017)\citenamefont {Hinder},
  \citenamefont {Kidder},\ and\ \citenamefont {Pfeiffer}}]{Hinder:2017sxy}%
  \BibitemOpen
  \bibfield  {author} {\bibinfo {author} {\bibfnamefont {Ian}\ \bibnamefont
  {Hinder}}, \bibinfo {author} {\bibfnamefont {Lawrence~E.}\ \bibnamefont
  {Kidder}}, \ and\ \bibinfo {author} {\bibfnamefont {Harald~P.}\ \bibnamefont
  {Pfeiffer}},\ }\bibfield  {title} {\enquote {\bibinfo {title} {{An eccentric
  binary black hole inspiral-merger-ringdown gravitational waveform model from
  numerical relativity and post-Newtonian theory}},}\ }\href@noop {} {\
  (\bibinfo {year} {2017})},\ \Eprint {http://arxiv.org/abs/1709.02007}
  {arXiv:1709.02007 [gr-qc]} \BibitemShut {NoStop}%
\bibitem [{\citenamefont {Islam}(2024)}]{Islam:2024tcs}%
  \BibitemOpen
  \bibfield  {author} {\bibinfo {author} {\bibfnamefont {Tousif}\ \bibnamefont
  {Islam}},\ }\bibfield  {title} {\enquote {\bibinfo {title} {{Study of
  eccentric binary black hole mergers using numerical relativity and an
  inspiral-merger-ringdown model}},}\ }\href@noop {} {\  (\bibinfo {year}
  {2024})},\ \Eprint {http://arxiv.org/abs/2403.03487} {arXiv:2403.03487
  [gr-qc]} \BibitemShut {NoStop}%
\bibitem [{\citenamefont {Islam}\ \emph {et~al.}(2021)\citenamefont {Islam},
  \citenamefont {Varma}, \citenamefont {Lodman}, \citenamefont {Field},
  \citenamefont {Khanna}, \citenamefont {Scheel}, \citenamefont {Pfeiffer},
  \citenamefont {Gerosa},\ and\ \citenamefont {Kidder}}]{Islam:2021mha}%
  \BibitemOpen
  \bibfield  {author} {\bibinfo {author} {\bibfnamefont {Tousif}\ \bibnamefont
  {Islam}}, \bibinfo {author} {\bibfnamefont {Vijay}\ \bibnamefont {Varma}},
  \bibinfo {author} {\bibfnamefont {Jackie}\ \bibnamefont {Lodman}}, \bibinfo
  {author} {\bibfnamefont {Scott~E.}\ \bibnamefont {Field}}, \bibinfo {author}
  {\bibfnamefont {Gaurav}\ \bibnamefont {Khanna}}, \bibinfo {author}
  {\bibfnamefont {Mark~A.}\ \bibnamefont {Scheel}}, \bibinfo {author}
  {\bibfnamefont {Harald~P.}\ \bibnamefont {Pfeiffer}}, \bibinfo {author}
  {\bibfnamefont {Davide}\ \bibnamefont {Gerosa}}, \ and\ \bibinfo {author}
  {\bibfnamefont {Lawrence~E.}\ \bibnamefont {Kidder}},\ }\bibfield  {title}
  {\enquote {\bibinfo {title} {{Eccentric binary black hole surrogate models
  for the gravitational waveform and remnant properties: comparable mass,
  nonspinning case}},}\ }\href {\doibase 10.1103/PhysRevD.103.064022}
  {\bibfield  {journal} {\bibinfo  {journal} {Phys. Rev. D}\ }\textbf {\bibinfo
  {volume} {103}},\ \bibinfo {pages} {064022} (\bibinfo {year} {2021})},\
  \Eprint {http://arxiv.org/abs/2101.11798} {arXiv:2101.11798 [gr-qc]}
  \BibitemShut {NoStop}%
\bibitem [{\citenamefont {Damour}\ \emph {et~al.}(2014)\citenamefont {Damour},
  \citenamefont {Guercilena}, \citenamefont {Hinder}, \citenamefont {Hopper},
  \citenamefont {Nagar},\ and\ \citenamefont {Rezzolla}}]{Damour:2014afa}%
  \BibitemOpen
  \bibfield  {author} {\bibinfo {author} {\bibfnamefont {Thibault}\
  \bibnamefont {Damour}}, \bibinfo {author} {\bibfnamefont {Federico}\
  \bibnamefont {Guercilena}}, \bibinfo {author} {\bibfnamefont {Ian}\
  \bibnamefont {Hinder}}, \bibinfo {author} {\bibfnamefont {Seth}\ \bibnamefont
  {Hopper}}, \bibinfo {author} {\bibfnamefont {Alessandro}\ \bibnamefont
  {Nagar}}, \ and\ \bibinfo {author} {\bibfnamefont {Luciano}\ \bibnamefont
  {Rezzolla}},\ }\bibfield  {title} {\enquote {\bibinfo {title} {{Strong-Field
  Scattering of Two Black Holes: Numerics Versus Analytics}},}\ }\href
  {\doibase 10.1103/PhysRevD.89.081503} {\bibfield  {journal} {\bibinfo
  {journal} {Phys. Rev. D}\ }\textbf {\bibinfo {volume} {89}},\ \bibinfo
  {pages} {081503} (\bibinfo {year} {2014})},\ \Eprint
  {http://arxiv.org/abs/1402.7307} {arXiv:1402.7307 [gr-qc]} \BibitemShut
  {NoStop}%
\bibitem [{\citenamefont {Hinderer}\ and\ \citenamefont
  {Babak}(2017)}]{Hinderer:2017jcs}%
  \BibitemOpen
  \bibfield  {author} {\bibinfo {author} {\bibfnamefont {Tanja}\ \bibnamefont
  {Hinderer}}\ and\ \bibinfo {author} {\bibfnamefont {Stanislav}\ \bibnamefont
  {Babak}},\ }\bibfield  {title} {\enquote {\bibinfo {title} {{Foundations of
  an effective-one-body model for coalescing binaries on eccentric orbits}},}\
  }\href {\doibase 10.1103/PhysRevD.96.104048} {\bibfield  {journal} {\bibinfo
  {journal} {Phys. Rev.}\ }\textbf {\bibinfo {volume} {D96}},\ \bibinfo {pages}
  {104048} (\bibinfo {year} {2017})},\ \Eprint
  {http://arxiv.org/abs/1707.08426} {arXiv:1707.08426 [gr-qc]} \BibitemShut
  {NoStop}%
\bibitem [{\citenamefont {Cao}\ and\ \citenamefont {Han}(2017)}]{Cao:2017ndf}%
  \BibitemOpen
  \bibfield  {author} {\bibinfo {author} {\bibfnamefont {Zhoujian}\
  \bibnamefont {Cao}}\ and\ \bibinfo {author} {\bibfnamefont {Wen-Biao}\
  \bibnamefont {Han}},\ }\bibfield  {title} {\enquote {\bibinfo {title}
  {{Waveform model for an eccentric binary black hole based on the
  effective-one-body-numerical-relativity formalism}},}\ }\href {\doibase
  10.1103/PhysRevD.96.044028} {\bibfield  {journal} {\bibinfo  {journal} {Phys.
  Rev.}\ }\textbf {\bibinfo {volume} {D96}},\ \bibinfo {pages} {044028}
  (\bibinfo {year} {2017})},\ \Eprint {http://arxiv.org/abs/1708.00166}
  {arXiv:1708.00166 [gr-qc]} \BibitemShut {NoStop}%
\bibitem [{\citenamefont {Liu}\ \emph {et~al.}(2021)\citenamefont {Liu},
  \citenamefont {Cao},\ and\ \citenamefont {Zhu}}]{Liu:2021pkr}%
  \BibitemOpen
  \bibfield  {author} {\bibinfo {author} {\bibfnamefont {Xiaolin}\ \bibnamefont
  {Liu}}, \bibinfo {author} {\bibfnamefont {Zhoujian}\ \bibnamefont {Cao}}, \
  and\ \bibinfo {author} {\bibfnamefont {Zong-Hong}\ \bibnamefont {Zhu}},\
  }\bibfield  {title} {\enquote {\bibinfo {title} {{A higher-multipole
  gravitational waveform model for an eccentric binary black holes based on the
  effective-one-body-numerical-relativity formalism}},}\ }\href@noop {} {\
  (\bibinfo {year} {2021})},\ \Eprint {http://arxiv.org/abs/2102.08614}
  {arXiv:2102.08614 [gr-qc]} \BibitemShut {NoStop}%
\bibitem [{\citenamefont {Liu}\ \emph {et~al.}(2023{\natexlab{a}})\citenamefont
  {Liu}, \citenamefont {Cao},\ and\ \citenamefont {Shao}}]{Liu:2023dgl}%
  \BibitemOpen
  \bibfield  {author} {\bibinfo {author} {\bibfnamefont {Xiaolin}\ \bibnamefont
  {Liu}}, \bibinfo {author} {\bibfnamefont {Zhoujian}\ \bibnamefont {Cao}}, \
  and\ \bibinfo {author} {\bibfnamefont {Lijing}\ \bibnamefont {Shao}},\
  }\bibfield  {title} {\enquote {\bibinfo {title} {{Upgraded waveform model of
  eccentric binary black hole based on effective-one-body-numerical-relativity
  for spin-aligned binary black holes}},}\ }\href {\doibase
  10.1142/S0218271823500153} {\bibfield  {journal} {\bibinfo  {journal} {Int.
  J. Mod. Phys. D}\ }\textbf {\bibinfo {volume} {32}},\ \bibinfo {pages}
  {2350015} (\bibinfo {year} {2023}{\natexlab{a}})},\ \Eprint
  {http://arxiv.org/abs/2306.15277} {arXiv:2306.15277 [gr-qc]} \BibitemShut
  {NoStop}%
\bibitem [{\citenamefont {Chiaramello}\ and\ \citenamefont
  {Nagar}(2020)}]{Chiaramello:2020ehz}%
  \BibitemOpen
  \bibfield  {author} {\bibinfo {author} {\bibfnamefont {Danilo}\ \bibnamefont
  {Chiaramello}}\ and\ \bibinfo {author} {\bibfnamefont {Alessandro}\
  \bibnamefont {Nagar}},\ }\bibfield  {title} {\enquote {\bibinfo {title}
  {{Faithful analytical effective-one-body waveform model for spin-aligned,
  moderately eccentric, coalescing black hole binaries}},}\ }\href {\doibase
  10.1103/PhysRevD.101.101501} {\bibfield  {journal} {\bibinfo  {journal}
  {Phys. Rev. D}\ }\textbf {\bibinfo {volume} {101}},\ \bibinfo {pages}
  {101501} (\bibinfo {year} {2020})},\ \Eprint
  {http://arxiv.org/abs/2001.11736} {arXiv:2001.11736 [gr-qc]} \BibitemShut
  {NoStop}%
\bibitem [{\citenamefont {Nagar}\ \emph {et~al.}(2021)\citenamefont {Nagar},
  \citenamefont {Bonino},\ and\ \citenamefont {Rettegno}}]{Nagar:2021gss}%
  \BibitemOpen
  \bibfield  {author} {\bibinfo {author} {\bibfnamefont {Alessandro}\
  \bibnamefont {Nagar}}, \bibinfo {author} {\bibfnamefont {Alice}\ \bibnamefont
  {Bonino}}, \ and\ \bibinfo {author} {\bibfnamefont {Piero}\ \bibnamefont
  {Rettegno}},\ }\bibfield  {title} {\enquote {\bibinfo {title} {{Effective
  one-body multipolar waveform model for spin-aligned, quasicircular,
  eccentric, hyperbolic black hole binaries}},}\ }\href {\doibase
  10.1103/PhysRevD.103.104021} {\bibfield  {journal} {\bibinfo  {journal}
  {Phys. Rev. D}\ }\textbf {\bibinfo {volume} {103}},\ \bibinfo {pages}
  {104021} (\bibinfo {year} {2021})},\ \Eprint
  {http://arxiv.org/abs/2101.08624} {arXiv:2101.08624 [gr-qc]} \BibitemShut
  {NoStop}%
\bibitem [{\citenamefont {Nagar}\ and\ \citenamefont
  {Rettegno}(2021)}]{Nagar:2021xnh}%
  \BibitemOpen
  \bibfield  {author} {\bibinfo {author} {\bibfnamefont {Alessandro}\
  \bibnamefont {Nagar}}\ and\ \bibinfo {author} {\bibfnamefont {Piero}\
  \bibnamefont {Rettegno}},\ }\bibfield  {title} {\enquote {\bibinfo {title}
  {{Next generation: Impact of high-order analytical information on effective
  one body waveform models for noncircularized, spin-aligned black hole
  binaries}},}\ }\href {\doibase 10.1103/PhysRevD.104.104004} {\bibfield
  {journal} {\bibinfo  {journal} {Phys. Rev. D}\ }\textbf {\bibinfo {volume}
  {104}},\ \bibinfo {pages} {104004} (\bibinfo {year} {2021})},\ \Eprint
  {http://arxiv.org/abs/2108.02043} {arXiv:2108.02043 [gr-qc]} \BibitemShut
  {NoStop}%
\bibitem [{\citenamefont {Nagar}\ \emph {et~al.}(2023)\citenamefont {Nagar},
  \citenamefont {Rettegno}, \citenamefont {Gamba}, \citenamefont {Albanesi},
  \citenamefont {Albertini},\ and\ \citenamefont {Bernuzzi}}]{Nagar:2023zxh}%
  \BibitemOpen
  \bibfield  {author} {\bibinfo {author} {\bibfnamefont {Alessandro}\
  \bibnamefont {Nagar}}, \bibinfo {author} {\bibfnamefont {Piero}\ \bibnamefont
  {Rettegno}}, \bibinfo {author} {\bibfnamefont {Rossella}\ \bibnamefont
  {Gamba}}, \bibinfo {author} {\bibfnamefont {Simone}\ \bibnamefont
  {Albanesi}}, \bibinfo {author} {\bibfnamefont {Angelica}\ \bibnamefont
  {Albertini}}, \ and\ \bibinfo {author} {\bibfnamefont {Sebastiano}\
  \bibnamefont {Bernuzzi}},\ }\bibfield  {title} {\enquote {\bibinfo {title}
  {{Analytic systematics in next generation of effective-one-body gravitational
  waveform models for future observations}},}\ }\href {\doibase
  10.1103/PhysRevD.108.124018} {\bibfield  {journal} {\bibinfo  {journal}
  {Phys. Rev. D}\ }\textbf {\bibinfo {volume} {108}},\ \bibinfo {pages}
  {124018} (\bibinfo {year} {2023})},\ \Eprint
  {http://arxiv.org/abs/2304.09662} {arXiv:2304.09662 [gr-qc]} \BibitemShut
  {NoStop}%
\bibitem [{\citenamefont {Ramos-Buades}\ \emph {et~al.}(2022)\citenamefont
  {Ramos-Buades}, \citenamefont {Buonanno}, \citenamefont {Khalil},\ and\
  \citenamefont {Ossokine}}]{Ramos-Buades:2021adz}%
  \BibitemOpen
  \bibfield  {author} {\bibinfo {author} {\bibfnamefont {Antoni}\ \bibnamefont
  {Ramos-Buades}}, \bibinfo {author} {\bibfnamefont {Alessandra}\ \bibnamefont
  {Buonanno}}, \bibinfo {author} {\bibfnamefont {Mohammed}\ \bibnamefont
  {Khalil}}, \ and\ \bibinfo {author} {\bibfnamefont {Serguei}\ \bibnamefont
  {Ossokine}},\ }\bibfield  {title} {\enquote {\bibinfo {title}
  {{Effective-one-body multipolar waveforms for eccentric binary black holes
  with nonprecessing spins}},}\ }\href {\doibase 10.1103/PhysRevD.105.044035}
  {\bibfield  {journal} {\bibinfo  {journal} {Phys. Rev. D}\ }\textbf {\bibinfo
  {volume} {105}},\ \bibinfo {pages} {044035} (\bibinfo {year} {2022})},\
  \Eprint {http://arxiv.org/abs/2112.06952} {arXiv:2112.06952 [gr-qc]}
  \BibitemShut {NoStop}%
\bibitem [{\citenamefont {Khalil}\ \emph {et~al.}(2021)\citenamefont {Khalil},
  \citenamefont {Buonanno}, \citenamefont {Steinhoff},\ and\ \citenamefont
  {Vines}}]{Khalil:2021txt}%
  \BibitemOpen
  \bibfield  {author} {\bibinfo {author} {\bibfnamefont {Mohammed}\
  \bibnamefont {Khalil}}, \bibinfo {author} {\bibfnamefont {Alessandra}\
  \bibnamefont {Buonanno}}, \bibinfo {author} {\bibfnamefont {Jan}\
  \bibnamefont {Steinhoff}}, \ and\ \bibinfo {author} {\bibfnamefont {Justin}\
  \bibnamefont {Vines}},\ }\bibfield  {title} {\enquote {\bibinfo {title}
  {{Radiation-reaction force and multipolar waveforms for eccentric,
  spin-aligned binaries in the effective-one-body formalism}},}\ }\href
  {\doibase 10.1103/PhysRevD.104.024046} {\bibfield  {journal} {\bibinfo
  {journal} {Phys. Rev. D}\ }\textbf {\bibinfo {volume} {104}},\ \bibinfo
  {pages} {024046} (\bibinfo {year} {2021})},\ \Eprint
  {http://arxiv.org/abs/2104.11705} {arXiv:2104.11705 [gr-qc]} \BibitemShut
  {NoStop}%
\bibitem [{\citenamefont {Placidi}\ \emph {et~al.}(2022)\citenamefont
  {Placidi}, \citenamefont {Albanesi}, \citenamefont {Nagar}, \citenamefont
  {Orselli}, \citenamefont {Bernuzzi},\ and\ \citenamefont
  {Grignani}}]{Placidi:2021rkh}%
  \BibitemOpen
  \bibfield  {author} {\bibinfo {author} {\bibfnamefont {Andrea}\ \bibnamefont
  {Placidi}}, \bibinfo {author} {\bibfnamefont {Simone}\ \bibnamefont
  {Albanesi}}, \bibinfo {author} {\bibfnamefont {Alessandro}\ \bibnamefont
  {Nagar}}, \bibinfo {author} {\bibfnamefont {Marta}\ \bibnamefont {Orselli}},
  \bibinfo {author} {\bibfnamefont {Sebastiano}\ \bibnamefont {Bernuzzi}}, \
  and\ \bibinfo {author} {\bibfnamefont {Gianluca}\ \bibnamefont {Grignani}},\
  }\bibfield  {title} {\enquote {\bibinfo {title} {{Exploiting
  Newton-factorized, 2PN-accurate waveform multipoles in effective-one-body
  models for spin-aligned noncircularized binaries}},}\ }\href {\doibase
  10.1103/PhysRevD.105.104030} {\bibfield  {journal} {\bibinfo  {journal}
  {Phys. Rev. D}\ }\textbf {\bibinfo {volume} {105}},\ \bibinfo {pages}
  {104030} (\bibinfo {year} {2022})},\ \Eprint
  {http://arxiv.org/abs/2112.05448} {arXiv:2112.05448 [gr-qc]} \BibitemShut
  {NoStop}%
\bibitem [{\citenamefont {Albanesi}\ \emph {et~al.}(2022)\citenamefont
  {Albanesi}, \citenamefont {Nagar}, \citenamefont {Bernuzzi}, \citenamefont
  {Placidi},\ and\ \citenamefont {Orselli}}]{Albanesi:2022ywx}%
  \BibitemOpen
  \bibfield  {author} {\bibinfo {author} {\bibfnamefont {Simone}\ \bibnamefont
  {Albanesi}}, \bibinfo {author} {\bibfnamefont {Alessandro}\ \bibnamefont
  {Nagar}}, \bibinfo {author} {\bibfnamefont {Sebastiano}\ \bibnamefont
  {Bernuzzi}}, \bibinfo {author} {\bibfnamefont {Andrea}\ \bibnamefont
  {Placidi}}, \ and\ \bibinfo {author} {\bibfnamefont {Marta}\ \bibnamefont
  {Orselli}},\ }\bibfield  {title} {\enquote {\bibinfo {title} {{Assessment of
  effective-one-body radiation reactions for generic planar orbits}},}\ }\href
  {\doibase 10.1103/PhysRevD.105.104031} {\bibfield  {journal} {\bibinfo
  {journal} {Phys. Rev. D}\ }\textbf {\bibinfo {volume} {105}},\ \bibinfo
  {pages} {104031} (\bibinfo {year} {2022})},\ \Eprint
  {http://arxiv.org/abs/2202.10063} {arXiv:2202.10063 [gr-qc]} \BibitemShut
  {NoStop}%
\bibitem [{\citenamefont {Klein}\ \emph {et~al.}(2018)\citenamefont {Klein},
  \citenamefont {Boetzel}, \citenamefont {Gopakumar}, \citenamefont {Jetzer},\
  and\ \citenamefont {de~Vittori}}]{Klein:2018ybm}%
  \BibitemOpen
  \bibfield  {author} {\bibinfo {author} {\bibfnamefont {Antoine}\ \bibnamefont
  {Klein}}, \bibinfo {author} {\bibfnamefont {Yannick}\ \bibnamefont
  {Boetzel}}, \bibinfo {author} {\bibfnamefont {Achamveedu}\ \bibnamefont
  {Gopakumar}}, \bibinfo {author} {\bibfnamefont {Philippe}\ \bibnamefont
  {Jetzer}}, \ and\ \bibinfo {author} {\bibfnamefont {Lorenzo}\ \bibnamefont
  {de~Vittori}},\ }\bibfield  {title} {\enquote {\bibinfo {title} {{Fourier
  domain gravitational waveforms for precessing eccentric binaries}},}\ }\href
  {\doibase 10.1103/PhysRevD.98.104043} {\bibfield  {journal} {\bibinfo
  {journal} {Phys. Rev.}\ }\textbf {\bibinfo {volume} {D98}},\ \bibinfo {pages}
  {104043} (\bibinfo {year} {2018})},\ \Eprint
  {http://arxiv.org/abs/1801.08542} {arXiv:1801.08542 [gr-qc]} \BibitemShut
  {NoStop}%
\bibitem [{\citenamefont {Phukon}\ \emph {et~al.}(2019)\citenamefont {Phukon},
  \citenamefont {Gupta}, \citenamefont {Bose},\ and\ \citenamefont
  {Jain}}]{Phukon:2019gfh}%
  \BibitemOpen
  \bibfield  {author} {\bibinfo {author} {\bibfnamefont {Khun~Sang}\
  \bibnamefont {Phukon}}, \bibinfo {author} {\bibfnamefont {Anuradha}\
  \bibnamefont {Gupta}}, \bibinfo {author} {\bibfnamefont {Sukanta}\
  \bibnamefont {Bose}}, \ and\ \bibinfo {author} {\bibfnamefont {Pankaj}\
  \bibnamefont {Jain}},\ }\bibfield  {title} {\enquote {\bibinfo {title}
  {{Effect of orbital eccentricity on the dynamics of precessing compact
  binaries}},}\ }\href {\doibase 10.1103/PhysRevD.100.124008} {\bibfield
  {journal} {\bibinfo  {journal} {Phys. Rev. D}\ }\textbf {\bibinfo {volume}
  {100}},\ \bibinfo {pages} {124008} (\bibinfo {year} {2019})},\ \Eprint
  {http://arxiv.org/abs/1904.03985} {arXiv:1904.03985 [gr-qc]} \BibitemShut
  {NoStop}%
\bibitem [{\citenamefont {Klein}(2021)}]{Klein:2021jtd}%
  \BibitemOpen
  \bibfield  {author} {\bibinfo {author} {\bibfnamefont {Antoine}\ \bibnamefont
  {Klein}},\ }\bibfield  {title} {\enquote {\bibinfo {title} {{EFPE: Efficient
  fully precessing eccentric gravitational waveforms for binaries with long
  inspirals}},}\ }\href@noop {} {\  (\bibinfo {year} {2021})},\ \Eprint
  {http://arxiv.org/abs/2106.10291} {arXiv:2106.10291 [gr-qc]} \BibitemShut
  {NoStop}%
\bibitem [{\citenamefont {Ireland}\ \emph {et~al.}(2019)\citenamefont
  {Ireland}, \citenamefont {Birnholtz}, \citenamefont {Nakano}, \citenamefont
  {West},\ and\ \citenamefont {Campanelli}}]{Ireland:2019tao}%
  \BibitemOpen
  \bibfield  {author} {\bibinfo {author} {\bibfnamefont {Brennan}\ \bibnamefont
  {Ireland}}, \bibinfo {author} {\bibfnamefont {Ofek}\ \bibnamefont
  {Birnholtz}}, \bibinfo {author} {\bibfnamefont {Hiroyuki}\ \bibnamefont
  {Nakano}}, \bibinfo {author} {\bibfnamefont {Eric}\ \bibnamefont {West}}, \
  and\ \bibinfo {author} {\bibfnamefont {Manuela}\ \bibnamefont {Campanelli}},\
  }\bibfield  {title} {\enquote {\bibinfo {title} {{Eccentric Binary Black
  Holes with Spin via the Direct Integration of the Post-Newtonian Equations of
  Motion}},}\ }\href {\doibase 10.1103/PhysRevD.100.024015} {\bibfield
  {journal} {\bibinfo  {journal} {Phys. Rev. D}\ }\textbf {\bibinfo {volume}
  {100}},\ \bibinfo {pages} {024015} (\bibinfo {year} {2019})},\ \Eprint
  {http://arxiv.org/abs/1904.03443} {arXiv:1904.03443 [gr-qc]} \BibitemShut
  {NoStop}%
\bibitem [{\citenamefont {Fumagalli}\ and\ \citenamefont
  {Gerosa}(2023)}]{Fumagalli:2023hde}%
  \BibitemOpen
  \bibfield  {author} {\bibinfo {author} {\bibfnamefont {Giulia}\ \bibnamefont
  {Fumagalli}}\ and\ \bibinfo {author} {\bibfnamefont {Davide}\ \bibnamefont
  {Gerosa}},\ }\bibfield  {title} {\enquote {\bibinfo {title}
  {{Spin-eccentricity interplay in merging binary black holes}},}\ }\href
  {\doibase 10.1103/PhysRevD.108.124055} {\bibfield  {journal} {\bibinfo
  {journal} {Phys. Rev. D}\ }\textbf {\bibinfo {volume} {108}},\ \bibinfo
  {pages} {124055} (\bibinfo {year} {2023})},\ \Eprint
  {http://arxiv.org/abs/2310.16893} {arXiv:2310.16893 [gr-qc]} \BibitemShut
  {NoStop}%
\bibitem [{\citenamefont {Arredondo}\ \emph {et~al.}(2024)\citenamefont
  {Arredondo}, \citenamefont {Klein},\ and\ \citenamefont
  {Yunes}}]{Arredondo:2024nsl}%
  \BibitemOpen
  \bibfield  {author} {\bibinfo {author} {\bibfnamefont {J.~Nijaid}\
  \bibnamefont {Arredondo}}, \bibinfo {author} {\bibfnamefont {Antoine}\
  \bibnamefont {Klein}}, \ and\ \bibinfo {author} {\bibfnamefont {Nicol\'as}\
  \bibnamefont {Yunes}},\ }\bibfield  {title} {\enquote {\bibinfo {title}
  {{Efficient Gravitational-Wave Model for Fully-Precessing and
  Moderately-Eccentric, Compact Binary Inspirals}},}\ }\href@noop {} {\
  (\bibinfo {year} {2024})},\ \Eprint {http://arxiv.org/abs/2402.06804}
  {arXiv:2402.06804 [gr-qc]} \BibitemShut {NoStop}%
\bibitem [{\citenamefont {Liu}\ \emph {et~al.}(2023{\natexlab{b}})\citenamefont
  {Liu}, \citenamefont {Cao},\ and\ \citenamefont {Zhu}}]{Liu:2023ldr}%
  \BibitemOpen
  \bibfield  {author} {\bibinfo {author} {\bibfnamefont {Xiaolin}\ \bibnamefont
  {Liu}}, \bibinfo {author} {\bibfnamefont {Zhoujian}\ \bibnamefont {Cao}}, \
  and\ \bibinfo {author} {\bibfnamefont {Zong-Hong}\ \bibnamefont {Zhu}},\
  }\bibfield  {title} {\enquote {\bibinfo {title} {{Effective-One-Body
  Numerical-Relativity waveform model for Eccentric spin-precessing binary
  black hole coalescence}},}\ }\href@noop {} {\  (\bibinfo {year}
  {2023}{\natexlab{b}})},\ \Eprint {http://arxiv.org/abs/2310.04552}
  {arXiv:2310.04552 [gr-qc]} \BibitemShut {NoStop}%
\bibitem [{\citenamefont {Boh{\'e}}\ \emph {et~al.}(2017)\citenamefont
  {Boh{\'e}} \emph {et~al.}}]{Bohe:2016gbl}%
  \BibitemOpen
  \bibfield  {author} {\bibinfo {author} {\bibfnamefont {Alejandro}\
  \bibnamefont {Boh{\'e}}} \emph {et~al.},\ }\bibfield  {title} {\enquote
  {\bibinfo {title} {{Improved effective-one-body model of spinning,
  nonprecessing binary black holes for the era of gravitational-wave
  astrophysics with advanced detectors}},}\ }\href {\doibase
  10.1103/PhysRevD.95.044028} {\bibfield  {journal} {\bibinfo  {journal} {Phys.
  Rev.}\ }\textbf {\bibinfo {volume} {D95}},\ \bibinfo {pages} {044028}
  (\bibinfo {year} {2017})},\ \Eprint {http://arxiv.org/abs/1611.03703}
  {arXiv:1611.03703 [gr-qc]} \BibitemShut {NoStop}%
\bibitem [{\citenamefont {Cotesta}\ \emph {et~al.}(2018)\citenamefont
  {Cotesta}, \citenamefont {Buonanno}, \citenamefont {Boh\'e}, \citenamefont
  {Taracchini}, \citenamefont {Hinder},\ and\ \citenamefont
  {Ossokine}}]{Cotesta:2018fcv}%
  \BibitemOpen
  \bibfield  {author} {\bibinfo {author} {\bibfnamefont {Roberto}\ \bibnamefont
  {Cotesta}}, \bibinfo {author} {\bibfnamefont {Alessandra}\ \bibnamefont
  {Buonanno}}, \bibinfo {author} {\bibfnamefont {Alejandro}\ \bibnamefont
  {Boh\'e}}, \bibinfo {author} {\bibfnamefont {Andrea}\ \bibnamefont
  {Taracchini}}, \bibinfo {author} {\bibfnamefont {Ian}\ \bibnamefont
  {Hinder}}, \ and\ \bibinfo {author} {\bibfnamefont {Serguei}\ \bibnamefont
  {Ossokine}},\ }\bibfield  {title} {\enquote {\bibinfo {title} {{Enriching the
  Symphony of Gravitational Waves from Binary Black Holes by Tuning Higher
  Harmonics}},}\ }\href {\doibase 10.1103/PhysRevD.98.084028} {\bibfield
  {journal} {\bibinfo  {journal} {Phys. Rev.}\ }\textbf {\bibinfo {volume}
  {D98}},\ \bibinfo {pages} {084028} (\bibinfo {year} {2018})},\ \Eprint
  {http://arxiv.org/abs/1803.10701} {arXiv:1803.10701 [gr-qc]} \BibitemShut
  {NoStop}%
\bibitem [{\citenamefont {Healy}\ and\ \citenamefont
  {Lousto}(2022)}]{Healy:2022wdn}%
  \BibitemOpen
  \bibfield  {author} {\bibinfo {author} {\bibfnamefont {James}\ \bibnamefont
  {Healy}}\ and\ \bibinfo {author} {\bibfnamefont {Carlos~O.}\ \bibnamefont
  {Lousto}},\ }\bibfield  {title} {\enquote {\bibinfo {title} {{Fourth RIT
  binary black hole simulations catalog: Extension to eccentric orbits}},}\
  }\href {\doibase 10.1103/PhysRevD.105.124010} {\bibfield  {journal} {\bibinfo
   {journal} {Phys. Rev. D}\ }\textbf {\bibinfo {volume} {105}},\ \bibinfo
  {pages} {124010} (\bibinfo {year} {2022})},\ \Eprint
  {http://arxiv.org/abs/2202.00018} {arXiv:2202.00018 [gr-qc]} \BibitemShut
  {NoStop}%
\bibitem [{\citenamefont {Gamba}\ \emph {et~al.}(2022)\citenamefont {Gamba},
  \citenamefont {Ak\c{c}ay}, \citenamefont {Bernuzzi},\ and\ \citenamefont
  {Williams}}]{Gamba:2021ydi}%
  \BibitemOpen
  \bibfield  {author} {\bibinfo {author} {\bibfnamefont {Rossella}\
  \bibnamefont {Gamba}}, \bibinfo {author} {\bibfnamefont {Sarp}\ \bibnamefont
  {Ak\c{c}ay}}, \bibinfo {author} {\bibfnamefont {Sebastiano}\ \bibnamefont
  {Bernuzzi}}, \ and\ \bibinfo {author} {\bibfnamefont {Jake}\ \bibnamefont
  {Williams}},\ }\bibfield  {title} {\enquote {\bibinfo {title}
  {{Effective-one-body waveforms for precessing coalescing compact binaries
  with post-Newtonian twist}},}\ }\href {\doibase 10.1103/PhysRevD.106.024020}
  {\bibfield  {journal} {\bibinfo  {journal} {Phys. Rev. D}\ }\textbf {\bibinfo
  {volume} {106}},\ \bibinfo {pages} {024020} (\bibinfo {year} {2022})},\
  \Eprint {http://arxiv.org/abs/2111.03675} {arXiv:2111.03675 [gr-qc]}
  \BibitemShut {NoStop}%
\bibitem [{\citenamefont {Bohe}\ \emph {et~al.}(2013)\citenamefont {Bohe},
  \citenamefont {Marsat}, \citenamefont {Faye},\ and\ \citenamefont
  {Blanchet}}]{Bohe:2012mr}%
  \BibitemOpen
  \bibfield  {author} {\bibinfo {author} {\bibfnamefont {Alejandro}\
  \bibnamefont {Bohe}}, \bibinfo {author} {\bibfnamefont {Sylvain}\
  \bibnamefont {Marsat}}, \bibinfo {author} {\bibfnamefont {Guillaume}\
  \bibnamefont {Faye}}, \ and\ \bibinfo {author} {\bibfnamefont {Luc}\
  \bibnamefont {Blanchet}},\ }\bibfield  {title} {\enquote {\bibinfo {title}
  {{Next-to-next-to-leading order spin-orbit effects in the near-zone metric
  and precession equations of compact binaries}},}\ }\href {\doibase
  10.1088/0264-9381/30/7/075017} {\bibfield  {journal} {\bibinfo  {journal}
  {Class. Quant. Grav.}\ }\textbf {\bibinfo {volume} {30}},\ \bibinfo {pages}
  {075017} (\bibinfo {year} {2013})},\ \Eprint {http://arxiv.org/abs/1212.5520}
  {arXiv:1212.5520 [gr-qc]} \BibitemShut {NoStop}%
\bibitem [{\citenamefont {Blanchet}(2014)}]{Blanchet:2013haa}%
  \BibitemOpen
  \bibfield  {author} {\bibinfo {author} {\bibfnamefont {Luc}\ \bibnamefont
  {Blanchet}},\ }\bibfield  {title} {\enquote {\bibinfo {title} {{Gravitational
  Radiation from Post-Newtonian Sources and Inspiralling Compact Binaries}},}\
  }\href {\doibase 10.12942/lrr-2014-2} {\bibfield  {journal} {\bibinfo
  {journal} {Living Rev. Relativity}\ }\textbf {\bibinfo {volume} {17}},\
  \bibinfo {pages} {2} (\bibinfo {year} {2014})},\ \Eprint
  {http://arxiv.org/abs/1310.1528} {arXiv:1310.1528 [gr-qc]} \BibitemShut
  {NoStop}%
\bibitem [{\citenamefont {Tacik}\ \emph {et~al.}(2015)\citenamefont {Tacik}
  \emph {et~al.}}]{Tacik:2015tja}%
  \BibitemOpen
  \bibfield  {author} {\bibinfo {author} {\bibfnamefont {Nick}\ \bibnamefont
  {Tacik}} \emph {et~al.},\ }\bibfield  {title} {\enquote {\bibinfo {title}
  {{Binary Neutron Stars with Arbitrary Spins in Numerical Relativity}},}\
  }\href {\doibase 10.1103/PhysRevD.94.049903, 10.1103/PhysRevD.92.124012}
  {\bibfield  {journal} {\bibinfo  {journal} {Phys. Rev.}\ }\textbf {\bibinfo
  {volume} {D92}},\ \bibinfo {pages} {124012} (\bibinfo {year} {2015})},\
  \bibinfo {note} {[Erratum: Phys. Rev.D94,no.4,049903(2016)]},\ \Eprint
  {http://arxiv.org/abs/1508.06986} {arXiv:1508.06986 [gr-qc]} \BibitemShut
  {NoStop}%
\bibitem [{\citenamefont {Tichy}\ \emph {et~al.}(2019)\citenamefont {Tichy},
  \citenamefont {Rashti}, \citenamefont {Dietrich}, \citenamefont {Dudi},\ and\
  \citenamefont {Brügmann}}]{Tichy:2019ouu}%
  \BibitemOpen
  \bibfield  {author} {\bibinfo {author} {\bibfnamefont {Wolfgang}\
  \bibnamefont {Tichy}}, \bibinfo {author} {\bibfnamefont {Alireza}\
  \bibnamefont {Rashti}}, \bibinfo {author} {\bibfnamefont {Tim}\ \bibnamefont
  {Dietrich}}, \bibinfo {author} {\bibfnamefont {Reetika}\ \bibnamefont
  {Dudi}}, \ and\ \bibinfo {author} {\bibfnamefont {Bernd}\ \bibnamefont
  {Brügmann}},\ }\bibfield  {title} {\enquote {\bibinfo {title} {{Constructing
  Binary Neutron Star Initial Data with High Spins, High Compactness, and High
  Mass-Ratios}},}\ }\href {\doibase 10.1103/PhysRevD.100.124046} {\bibfield
  {journal} {\bibinfo  {journal} {Phys. Rev.}\ }\textbf {\bibinfo {volume}
  {D100}},\ \bibinfo {pages} {124046} (\bibinfo {year} {2019})},\ \Eprint
  {http://arxiv.org/abs/1910.09690} {arXiv:1910.09690 [gr-qc]} \BibitemShut
  {NoStop}%
\bibitem [{\citenamefont {Racine}(2008)}]{Racine:2008qv}%
  \BibitemOpen
  \bibfield  {author} {\bibinfo {author} {\bibfnamefont {Etienne}\ \bibnamefont
  {Racine}},\ }\bibfield  {title} {\enquote {\bibinfo {title} {{Analysis of
  spin precession in binary black hole systems including quadrupole-monopole
  interaction}},}\ }\href {\doibase 10.1103/PhysRevD.78.044021} {\bibfield
  {journal} {\bibinfo  {journal} {Phys. Rev.}\ }\textbf {\bibinfo {volume}
  {D78}},\ \bibinfo {pages} {044021} (\bibinfo {year} {2008})},\ \Eprint
  {http://arxiv.org/abs/0803.1820} {arXiv:0803.1820 [gr-qc]} \BibitemShut
  {NoStop}%
\bibitem [{\citenamefont {Thomas}\ \emph {et~al.}(2021)\citenamefont {Thomas},
  \citenamefont {Schmidt},\ and\ \citenamefont {Pratten}}]{Thomas:2020uqj}%
  \BibitemOpen
  \bibfield  {author} {\bibinfo {author} {\bibfnamefont {Lucy~M.}\ \bibnamefont
  {Thomas}}, \bibinfo {author} {\bibfnamefont {Patricia}\ \bibnamefont
  {Schmidt}}, \ and\ \bibinfo {author} {\bibfnamefont {Geraint}\ \bibnamefont
  {Pratten}},\ }\bibfield  {title} {\enquote {\bibinfo {title} {{New effective
  precession spin for modeling multimodal gravitational waveforms in the
  strong-field regime}},}\ }\href {\doibase 10.1103/PhysRevD.103.083022}
  {\bibfield  {journal} {\bibinfo  {journal} {Phys. Rev. D}\ }\textbf {\bibinfo
  {volume} {103}},\ \bibinfo {pages} {083022} (\bibinfo {year} {2021})},\
  \Eprint {http://arxiv.org/abs/2012.02209} {arXiv:2012.02209 [gr-qc]}
  \BibitemShut {NoStop}%
\bibitem [{\citenamefont {Hopper}\ \emph {et~al.}(2023)\citenamefont {Hopper},
  \citenamefont {Nagar},\ and\ \citenamefont {Rettegno}}]{Hopper:2022rwo}%
  \BibitemOpen
  \bibfield  {author} {\bibinfo {author} {\bibfnamefont {Seth}\ \bibnamefont
  {Hopper}}, \bibinfo {author} {\bibfnamefont {Alessandro}\ \bibnamefont
  {Nagar}}, \ and\ \bibinfo {author} {\bibfnamefont {Piero}\ \bibnamefont
  {Rettegno}},\ }\bibfield  {title} {\enquote {\bibinfo {title} {{Strong-field
  scattering of two spinning black holes: Numerics versus analytics}},}\ }\href
  {\doibase 10.1103/PhysRevD.107.124034} {\bibfield  {journal} {\bibinfo
  {journal} {Phys. Rev. D}\ }\textbf {\bibinfo {volume} {107}},\ \bibinfo
  {pages} {124034} (\bibinfo {year} {2023})},\ \Eprint
  {http://arxiv.org/abs/2204.10299} {arXiv:2204.10299 [gr-qc]} \BibitemShut
  {NoStop}%
\bibitem [{\citenamefont {Damour}\ and\ \citenamefont
  {Rettegno}(2023)}]{Damour:2022ybd}%
  \BibitemOpen
  \bibfield  {author} {\bibinfo {author} {\bibfnamefont {Thibault}\
  \bibnamefont {Damour}}\ and\ \bibinfo {author} {\bibfnamefont {Piero}\
  \bibnamefont {Rettegno}},\ }\bibfield  {title} {\enquote {\bibinfo {title}
  {{Strong-field scattering of two black holes: Numerical relativity meets
  post-Minkowskian gravity}},}\ }\href {\doibase 10.1103/PhysRevD.107.064051}
  {\bibfield  {journal} {\bibinfo  {journal} {Phys. Rev. D}\ }\textbf {\bibinfo
  {volume} {107}},\ \bibinfo {pages} {064051} (\bibinfo {year} {2023})},\
  \Eprint {http://arxiv.org/abs/2211.01399} {arXiv:2211.01399 [gr-qc]}
  \BibitemShut {NoStop}%
\bibitem [{\citenamefont {Rettegno}\ \emph {et~al.}(2023)\citenamefont
  {Rettegno}, \citenamefont {Pratten}, \citenamefont {Thomas}, \citenamefont
  {Schmidt},\ and\ \citenamefont {Damour}}]{Rettegno:2023ghr}%
  \BibitemOpen
  \bibfield  {author} {\bibinfo {author} {\bibfnamefont {Piero}\ \bibnamefont
  {Rettegno}}, \bibinfo {author} {\bibfnamefont {Geraint}\ \bibnamefont
  {Pratten}}, \bibinfo {author} {\bibfnamefont {Lucy~M.}\ \bibnamefont
  {Thomas}}, \bibinfo {author} {\bibfnamefont {Patricia}\ \bibnamefont
  {Schmidt}}, \ and\ \bibinfo {author} {\bibfnamefont {Thibault}\ \bibnamefont
  {Damour}},\ }\bibfield  {title} {\enquote {\bibinfo {title} {{Strong-field
  scattering of two spinning black holes: Numerical relativity versus
  post-Minkowskian gravity}},}\ }\href {\doibase 10.1103/PhysRevD.108.124016}
  {\bibfield  {journal} {\bibinfo  {journal} {Phys. Rev. D}\ }\textbf {\bibinfo
  {volume} {108}},\ \bibinfo {pages} {124016} (\bibinfo {year} {2023})},\
  \Eprint {http://arxiv.org/abs/2307.06999} {arXiv:2307.06999 [gr-qc]}
  \BibitemShut {NoStop}%
\bibitem [{\citenamefont {Bern}\ \emph {et~al.}(2019)\citenamefont {Bern},
  \citenamefont {Cheung}, \citenamefont {Roiban}, \citenamefont {Shen},
  \citenamefont {Solon},\ and\ \citenamefont {Zeng}}]{Bern:2019nnu}%
  \BibitemOpen
  \bibfield  {author} {\bibinfo {author} {\bibfnamefont {Zvi}\ \bibnamefont
  {Bern}}, \bibinfo {author} {\bibfnamefont {Clifford}\ \bibnamefont {Cheung}},
  \bibinfo {author} {\bibfnamefont {Radu}\ \bibnamefont {Roiban}}, \bibinfo
  {author} {\bibfnamefont {Chia-Hsien}\ \bibnamefont {Shen}}, \bibinfo {author}
  {\bibfnamefont {Mikhail~P.}\ \bibnamefont {Solon}}, \ and\ \bibinfo {author}
  {\bibfnamefont {Mao}\ \bibnamefont {Zeng}},\ }\bibfield  {title} {\enquote
  {\bibinfo {title} {{Scattering Amplitudes and the Conservative Hamiltonian
  for Binary Systems at Third Post-Minkowskian Order}},}\ }\href {\doibase
  10.1103/PhysRevLett.122.201603} {\bibfield  {journal} {\bibinfo  {journal}
  {Phys. Rev. Lett.}\ }\textbf {\bibinfo {volume} {122}},\ \bibinfo {pages}
  {201603} (\bibinfo {year} {2019})},\ \Eprint
  {http://arxiv.org/abs/1901.04424} {arXiv:1901.04424 [hep-th]} \BibitemShut
  {NoStop}%
\bibitem [{\citenamefont {K\"alin}\ \emph {et~al.}(2020)\citenamefont
  {K\"alin}, \citenamefont {Liu},\ and\ \citenamefont {Porto}}]{Kalin:2020fhe}%
  \BibitemOpen
  \bibfield  {author} {\bibinfo {author} {\bibfnamefont {Gregor}\ \bibnamefont
  {K\"alin}}, \bibinfo {author} {\bibfnamefont {Zhengwen}\ \bibnamefont {Liu}},
  \ and\ \bibinfo {author} {\bibfnamefont {Rafael~A.}\ \bibnamefont {Porto}},\
  }\bibfield  {title} {\enquote {\bibinfo {title} {{Conservative Dynamics of
  Binary Systems to Third Post-Minkowskian Order from the Effective Field
  Theory Approach}},}\ }\href {\doibase 10.1103/PhysRevLett.125.261103}
  {\bibfield  {journal} {\bibinfo  {journal} {Phys. Rev. Lett.}\ }\textbf
  {\bibinfo {volume} {125}},\ \bibinfo {pages} {261103} (\bibinfo {year}
  {2020})},\ \Eprint {http://arxiv.org/abs/2007.04977} {arXiv:2007.04977
  [hep-th]} \BibitemShut {NoStop}%
\bibitem [{\citenamefont {Bjerrum-Bohr}\ \emph {et~al.}(2021)\citenamefont
  {Bjerrum-Bohr}, \citenamefont {Damgaard}, \citenamefont {Plant\'e},\ and\
  \citenamefont {Vanhove}}]{Bjerrum-Bohr:2021din}%
  \BibitemOpen
  \bibfield  {author} {\bibinfo {author} {\bibfnamefont {N.~Emil~J.}\
  \bibnamefont {Bjerrum-Bohr}}, \bibinfo {author} {\bibfnamefont {Poul~H.}\
  \bibnamefont {Damgaard}}, \bibinfo {author} {\bibfnamefont {Ludovic}\
  \bibnamefont {Plant\'e}}, \ and\ \bibinfo {author} {\bibfnamefont {Pierre}\
  \bibnamefont {Vanhove}},\ }\bibfield  {title} {\enquote {\bibinfo {title}
  {{The amplitude for classical gravitational scattering at third
  Post-Minkowskian order}},}\ }\href {\doibase 10.1007/JHEP08(2021)172}
  {\bibfield  {journal} {\bibinfo  {journal} {JHEP}\ }\textbf {\bibinfo
  {volume} {08}},\ \bibinfo {pages} {172} (\bibinfo {year} {2021})},\ \Eprint
  {http://arxiv.org/abs/2105.05218} {arXiv:2105.05218 [hep-th]} \BibitemShut
  {NoStop}%
\bibitem [{\citenamefont {Bern}\ \emph {et~al.}(2022)\citenamefont {Bern},
  \citenamefont {Parra-Martinez}, \citenamefont {Roiban}, \citenamefont {Ruf},
  \citenamefont {Shen}, \citenamefont {Solon},\ and\ \citenamefont
  {Zeng}}]{Bern:2021yeh}%
  \BibitemOpen
  \bibfield  {author} {\bibinfo {author} {\bibfnamefont {Zvi}\ \bibnamefont
  {Bern}}, \bibinfo {author} {\bibfnamefont {Julio}\ \bibnamefont
  {Parra-Martinez}}, \bibinfo {author} {\bibfnamefont {Radu}\ \bibnamefont
  {Roiban}}, \bibinfo {author} {\bibfnamefont {Michael~S.}\ \bibnamefont
  {Ruf}}, \bibinfo {author} {\bibfnamefont {Chia-Hsien}\ \bibnamefont {Shen}},
  \bibinfo {author} {\bibfnamefont {Mikhail~P.}\ \bibnamefont {Solon}}, \ and\
  \bibinfo {author} {\bibfnamefont {Mao}\ \bibnamefont {Zeng}},\ }\bibfield
  {title} {\enquote {\bibinfo {title} {{Scattering Amplitudes, the Tail Effect,
  and Conservative Binary Dynamics at O(G4)}},}\ }\href {\doibase
  10.1103/PhysRevLett.128.161103} {\bibfield  {journal} {\bibinfo  {journal}
  {Phys. Rev. Lett.}\ }\textbf {\bibinfo {volume} {128}},\ \bibinfo {pages}
  {161103} (\bibinfo {year} {2022})},\ \Eprint
  {http://arxiv.org/abs/2112.10750} {arXiv:2112.10750 [hep-th]} \BibitemShut
  {NoStop}%
\bibitem [{\citenamefont {Dlapa}\ \emph {et~al.}(2022)\citenamefont {Dlapa},
  \citenamefont {K\"alin}, \citenamefont {Liu},\ and\ \citenamefont
  {Porto}}]{Dlapa:2021vgp}%
  \BibitemOpen
  \bibfield  {author} {\bibinfo {author} {\bibfnamefont {Christoph}\
  \bibnamefont {Dlapa}}, \bibinfo {author} {\bibfnamefont {Gregor}\
  \bibnamefont {K\"alin}}, \bibinfo {author} {\bibfnamefont {Zhengwen}\
  \bibnamefont {Liu}}, \ and\ \bibinfo {author} {\bibfnamefont {Rafael~A.}\
  \bibnamefont {Porto}},\ }\bibfield  {title} {\enquote {\bibinfo {title}
  {{Conservative Dynamics of Binary Systems at Fourth Post-Minkowskian Order in
  the Large-Eccentricity Expansion}},}\ }\href {\doibase
  10.1103/PhysRevLett.128.161104} {\bibfield  {journal} {\bibinfo  {journal}
  {Phys. Rev. Lett.}\ }\textbf {\bibinfo {volume} {128}},\ \bibinfo {pages}
  {161104} (\bibinfo {year} {2022})},\ \Eprint
  {http://arxiv.org/abs/2112.11296} {arXiv:2112.11296 [hep-th]} \BibitemShut
  {NoStop}%
\bibitem [{\citenamefont {Damour}(2020)}]{Damour:2020tta}%
  \BibitemOpen
  \bibfield  {author} {\bibinfo {author} {\bibfnamefont {Thibault}\
  \bibnamefont {Damour}},\ }\bibfield  {title} {\enquote {\bibinfo {title}
  {{Radiative contribution to classical gravitational scattering at the third
  order in $G$}},}\ }\href {\doibase 10.1103/PhysRevD.102.124008} {\bibfield
  {journal} {\bibinfo  {journal} {Phys. Rev. D}\ }\textbf {\bibinfo {volume}
  {102}},\ \bibinfo {pages} {124008} (\bibinfo {year} {2020})},\ \Eprint
  {http://arxiv.org/abs/2010.01641} {arXiv:2010.01641 [gr-qc]} \BibitemShut
  {NoStop}%
\bibitem [{\citenamefont {Di~Vecchia}\ \emph
  {et~al.}(2021{\natexlab{a}})\citenamefont {Di~Vecchia}, \citenamefont
  {Heissenberg}, \citenamefont {Russo},\ and\ \citenamefont
  {Veneziano}}]{DiVecchia:2021ndb}%
  \BibitemOpen
  \bibfield  {author} {\bibinfo {author} {\bibfnamefont {Paolo}\ \bibnamefont
  {Di~Vecchia}}, \bibinfo {author} {\bibfnamefont {Carlo}\ \bibnamefont
  {Heissenberg}}, \bibinfo {author} {\bibfnamefont {Rodolfo}\ \bibnamefont
  {Russo}}, \ and\ \bibinfo {author} {\bibfnamefont {Gabriele}\ \bibnamefont
  {Veneziano}},\ }\bibfield  {title} {\enquote {\bibinfo {title} {{Radiation
  Reaction from Soft Theorems}},}\ }\href {\doibase
  10.1016/j.physletb.2021.136379} {\bibfield  {journal} {\bibinfo  {journal}
  {Phys. Lett. B}\ }\textbf {\bibinfo {volume} {818}},\ \bibinfo {pages}
  {136379} (\bibinfo {year} {2021}{\natexlab{a}})},\ \Eprint
  {http://arxiv.org/abs/2101.05772} {arXiv:2101.05772 [hep-th]} \BibitemShut
  {NoStop}%
\bibitem [{\citenamefont {Cho}\ \emph {et~al.}(2022)\citenamefont {Cho},
  \citenamefont {K\"alin},\ and\ \citenamefont {Porto}}]{Cho:2021arx}%
  \BibitemOpen
  \bibfield  {author} {\bibinfo {author} {\bibfnamefont {Gihyuk}\ \bibnamefont
  {Cho}}, \bibinfo {author} {\bibfnamefont {Gregor}\ \bibnamefont {K\"alin}}, \
  and\ \bibinfo {author} {\bibfnamefont {Rafael~A.}\ \bibnamefont {Porto}},\
  }\bibfield  {title} {\enquote {\bibinfo {title} {{From boundary data to bound
  states. Part III. Radiative effects}},}\ }\href {\doibase
  10.1007/JHEP04(2022)154} {\bibfield  {journal} {\bibinfo  {journal} {JHEP}\
  }\textbf {\bibinfo {volume} {04}},\ \bibinfo {pages} {154} (\bibinfo {year}
  {2022})},\ \bibinfo {note} {[Erratum: JHEP 07, 002 (2022)]},\ \Eprint
  {http://arxiv.org/abs/2112.03976} {arXiv:2112.03976 [hep-th]} \BibitemShut
  {NoStop}%
\bibitem [{\citenamefont {Di~Vecchia}\ \emph
  {et~al.}(2021{\natexlab{b}})\citenamefont {Di~Vecchia}, \citenamefont
  {Heissenberg}, \citenamefont {Russo},\ and\ \citenamefont
  {Veneziano}}]{DiVecchia:2021bdo}%
  \BibitemOpen
  \bibfield  {author} {\bibinfo {author} {\bibfnamefont {Paolo}\ \bibnamefont
  {Di~Vecchia}}, \bibinfo {author} {\bibfnamefont {Carlo}\ \bibnamefont
  {Heissenberg}}, \bibinfo {author} {\bibfnamefont {Rodolfo}\ \bibnamefont
  {Russo}}, \ and\ \bibinfo {author} {\bibfnamefont {Gabriele}\ \bibnamefont
  {Veneziano}},\ }\bibfield  {title} {\enquote {\bibinfo {title} {{The eikonal
  approach to gravitational scattering and radiation at $ \mathcal{O}
  $(G$^{3}$)}},}\ }\href {\doibase 10.1007/JHEP07(2021)169} {\bibfield
  {journal} {\bibinfo  {journal} {JHEP}\ }\textbf {\bibinfo {volume} {07}},\
  \bibinfo {pages} {169} (\bibinfo {year} {2021}{\natexlab{b}})},\ \Eprint
  {http://arxiv.org/abs/2104.03256} {arXiv:2104.03256 [hep-th]} \BibitemShut
  {NoStop}%
\bibitem [{\citenamefont {Herrmann}\ \emph {et~al.}(2021)\citenamefont
  {Herrmann}, \citenamefont {Parra-Martinez}, \citenamefont {Ruf},\ and\
  \citenamefont {Zeng}}]{Herrmann:2021tct}%
  \BibitemOpen
  \bibfield  {author} {\bibinfo {author} {\bibfnamefont {Enrico}\ \bibnamefont
  {Herrmann}}, \bibinfo {author} {\bibfnamefont {Julio}\ \bibnamefont
  {Parra-Martinez}}, \bibinfo {author} {\bibfnamefont {Michael~S.}\
  \bibnamefont {Ruf}}, \ and\ \bibinfo {author} {\bibfnamefont {Mao}\
  \bibnamefont {Zeng}},\ }\bibfield  {title} {\enquote {\bibinfo {title}
  {{Radiative classical gravitational observables at $ \mathcal{O} $(G$^{3}$)
  from scattering amplitudes}},}\ }\href {\doibase 10.1007/JHEP10(2021)148}
  {\bibfield  {journal} {\bibinfo  {journal} {JHEP}\ }\textbf {\bibinfo
  {volume} {10}},\ \bibinfo {pages} {148} (\bibinfo {year} {2021})},\ \Eprint
  {http://arxiv.org/abs/2104.03957} {arXiv:2104.03957 [hep-th]} \BibitemShut
  {NoStop}%
\bibitem [{\citenamefont {Bini}\ \emph {et~al.}(2021)\citenamefont {Bini},
  \citenamefont {Damour},\ and\ \citenamefont {Geralico}}]{Bini:2021gat}%
  \BibitemOpen
  \bibfield  {author} {\bibinfo {author} {\bibfnamefont {Donato}\ \bibnamefont
  {Bini}}, \bibinfo {author} {\bibfnamefont {Thibault}\ \bibnamefont {Damour}},
  \ and\ \bibinfo {author} {\bibfnamefont {Andrea}\ \bibnamefont {Geralico}},\
  }\bibfield  {title} {\enquote {\bibinfo {title} {{Radiative contributions to
  gravitational scattering}},}\ }\href {\doibase 10.1103/PhysRevD.104.084031}
  {\bibfield  {journal} {\bibinfo  {journal} {Phys. Rev. D}\ }\textbf {\bibinfo
  {volume} {104}},\ \bibinfo {pages} {084031} (\bibinfo {year} {2021})},\
  \Eprint {http://arxiv.org/abs/2107.08896} {arXiv:2107.08896 [gr-qc]}
  \BibitemShut {NoStop}%
\bibitem [{\citenamefont {Bini}\ and\ \citenamefont
  {Geralico}(2021)}]{Bini:2021qvf}%
  \BibitemOpen
  \bibfield  {author} {\bibinfo {author} {\bibfnamefont {Donato}\ \bibnamefont
  {Bini}}\ and\ \bibinfo {author} {\bibfnamefont {Andrea}\ \bibnamefont
  {Geralico}},\ }\bibfield  {title} {\enquote {\bibinfo {title} {{Higher-order
  tail contributions to the energy and angular momentum fluxes in a two-body
  scattering process}},}\ }\href {\doibase 10.1103/PhysRevD.104.104020}
  {\bibfield  {journal} {\bibinfo  {journal} {Phys. Rev. D}\ }\textbf {\bibinfo
  {volume} {104}},\ \bibinfo {pages} {104020} (\bibinfo {year} {2021})},\
  \Eprint {http://arxiv.org/abs/2108.05445} {arXiv:2108.05445 [gr-qc]}
  \BibitemShut {NoStop}%
\bibitem [{\citenamefont {Manohar}\ \emph {et~al.}(2022)\citenamefont
  {Manohar}, \citenamefont {Ridgway},\ and\ \citenamefont
  {Shen}}]{Manohar:2022dea}%
  \BibitemOpen
  \bibfield  {author} {\bibinfo {author} {\bibfnamefont {Aneesh~V.}\
  \bibnamefont {Manohar}}, \bibinfo {author} {\bibfnamefont {Alexander~K.}\
  \bibnamefont {Ridgway}}, \ and\ \bibinfo {author} {\bibfnamefont
  {Chia-Hsien}\ \bibnamefont {Shen}},\ }\bibfield  {title} {\enquote {\bibinfo
  {title} {{Radiated Angular Momentum and Dissipative Effects in Classical
  Scattering}},}\ }\href {\doibase 10.1103/PhysRevLett.129.121601} {\bibfield
  {journal} {\bibinfo  {journal} {Phys. Rev. Lett.}\ }\textbf {\bibinfo
  {volume} {129}},\ \bibinfo {pages} {121601} (\bibinfo {year} {2022})},\
  \Eprint {http://arxiv.org/abs/2203.04283} {arXiv:2203.04283 [hep-th]}
  \BibitemShut {NoStop}%
\bibitem [{\citenamefont {Dlapa}\ \emph {et~al.}(2023)\citenamefont {Dlapa},
  \citenamefont {K\"alin}, \citenamefont {Liu}, \citenamefont {Neef},\ and\
  \citenamefont {Porto}}]{Dlapa:2022lmu}%
  \BibitemOpen
  \bibfield  {author} {\bibinfo {author} {\bibfnamefont {Christoph}\
  \bibnamefont {Dlapa}}, \bibinfo {author} {\bibfnamefont {Gregor}\
  \bibnamefont {K\"alin}}, \bibinfo {author} {\bibfnamefont {Zhengwen}\
  \bibnamefont {Liu}}, \bibinfo {author} {\bibfnamefont {Jakob}\ \bibnamefont
  {Neef}}, \ and\ \bibinfo {author} {\bibfnamefont {Rafael~A.}\ \bibnamefont
  {Porto}},\ }\bibfield  {title} {\enquote {\bibinfo {title} {{Radiation
  Reaction and Gravitational Waves at Fourth Post-Minkowskian Order}},}\ }\href
  {\doibase 10.1103/PhysRevLett.130.101401} {\bibfield  {journal} {\bibinfo
  {journal} {Phys. Rev. Lett.}\ }\textbf {\bibinfo {volume} {130}},\ \bibinfo
  {pages} {101401} (\bibinfo {year} {2023})},\ \Eprint
  {http://arxiv.org/abs/2210.05541} {arXiv:2210.05541 [hep-th]} \BibitemShut
  {NoStop}%
\bibitem [{\citenamefont {Ferguson}\ \emph
  {et~al.}(2023{\natexlab{a}})\citenamefont {Ferguson} \emph
  {et~al.}}]{Ferguson:2023vta}%
  \BibitemOpen
  \bibfield  {author} {\bibinfo {author} {\bibfnamefont {Deborah}\ \bibnamefont
  {Ferguson}} \emph {et~al.},\ }\bibfield  {title} {\enquote {\bibinfo {title}
  {{Second MAYA Catalog of Binary Black Hole Numerical Relativity
  Waveforms}},}\ }\href@noop {} {\  (\bibinfo {year} {2023}{\natexlab{a}})},\
  \Eprint {http://arxiv.org/abs/2309.00262} {arXiv:2309.00262 [gr-qc]}
  \BibitemShut {NoStop}%
\bibitem [{\citenamefont {Ferguson}\ \emph
  {et~al.}(2023{\natexlab{b}})\citenamefont {Ferguson} \emph
  {et~al.}}]{Ferguson:2023mks}%
  \BibitemOpen
  \bibfield  {author} {\bibinfo {author} {\bibfnamefont {Deborah}\ \bibnamefont
  {Ferguson}} \emph {et~al.},\ }\bibfield  {title} {\enquote {\bibinfo {title}
  {{Mayawaves: Python Library for Interacting with the Einstein Toolkit and the
  MAYA Catalog}},}\ }\href@noop {} {\  (\bibinfo {year}
  {2023}{\natexlab{b}})},\ \Eprint {http://arxiv.org/abs/2309.00653}
  {arXiv:2309.00653 [astro-ph.IM]} \BibitemShut {NoStop}%
\bibitem [{\citenamefont {Carullo}\ \emph {et~al.}(2024)\citenamefont
  {Carullo}, \citenamefont {Albanesi}, \citenamefont {Nagar}, \citenamefont
  {Gamba}, \citenamefont {Bernuzzi}, \citenamefont {Andrade},\ and\
  \citenamefont {Trenado}}]{Carullo:2023kvj}%
  \BibitemOpen
  \bibfield  {author} {\bibinfo {author} {\bibfnamefont {Gregorio}\
  \bibnamefont {Carullo}}, \bibinfo {author} {\bibfnamefont {Simone}\
  \bibnamefont {Albanesi}}, \bibinfo {author} {\bibfnamefont {Alessandro}\
  \bibnamefont {Nagar}}, \bibinfo {author} {\bibfnamefont {Rossella}\
  \bibnamefont {Gamba}}, \bibinfo {author} {\bibfnamefont {Sebastiano}\
  \bibnamefont {Bernuzzi}}, \bibinfo {author} {\bibfnamefont {Tomas}\
  \bibnamefont {Andrade}}, \ and\ \bibinfo {author} {\bibfnamefont {Juan}\
  \bibnamefont {Trenado}},\ }\bibfield  {title} {\enquote {\bibinfo {title}
  {{Unveiling the Merger Structure of Black Hole Binaries in Generic Planar
  Orbits}},}\ }\href {\doibase 10.1103/PhysRevLett.132.101401} {\bibfield
  {journal} {\bibinfo  {journal} {Phys. Rev. Lett.}\ }\textbf {\bibinfo
  {volume} {132}},\ \bibinfo {pages} {101401} (\bibinfo {year} {2024})},\
  \Eprint {http://arxiv.org/abs/2309.07228} {arXiv:2309.07228 [gr-qc]}
  \BibitemShut {NoStop}%
\bibitem [{\citenamefont {Pekowsky}\ \emph {et~al.}(2013)\citenamefont
  {Pekowsky}, \citenamefont {O'Shaughnessy}, \citenamefont {Healy},\ and\
  \citenamefont {Shoemaker}}]{Pekowsky:2013ska}%
  \BibitemOpen
  \bibfield  {author} {\bibinfo {author} {\bibfnamefont {L.}~\bibnamefont
  {Pekowsky}}, \bibinfo {author} {\bibfnamefont {R.}~\bibnamefont
  {O'Shaughnessy}}, \bibinfo {author} {\bibfnamefont {J.}~\bibnamefont
  {Healy}}, \ and\ \bibinfo {author} {\bibfnamefont {D.}~\bibnamefont
  {Shoemaker}},\ }\bibfield  {title} {\enquote {\bibinfo {title} {{Comparing
  gravitational waves from nonprecessing and precessing black hole binaries in
  the corotating frame}},}\ }\href {\doibase 10.1103/PhysRevD.88.024040}
  {\bibfield  {journal} {\bibinfo  {journal} {Phys. Rev. D}\ }\textbf {\bibinfo
  {volume} {88}},\ \bibinfo {pages} {024040} (\bibinfo {year} {2013})},\
  \Eprint {http://arxiv.org/abs/1304.3176} {arXiv:1304.3176 [gr-qc]}
  \BibitemShut {NoStop}%
\bibitem [{\citenamefont {Ochsner}\ and\ \citenamefont
  {O'Shaughnessy}(2012)}]{Ochsner:2012dj}%
  \BibitemOpen
  \bibfield  {author} {\bibinfo {author} {\bibfnamefont {Evan}\ \bibnamefont
  {Ochsner}}\ and\ \bibinfo {author} {\bibfnamefont {Richard}\ \bibnamefont
  {O'Shaughnessy}},\ }\bibfield  {title} {\enquote {\bibinfo {title}
  {{Asymptotic frame selection for binary black hole spacetimes II:
  Post-Newtonian limit}},}\ }\href {\doibase 10.1103/PhysRevD.86.104037}
  {\bibfield  {journal} {\bibinfo  {journal} {Phys. Rev. D}\ }\textbf {\bibinfo
  {volume} {86}},\ \bibinfo {pages} {104037} (\bibinfo {year} {2012})},\
  \Eprint {http://arxiv.org/abs/1205.2287} {arXiv:1205.2287 [gr-qc]}
  \BibitemShut {NoStop}%
\bibitem [{\citenamefont {Nagar}\ \emph {et~al.}(2024)\citenamefont {Nagar},
  \citenamefont {Gamba}, \citenamefont {Rettegno}, \citenamefont {Fantini},\
  and\ \citenamefont {Bernuzzi}}]{Nagar:2024dzj}%
  \BibitemOpen
  \bibfield  {author} {\bibinfo {author} {\bibfnamefont {Alessandro}\
  \bibnamefont {Nagar}}, \bibinfo {author} {\bibfnamefont {Rossella}\
  \bibnamefont {Gamba}}, \bibinfo {author} {\bibfnamefont {Piero}\ \bibnamefont
  {Rettegno}}, \bibinfo {author} {\bibfnamefont {Veronica}\ \bibnamefont
  {Fantini}}, \ and\ \bibinfo {author} {\bibfnamefont {Sebastiano}\
  \bibnamefont {Bernuzzi}},\ }\bibfield  {title} {\enquote {\bibinfo {title}
  {{Effective-one-body waveform model for non-circularized, planar, coalescing
  black hole binaries: the importance of radiation reaction}},}\ }\href@noop {}
  {\  (\bibinfo {year} {2024})},\ \Eprint {http://arxiv.org/abs/2404.05288}
  {arXiv:2404.05288 [gr-qc]} \BibitemShut {NoStop}%
\bibitem [{\citenamefont {Bini}\ \emph {et~al.}(2019)\citenamefont {Bini},
  \citenamefont {Damour},\ and\ \citenamefont {Geralico}}]{Bini:2019nra}%
  \BibitemOpen
  \bibfield  {author} {\bibinfo {author} {\bibfnamefont {Donato}\ \bibnamefont
  {Bini}}, \bibinfo {author} {\bibfnamefont {Thibault}\ \bibnamefont {Damour}},
  \ and\ \bibinfo {author} {\bibfnamefont {Andrea}\ \bibnamefont {Geralico}},\
  }\bibfield  {title} {\enquote {\bibinfo {title} {{Novel approach to binary
  dynamics: application to the fifth post-Newtonian level}},}\ }\href {\doibase
  10.1103/PhysRevLett.123.231104} {\bibfield  {journal} {\bibinfo  {journal}
  {Phys. Rev. Lett.}\ }\textbf {\bibinfo {volume} {123}},\ \bibinfo {pages}
  {231104} (\bibinfo {year} {2019})},\ \Eprint
  {http://arxiv.org/abs/1909.02375} {arXiv:1909.02375 [gr-qc]} \BibitemShut
  {NoStop}%
\bibitem [{\citenamefont {Damour}\ and\ \citenamefont
  {Nagar}(2014)}]{Damour:2014sva}%
  \BibitemOpen
  \bibfield  {author} {\bibinfo {author} {\bibfnamefont {Thibault}\
  \bibnamefont {Damour}}\ and\ \bibinfo {author} {\bibfnamefont {Alessandro}\
  \bibnamefont {Nagar}},\ }\bibfield  {title} {\enquote {\bibinfo {title} {{New
  effective-one-body description of coalescing nonprecessing spinning
  black-hole binaries}},}\ }\href {\doibase 10.1103/PhysRevD.90.044018}
  {\bibfield  {journal} {\bibinfo  {journal} {Phys.Rev.}\ }\textbf {\bibinfo
  {volume} {D90}},\ \bibinfo {pages} {044018} (\bibinfo {year} {2014})},\
  \Eprint {http://arxiv.org/abs/1406.6913} {arXiv:1406.6913 [gr-qc]}
  \BibitemShut {NoStop}%
\bibitem [{\citenamefont {Nagar}\ \emph {et~al.}(2018)\citenamefont {Nagar}
  \emph {et~al.}}]{Nagar:2018zoe}%
  \BibitemOpen
  \bibfield  {author} {\bibinfo {author} {\bibfnamefont {Alessandro}\
  \bibnamefont {Nagar}} \emph {et~al.},\ }\bibfield  {title} {\enquote
  {\bibinfo {title} {{Time-domain effective-one-body gravitational waveforms
  for coalescing compact binaries with nonprecessing spins, tides and self-spin
  effects}},}\ }\href {\doibase 10.1103/PhysRevD.98.104052} {\bibfield
  {journal} {\bibinfo  {journal} {Phys. Rev.}\ }\textbf {\bibinfo {volume}
  {D98}},\ \bibinfo {pages} {104052} (\bibinfo {year} {2018})},\ \Eprint
  {http://arxiv.org/abs/1806.01772} {arXiv:1806.01772 [gr-qc]} \BibitemShut
  {NoStop}%
\bibitem [{\citenamefont {Andrade}\ \emph {et~al.}(2023)\citenamefont {Andrade}
  \emph {et~al.}}]{Andrade:2023trh}%
  \BibitemOpen
  \bibfield  {author} {\bibinfo {author} {\bibfnamefont {Tomas}\ \bibnamefont
  {Andrade}} \emph {et~al.},\ }\bibfield  {title} {\enquote {\bibinfo {title}
  {{Towards numerical-relativity informed effective-one-body waveforms for
  dynamical capture black hole binaries}},}\ }\href@noop {} {\  (\bibinfo
  {year} {2023})},\ \Eprint {http://arxiv.org/abs/2307.08697} {arXiv:2307.08697
  [gr-qc]} \BibitemShut {NoStop}%
\bibitem [{\citenamefont {Albanesi}\ \emph {et~al.}(2021)\citenamefont
  {Albanesi}, \citenamefont {Nagar},\ and\ \citenamefont
  {Bernuzzi}}]{Albanesi:2021rby}%
  \BibitemOpen
  \bibfield  {author} {\bibinfo {author} {\bibfnamefont {Simone}\ \bibnamefont
  {Albanesi}}, \bibinfo {author} {\bibfnamefont {Alessandro}\ \bibnamefont
  {Nagar}}, \ and\ \bibinfo {author} {\bibfnamefont {Sebastiano}\ \bibnamefont
  {Bernuzzi}},\ }\bibfield  {title} {\enquote {\bibinfo {title} {{Effective
  one-body model for extreme-mass-ratio spinning binaries on eccentric
  equatorial orbits: Testing radiation reaction and waveform}},}\ }\href
  {\doibase 10.1103/PhysRevD.104.024067} {\bibfield  {journal} {\bibinfo
  {journal} {Phys. Rev. D}\ }\textbf {\bibinfo {volume} {104}},\ \bibinfo
  {pages} {024067} (\bibinfo {year} {2021})},\ \Eprint
  {http://arxiv.org/abs/2104.10559} {arXiv:2104.10559 [gr-qc]} \BibitemShut
  {NoStop}%
\bibitem [{\citenamefont {Nagar}\ and\ \citenamefont
  {Rettegno}(2019)}]{Nagar:2018gnk}%
  \BibitemOpen
  \bibfield  {author} {\bibinfo {author} {\bibfnamefont {Alessandro}\
  \bibnamefont {Nagar}}\ and\ \bibinfo {author} {\bibfnamefont {Piero}\
  \bibnamefont {Rettegno}},\ }\bibfield  {title} {\enquote {\bibinfo {title}
  {{Efficient effective one body time-domain gravitational waveforms}},}\
  }\href {\doibase 10.1103/PhysRevD.99.021501} {\bibfield  {journal} {\bibinfo
  {journal} {Phys. Rev.}\ }\textbf {\bibinfo {volume} {D99}},\ \bibinfo {pages}
  {021501} (\bibinfo {year} {2019})},\ \Eprint
  {http://arxiv.org/abs/1805.03891} {arXiv:1805.03891 [gr-qc]} \BibitemShut
  {NoStop}%
\bibitem [{\citenamefont {Gamba}\ \emph {et~al.}(2021)\citenamefont {Gamba},
  \citenamefont {Bernuzzi},\ and\ \citenamefont {Nagar}}]{Gamba:2020ljo}%
  \BibitemOpen
  \bibfield  {author} {\bibinfo {author} {\bibfnamefont {Rossella}\
  \bibnamefont {Gamba}}, \bibinfo {author} {\bibfnamefont {Sebastiano}\
  \bibnamefont {Bernuzzi}}, \ and\ \bibinfo {author} {\bibfnamefont
  {Alessandro}\ \bibnamefont {Nagar}},\ }\bibfield  {title} {\enquote {\bibinfo
  {title} {{Fast, faithful, frequency-domain effective-one-body waveforms for
  compact binary coalescences}},}\ }\href {\doibase
  10.1103/PhysRevD.104.084058} {\bibfield  {journal} {\bibinfo  {journal}
  {Phys. Rev. D}\ }\textbf {\bibinfo {volume} {104}},\ \bibinfo {pages}
  {084058} (\bibinfo {year} {2021})},\ \Eprint
  {http://arxiv.org/abs/2012.00027} {arXiv:2012.00027 [gr-qc]} \BibitemShut
  {NoStop}%
\bibitem [{\citenamefont {Shaikh}\ \emph {et~al.}(2023)\citenamefont {Shaikh},
  \citenamefont {Varma}, \citenamefont {Pfeiffer}, \citenamefont
  {Ramos-Buades},\ and\ \citenamefont {van~de Meent}}]{Shaikh:2023ypz}%
  \BibitemOpen
  \bibfield  {author} {\bibinfo {author} {\bibfnamefont {Md~Arif}\ \bibnamefont
  {Shaikh}}, \bibinfo {author} {\bibfnamefont {Vijay}\ \bibnamefont {Varma}},
  \bibinfo {author} {\bibfnamefont {Harald~P.}\ \bibnamefont {Pfeiffer}},
  \bibinfo {author} {\bibfnamefont {Antoni}\ \bibnamefont {Ramos-Buades}}, \
  and\ \bibinfo {author} {\bibfnamefont {Maarten}\ \bibnamefont {van~de
  Meent}},\ }\bibfield  {title} {\enquote {\bibinfo {title} {{Defining
  eccentricity for gravitational wave astronomy}},}\ }\href {\doibase
  10.1103/PhysRevD.108.104007} {\bibfield  {journal} {\bibinfo  {journal}
  {Phys. Rev. D}\ }\textbf {\bibinfo {volume} {108}},\ \bibinfo {pages}
  {104007} (\bibinfo {year} {2023})},\ \Eprint
  {http://arxiv.org/abs/2302.11257} {arXiv:2302.11257 [gr-qc]} \BibitemShut
  {NoStop}%
\bibitem [{\citenamefont {Schnittman}(2004)}]{Schnittman:2004vq}%
  \BibitemOpen
  \bibfield  {author} {\bibinfo {author} {\bibfnamefont {Jeremy~D.}\
  \bibnamefont {Schnittman}},\ }\bibfield  {title} {\enquote {\bibinfo {title}
  {{Spin-orbit resonance and the evolution of compact binary systems}},}\
  }\href {\doibase 10.1103/PhysRevD.70.124020} {\bibfield  {journal} {\bibinfo
  {journal} {Phys. Rev. D}\ }\textbf {\bibinfo {volume} {70}},\ \bibinfo
  {pages} {124020} (\bibinfo {year} {2004})},\ \Eprint
  {http://arxiv.org/abs/astro-ph/0409174} {arXiv:astro-ph/0409174} \BibitemShut
  {NoStop}%
\bibitem [{\citenamefont {Gamba}\ \emph {et~al.}()\citenamefont {Gamba} \emph
  {et~al.}}]{Gamba:2024}%
  \BibitemOpen
  \bibfield  {author} {\bibinfo {author} {\bibfnamefont {Rossella}\
  \bibnamefont {Gamba}} \emph {et~al.},\ }\bibfield  {title} {\enquote
  {\bibinfo {title} {Highly accurate simulations of eccentric, non-planar
  binary black holes systems},}\ }\href@noop {} {\ }\BibitemShut {NoStop}%
\bibitem [{\citenamefont {Harry}\ \emph {et~al.}(2016)\citenamefont {Harry},
  \citenamefont {Privitera}, \citenamefont {Bohé},\ and\ \citenamefont
  {Buonanno}}]{Harry:2016ijz}%
  \BibitemOpen
  \bibfield  {author} {\bibinfo {author} {\bibfnamefont {Ian}\ \bibnamefont
  {Harry}}, \bibinfo {author} {\bibfnamefont {Stephen}\ \bibnamefont
  {Privitera}}, \bibinfo {author} {\bibfnamefont {Alejandro}\ \bibnamefont
  {Bohé}}, \ and\ \bibinfo {author} {\bibfnamefont {Alessandra}\ \bibnamefont
  {Buonanno}},\ }\bibfield  {title} {\enquote {\bibinfo {title} {{Searching for
  Gravitational Waves from Compact Binaries with Precessing Spins}},}\ }\href
  {\doibase 10.1103/PhysRevD.94.024012} {\bibfield  {journal} {\bibinfo
  {journal} {Phys. Rev.}\ }\textbf {\bibinfo {volume} {D94}},\ \bibinfo {pages}
  {024012} (\bibinfo {year} {2016})},\ \Eprint
  {http://arxiv.org/abs/1603.02444} {arXiv:1603.02444 [gr-qc]} \BibitemShut
  {NoStop}%
\bibitem [{\citenamefont {Harry}\ \emph {et~al.}(2018)\citenamefont {Harry},
  \citenamefont {Calderón~Bustillo},\ and\ \citenamefont
  {Nitz}}]{Harry:2017weg}%
  \BibitemOpen
  \bibfield  {author} {\bibinfo {author} {\bibfnamefont {Ian}\ \bibnamefont
  {Harry}}, \bibinfo {author} {\bibfnamefont {Juan}\ \bibnamefont
  {Calderón~Bustillo}}, \ and\ \bibinfo {author} {\bibfnamefont {Alex}\
  \bibnamefont {Nitz}},\ }\bibfield  {title} {\enquote {\bibinfo {title}
  {{Searching for the full symphony of black hole binary mergers}},}\ }\href
  {\doibase 10.1103/PhysRevD.97.023004} {\bibfield  {journal} {\bibinfo
  {journal} {Phys. Rev.}\ }\textbf {\bibinfo {volume} {D97}},\ \bibinfo {pages}
  {023004} (\bibinfo {year} {2018})},\ \Eprint
  {http://arxiv.org/abs/1709.09181} {arXiv:1709.09181 [gr-qc]} \BibitemShut
  {NoStop}%
\bibitem [{Sn:()}]{Sn:advLIGO}%
  \BibitemOpen
  \href@noop {} {\enquote {\bibinfo {title} {{LIGO Document T0900288-v3}},}\
  }\bibinfo {howpublished}
  {\url{https://dcc.ligo.org/cgi-bin/DocDB/ShowDocument?docid=2974}},\ \bibinfo
  {note} {{Advanced LIGO anticipated sensitivity curves}}\BibitemShut {NoStop}%
\bibitem [{\citenamefont {Mac~Uilliam}\ \emph {et~al.}(2024)\citenamefont
  {Mac~Uilliam}, \citenamefont {Akcay},\ and\ \citenamefont
  {Thompson}}]{MacUilliam:2024oif}%
  \BibitemOpen
  \bibfield  {author} {\bibinfo {author} {\bibfnamefont {Jake}\ \bibnamefont
  {Mac~Uilliam}}, \bibinfo {author} {\bibfnamefont {Sarp}\ \bibnamefont
  {Akcay}}, \ and\ \bibinfo {author} {\bibfnamefont {Jonathan~E.}\ \bibnamefont
  {Thompson}},\ }\bibfield  {title} {\enquote {\bibinfo {title} {{A Survey of
  Four Precessing Waveform Models for Binary Black Hole Systems}},}\
  }\href@noop {} {\  (\bibinfo {year} {2024})},\ \Eprint
  {http://arxiv.org/abs/2402.06781} {arXiv:2402.06781 [gr-qc]} \BibitemShut
  {NoStop}%
\bibitem [{\citenamefont {Memmesheimer}\ \emph {et~al.}(2004)\citenamefont
  {Memmesheimer}, \citenamefont {Gopakumar},\ and\ \citenamefont
  {Schaefer}}]{Memmesheimer:2004cv}%
  \BibitemOpen
  \bibfield  {author} {\bibinfo {author} {\bibfnamefont {Raoul-Martin}\
  \bibnamefont {Memmesheimer}}, \bibinfo {author} {\bibfnamefont {Achamveedu}\
  \bibnamefont {Gopakumar}}, \ and\ \bibinfo {author} {\bibfnamefont {Gerhard}\
  \bibnamefont {Schaefer}},\ }\bibfield  {title} {\enquote {\bibinfo {title}
  {{Third post-Newtonian accurate generalized quasi-Keplerian parametrization
  for compact binaries in eccentric orbits}},}\ }\href {\doibase
  10.1103/PhysRevD.70.104011} {\bibfield  {journal} {\bibinfo  {journal} {Phys.
  Rev. D}\ }\textbf {\bibinfo {volume} {70}},\ \bibinfo {pages} {104011}
  (\bibinfo {year} {2004})},\ \Eprint {http://arxiv.org/abs/gr-qc/0407049}
  {arXiv:gr-qc/0407049} \BibitemShut {NoStop}%
\end{thebibliography}
\end{document}